%% file: status.tex
\documentclass[11pt,a4paper]{article}
\usepackage{jheppub}
\usepackage{dcolumn}
\usepackage{microtype}
\usepackage{booktabs}
\usepackage{dcolumn}

\input{Commands.tex}

\begin{document}

\begin{flushright}
    IFIC/13-46 \\
     
\end{flushright}

\title{Present status and future perspectives\\ of the NEXT experiment}

\input{src2/Authors}

\abstract{NEXT is an experiment dedicated to neutrinoless double beta decay searches in xenon.  The detector is a TPC, holding 100 kg of high-pressure xenon enriched in the $^{136}$Xe isotope. It is under construction in the Laboratorio Subterr\'aneo de Canfranc in Spain, and it will begin operations in 2015. The NEXT detector concept provides an energy resolution better than 1\% FWHM and a topological signal that can be used to reduce the background. Furthermore, the NEXT technology can be extrapolated to a 1-ton scale experiment.}

\maketitle

\section{Introduction}
\input{src2/Intro.tex}

 This paper is organised as follows. Section \ref{sec:physics} describes the physics of Majorana neutrinos and \bbonu\ searches, reviews xenon experiments, and discusses their discovery potential. The NEXT detector is described with some detail in Section
\ref{sec:next}, while Section \ref{sec:bkgm} gives details of the NEXT background model. Section
\ref{sec:EL} describes our two electroluminescence prototypes, NEXT-DEMO and NEXT-DBDM. Finally, conclusions are presented in Section \ref{sec:conclu}.

\section{The physics of NEXT}
\label{sec:physics}
\input{src2/Physics.tex}

\section{The NEXT detector}
\label{sec:next}
\input{src2/NextDetector.tex}

\section{NEXT background model}
\label{sec:bkgm}
\input{src2/NextBKGM.tex}

\section{The NEXT EL  prototypes}
\label{sec:EL}
\input{src2/NextDemo.tex}

\input{src2/NextDBDM.tex}

\section{Conclusions}
\label{sec:conclu}

In this paper, the current status and future prospects of the NEXT project, and in particular of the NEXT-100 experiment at LSC has been described. NEXT has a large discovery potential, and the capability to offer a technique that can be extrapolated, at a very competitive cost, to the ton scale. 

The collaboration has developed the advanced technology of high-pressure chambers, and, in particular, NEXT-DEMO is the first large-scale HPXe TPC using the EL technology. The innovations include the use of SiPMs for the tracking plane, a technology that was in its infancy only five years ago. 

The collaboration has published results that illustrate the physics case and demonstrate the good performance of the EL technology. Very good energy resolution, better than 1\% FWHM, has been measured and the topological signature of electrons has been clearly established. 

The NEXT-100 detector is now in the initial stages of construction and is expected to be taking data in 2015. The scientific opportunity is extraordinary and the ratio costs to scientific impact is relatively modest. In addition to contributing to the current scientific development, NEXT could be the springboard for a future, large scale experiment that could finally demonstrate that Ettore Majorana insight was correct. 

\acknowledgments
This work was supported by the following agencies and institutions:
the Ministerio de Econom\'ia y Competitividad of Spain under grants
CONSOLIDER-Ingenio 2010 CSD2008-0037 (CUP), FPA2009-13697-C04-04 and FIS2012-37947-C04;
the Director, Office of Science, Office of Basic Energy Sciences, of
the US Department of Energy under contract no.\ DE-AC02-05CH11231; and
the Portuguese FCT and FEDER through the program COMPETE, projects
PTDC/FIS/103860/2008 and PTDC/FIS/112272/2009. J.~Renner (LBNL) acknowledges the support of a
US DOE NNSA Stewardship Science Graduate Fellowship under contract
no.\ DE-FC52-08NA28752.

\bibliographystyle{JHEP}
\bibliography{references}

\end{document}

%% file: Commands.tex
\newcommand{\bb}{\ensuremath{\beta\beta}}
\newcommand{\bbonu}{\ensuremath{\beta\beta0\nu}}
\newcommand{\bbtnu}{\ensuremath{\beta\beta2\nu}}

\newcommand{\mbb}{\ensuremath{m_{\beta\beta}}}

\newcommand{\ckky}{\ensuremath{\rm counts/(keV \cdot kg \cdot y)}}

\newcommand{\Qbb}{\ensuremath{Q_{\beta\beta}}}

\newcommand{\CS}{\ensuremath{^{137}}Cs}

\newcommand{\NA}{\ensuremath{^{22}}Na}

\newcommand{\Pb}{\ensuremath{^{208}}Pb}

\newcommand{\Po}{\ensuremath{^{214}}Po}

\newcommand{\XE}{\ensuremath{{}^{136}\rm Xe}}
\newcommand{\GE}{\ensuremath{{}^{76}\rm Ge}}

\newcommand{\TL}{\ensuremath{{}^{208}\rm{Tl}}}

\newcommand{\BI}{\ensuremath{{}^{214}}Bi}

%% file: src2/Authors.tex
\collaboration{The NEXT Collaboration}

\author[a,1]{J.J.~G\'omez-Cadenas,\note{Spokesperson (gomez@mail.cern.ch)}}
\author[a]{V.~\'Alvarez,}
\author[b]{F.I.G.~Borges,}
\author[a]{S.~C\'arcel,}
\author[c]{J.~Castel,}
\author[c]{S.~Cebri\'an,}
\author[a]{A.~Cervera,}
\author[b]{C.A.N.~Conde}
\author[c]{T.~Dafni,}
\author[b]{T.H.V.T.~Dias,}
\author[a]{J.~D\'iaz,}
\author[d]{M.~Egorov,}
\author[e]{R.~Esteve,}
\author[f]{P.~Evtoukhovitch,}
\author[b]{L.M.P.~Fernandes,}
\author[a,2]{P.~Ferrario,\note{Corresponding author}}
\author[g]{A.L.~Ferreira}
\author[b]{E.D.C.~Freitas}
\author[d]{V.M.~Gehman,}
\author[a]{A.~Gil,}
\author[d]{A.~Goldschmidt,}
\author[c]{H.~G\'omez,}
\author[c]{D.~Gonz\'alez-D\'iaz,}
\author[h]{R.M.~Guti\'errez,}
\author[i]{J.~Hauptman,}
\author[j]{J.A.~Hernando Morata,}
\author[c]{D.C.~Herrera,}
\author[c]{F.J.~Iguaz,}
\author[c]{I.G.~Irastorza,}
\author[h]{M.A.~Jinete,}
\author[k]{L.~Labarga,}
\author[a]{A.~Laing,}
\author[a]{I.~Liubarsky,}
\author[b]{J.A.M.~Lopes,}
\author[a]{D.~Lorca,}
\author[h]{M.~Losada,}
\author[c]{G.~Luz\'on,}
\author[e]{A.~Mar\'i,}
\author[a]{J.~Mart\'in-Albo}
\author[a]{A.~Mart\'inez,}
\author[d]{T.~Miller,}
\author[f]{A.~Moiseenko,}
\author[a]{F.~Monrabal,}
\author[b]{C.M.B.~Monteiro,}
\author[e]{F.J.~Mora,}
\author[g]{L.M. Moutinho,}
\author[a]{J.~Mu\~noz~Vidal,}
\author[b]{H.~Natal~da~Luz,}
\author[h]{G.~Navarro,}
\author[a]{M.~Nebot-Guinot,}
\author[d]{D.~Nygren,}
\author[d]{C.A.B.~Oliveira,}
\author[l]{R.~Palma,}
\author[m]{J.~P\'erez,}
\author[l]{J.L.~P\'erez~Aparicio,}
\author[d]{J.~Renner,}
\author[n]{L.~Ripoll,}
\author[c]{A.~Rodr\'iguez,}
\author[a]{J.~Rodr\'iguez,}
\author[b]{F.P.~Santos,}
\author[b]{J.M.F.~dos~Santos,}
\author[c]{L.~Segu\'i,}
\author[a]{L.~Serra,}
\author[d]{D.~Shuman,}
\author[a]{A.~Sim\'on,}
\author[o]{C.~Sofka,}
\author[a]{M.~Sorel,}
\author[e]{J.F.~Toledo,}
\author[c]{A.~Tom\'as,}
\author[n]{J.~Torrent,}
\author[f]{Z.~Tsamalaidze,}
\author[g]{J.F.C.A.~Veloso,}
\author[c]{J.A.~Villar,}
\author[o]{R.~Webb,}
\author[o]{J.T.~White,}
\author[a]{N.~Yahlali}

\emailAdd{paola.ferrario@ific.uv.es}
\affiliation[a]{
Instituto de F\'isica Corpuscular (IFIC), CSIC \& Universitat de Val\`encia\\
Calle Catedr\'atico Jos\'e Beltr\'an, 2, 46980 Paterna, Valencia, Spain}
\affiliation[b]{
Departamento de Fisica, Universidade de Coimbra\\
Rua Larga, 3004-516 Coimbra, Portugal}
\affiliation[c]{
Laboratorio de F\'isica Nuclear y Astropart\'iculas, Universidad de Zaragoza\\ 
Calle Pedro Cerbuna, 12, 50009 Zaragoza, Spain}
\affiliation[d]{
Lawrence Berkeley National Laboratory (LBNL)\\
1 Cyclotron Road, Berkeley, California 94720, USA}
\affiliation[e]{
Instituto de Instrumentaci\'on para Imagen Molecular (I3M), Universitat Polit\`ecnica de Val\`encia\\ 
Camino de Vera, s/n, Edificio 8B, 46022 Valencia, Spain}
\affiliation[f]{
Joint Institute for Nuclear Research (JINR)\\
Joliot-Curie 6, 141980 Dubna, Russia}
\affiliation[g]{
Institute of Nanostructures, Nanomodelling and Nanofabrication (i3N), Universidade de Aveiro\\
Campus de Santiago, 3810-193 Aveiro, Portugal}
\affiliation[h]
{Centro de Investigaciones, Universidad Antonio Nari\~no\\ 
Carretera 3 este No.\ 47A-15, Bogot\'a, Colombia}
\affiliation[i]{
Department of Physics and Astronomy, Iowa State University\\
12 Physics Hall, Ames, Iowa 50011-3160, USA}
\affiliation[j]{
Instituto Gallego de F\'isica de Altas Energ\'ias (IGFAE), Univ.\ de Santiago de Compostela\\
Campus sur, R\'ua Xos\'e Mar\'ia Su\'arez N\'u\~nez, s/n, 15782 Santiago de Compostela, Spain}
\affiliation[k]{
Departamento de F\'isica Te\'orica, Universidad Aut\'onoma de Madrid\\
Campus de Cantoblanco, 28049 Madrid, Spain}
\affiliation[l]{
Dpto.\ de Mec\'anica de Medios Continuos y Teor\'ia de Estructuras, Univ.\ Polit\`ecnica de Val\`encia\\
Camino de Vera, s/n, 46071 Valencia, Spain}
\affiliation[m]{
Instituto de F\'isica Te\'orica (IFT), UAM/CSIC\\
Campus de Cantoblanco, 28049 Madrid, Spain}
\affiliation[n]{
Escola Polit\`ecnica Superior, Universitat de Girona\\
Av.~Montilivi, s/n, 17071 Girona, Spain}
\affiliation[o]{
Department of Physics and Astronomy, Texas A\&M University\\
College Station, Texas 77843-4242, USA}


%% file: src2/Intro.tex
This article presents the current status and future prospects of the {\em Neutrino Experiment with a Xenon TPC} ({\bf NEXT})\footnote{\url{http://next.ific.uv.es/next}}. The primary goal of the project is the construction, commissioning and operation of the NEXT-100 detector, a high-pressure, xenon (HPXe) Time Projection Chamber (TPC). NEXT-100 will search for neutrinoless double beta decay  (\bbonu) events in \XE, using 100 kg of xenon enriched at 90\% in the isotope \XE.
The experiment will operate at the Canfranc Underground Laboratory (LSC), starting in 2015. The NEXT collaboration includes institutions from Spain, Portugal, USA, Russia and Colombia. 

The discovery potential of a HPXe TPC combines four desirable features that make it an almost-ideal experiment for \bbonu\ searches, namely: 
\begin{enumerate}
\item Excellent energy resolution (0.5--0.7\% FWHM in the region of interest).
\item A topological signature (the observation of the tracks of the two electrons).
\item A fully active, very radiopure apparatus of large mass.  
\item The capability of extending the technology to a ton-scale experiment..
\end{enumerate}

Currently, two xenon-based experiments, with a mass in the range of hundred kilograms, are dominating the field of \bbonu\ searches. These are: EXO-200 (a liquid xenon TPC) and KamLAND-Zen (a large, liquid scintillator calorimeter, where xenon is dissolved in the scintillator). NEXT features a better resolution and the extra handle of the identification of the two electrons, which could result in a discovery, in spite of a late start. If evidence is found by EXO-200 or KamLAND-Zen of the existence of a signal, NEXT would be ideally suited to confirm it in an unambiguous way, in particular given the discriminating power of the topological signature. 

The negative results of EXO-200 and  KamLAND-Zen \cite{Gando:2012zm} indicate that the effective neutrino mass (the quantity measured in \bb\ decays, as further discussed later in the text) must be smaller than 120-250 meV, where the mass range is due to uncertainties in the nuclear matrix elements.  On the other hand, recent measurements
of the cosmic microwave background (CMB) by the Planck experiment\cite{Ade:2013zuv,Haba:2013xwa} 
yield an upper limit for the sum of the three light neutrino masses of 230 meV. 
The latter result excludes most of the so-called degenerate spectrum, in which the three neutrino masses are relatively large. The current sensitivity of the \bbonu\ experiments is not enough to explore significantly the so-called inverse hierarchy, which require sensitivities to effective neutrino masses in the range of 20 meV (if Nature has chosen the so-called normal hierarchy as her preferred pattern for neutrino masses, the search for \bbonu\ processes becomes extremely difficult if not hopeless). It follows that the next generation of \bbonu\ experiments must improve their sensitivity by typically one order of magnitude in the effective neutrinos mass, or two orders of magnitude in the period of the \bbonu\ decay. This, in turn, requires to {\em increase by a factor 100} the exposure from the ``typical'' values of the current generation of experiments (thus going from $\sim$~100 kg per year to $\sim$~one ton per 10 years), while at the same time {\em decreasing by a factor 100} the residual backgrounds (e.g., going from few events per 100 kg to $\sim$~0.1 events per ton). 

This tremendous challenge requires a detector capable to deploy a large source mass of pure isotope at a reasonable cost. Currently, only xenon has demonstrated this capability. There is already more than one ton of enriched xenon in the world, owned by KamLAND-Zen (800 kg), EXO-200 (200 kg) and NEXT (100 kg). Furthermore, xenon detectors are fully active (the detection medium is the same as the isotope source) and scalable, being either TPCs (EXO-200, NEXT) or scintillating calorimeters with the xenon dissolved in the scintillator (KamLAND-Zen). Therefore, it is highly probable that the next generation of \bbonu\ experiments will use xenon detectors.

The physics case of a HPXe TPC is outstanding, given the combination of excellent energy resolution and the high background rejection power that the observation of the two electrons provides. In that respect, NEXT-100 will serve also as springboard for the next generation of ton-scale, HPXe experiments.

%% file: src2/Physics.tex
\subsection{Majorana neutrinos and \bbonu\ experiments}
\label{sec:majorana}

Neutrinos, unlike the other Standard Model fermions, could be truly neutral particles, that is, indistinguishable from their antiparticles. The existence of such 
{\em Majorana neutrinos}\footnote{The term Majorana neutrino honours the italian physicist E. Majorana, who, in 1937, published a fundamental paper \cite{Majorana:1937vz} in which he was able ``to build a substantially novel theory for the particles deprived of electric charge''. Even if in those times the only known ``charge'' was the electric charge, Majorana implicitly assumed particles deprived of all the possible charges. In modern language, neutrinos should not have any lepton number. In other words, Majorana theory describes completely neutral spin 1/2 particles, which are identical to their antiparticles.} would imply the existence of a new energy scale of physics that characterises new dynamics beyond the Standard Model and provides the simplest explanation of why neutrino masses are so much lighter than the charged fermions. Understanding the new physics that underlies neutrino masses is one of the most important open questions in particle physics. It could have profound implications in our understanding of the mechanism of symmetry breaking, the origin of mass and the flavour problem \cite{Hernandez:2010mi}.

Furthermore, the existence of Majorana neutrinos would imply that lepton number is not a conserved quantum number. This, in turn, could be the origin of the matter-antimatter asymmetry observed in the Universe. The new physics related to neutrino masses could provide a new mechanism to generate that asymmetry, called leptogenesis. Although the predictions are model dependent, two essential ingredients must be confirmed experimentally: 1) the violation of lepton number and 2) CP violation in the lepton sector.

The only practical way to establish experimentally that neutrinos are their own antiparticle is the detection of neutrinoless double beta decay (\bbonu). This is a postulated very slow radioactive process in which a nucleus with $Z$ protons decays into a nucleus with $Z+2$ protons and the same mass number $A$, emitting two electrons that carry essentially all the energy released (\Qbb). The process can occur if and only if neutrinos are massive, Majorana particles.

Several underlying mechanisms --- involving physics beyond the Standard Model  --- have been proposed for \bbonu, the simplest one being the virtual exchange of light Majorana neutrinos. Assuming this to be the dominant process at low energies, the half-life of \bbonu\ can be written as:

\begin{equation}
(T^{0\nu}_{1/2})^{-1} = G^{0\nu} \ \big|M^{0\nu}\big|^{2} \ \mbb^{2} \, .
\label{eq:Tonu}
\end{equation}

In this equation, $G^{0\nu}$ is an exactly-calculable phase-space integral for the emission of two electrons; $M^{0\nu}$ is the nuclear matrix element (NME) of the transition, which has to be evaluated theoretically; and \mbb\ is the \emph{effective Majorana mass} of the electron neutrino:

\begin{equation}
\mbb = \Big| \sum_{i} U^{2}_{ei} \ m_{i} \Big| \, ,
\label{eq:mbb}
\end{equation}
where $m_{i}$ are the neutrino mass eigenstates and $U_{ei}$~ are elements of the neutrino mixing matrix. Therefore, a measurement of the decay rate of \bbonu\ would provide direct information on neutrino masses.

\begin{figure}[tbh]
\centering
\includegraphics[width=0.99\textwidth]{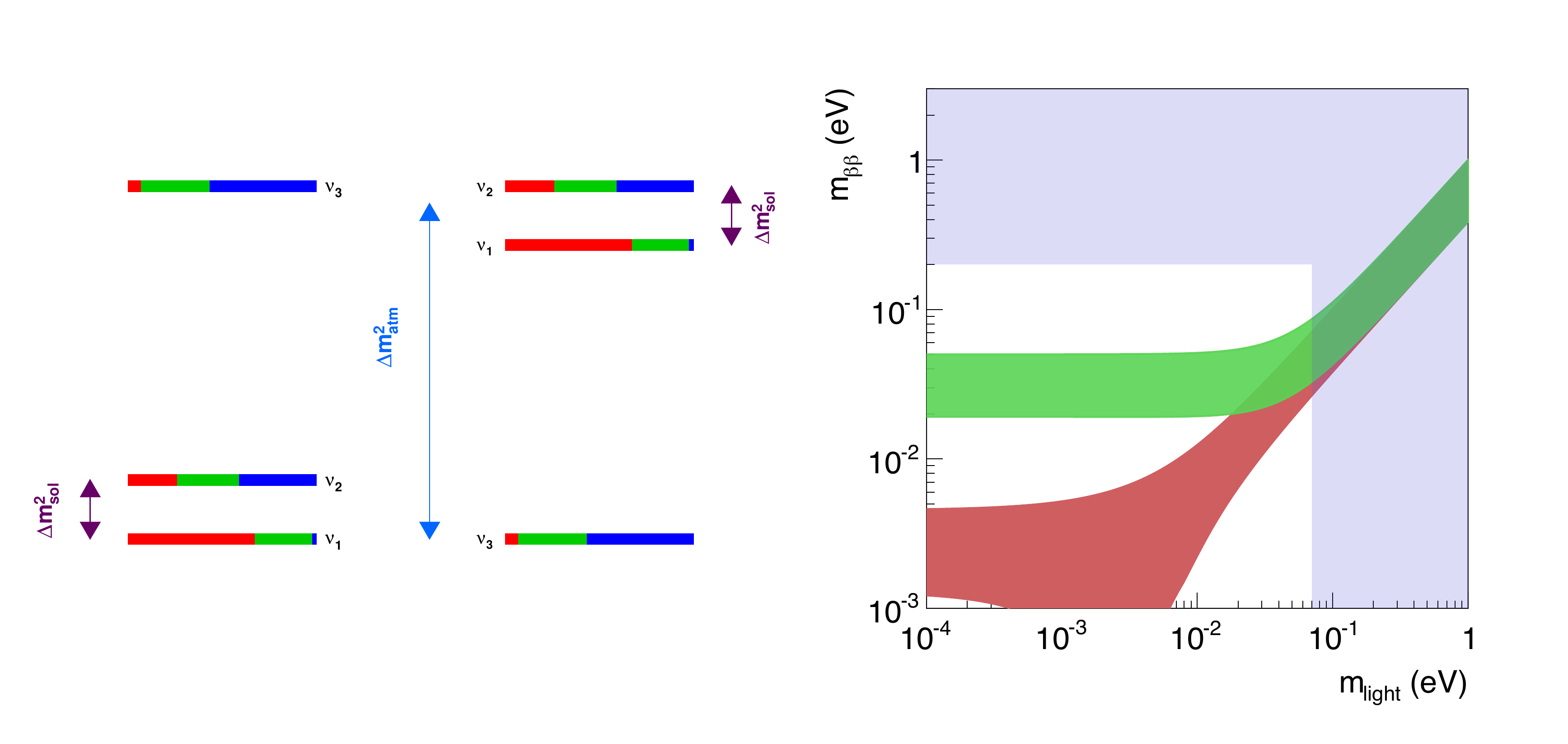}
\caption{\small The left panel shows the normal (left) and inverted (right) mass orderings. The electron, muon and tau flavor content of each neutrino mass eigenstate is shown via the red, green and blue fractions, respectively. The right panel shows the effective neutrino Majorana mass, \mbb, as a function of the lightest neutrino mass, $m_{\rm light}$. The green band corresponds to the inverse hierarchy of neutrino masses, whereas the red corresponds to the normal ordering. The upper bound on the lightest neutrino mass comes from cosmological bounds; the bound on the effective Majorana mass from \bbonu\ constraints.} \label{fig:numass_ordering}
\end{figure}

The relationship between \mbb\ and the actual neutrino masses $m_i$ is affected by the uncertainties in the measured oscillation parameters, the unknown neutrino mass ordering (normal or inverted), and the unknown phases in the neutrino mixing matrix. The current knowledge on neutrino masses and mixings provided by neutrino oscillation experiments is summarized in the left panel of Figure~\ref{fig:numass_ordering}. The diagram shows the two possible mass orderings that are compatible with neutrino oscillation data, with increasing neutrino masses from bottom to top. The relationship between \mbb\ and the lightest neutrino mass $m_{\rm light}$ (which is equal to $m_1$ or $m_3$ in the normal and inverted mass orderings, respectively) is illustrated in the right panel of Figure~\ref{fig:numass_ordering}.

The upper bound on the effective Majorana mass corresponds to the experimental constraint set by the Heidelberg-Moscow (HM) experiment, which was until very recently the most sensitive limit to the half-life of \bbonu: $T^{0\nu}_{1/2}(\GE) \ge 1.9\times10^{25}$ years at 90\% CL \cite{KlapdorKleingrothaus:2000sn}.  A subgroup of the HM experiment interpreted the data as {\em evidence} of a positive signal, with a best value for the half-life of $2.23\times10^{25}$ years, corresponding to an effective Majorana mass of about 300 meV \cite{KlapdorKleingrothaus:2006ff}. This claim  was very controversial and the experimental effort of the last decade has been focused in confirming or refuting it. The recent results from the KamLAND-Zen and EXO experiments have almost excluded the claim, and new data from other experiments such as GERDA, {\sc Majorana}  and CUORE will definitively settle the question shortly. 


\subsection{The current generation of \bbonu\ experiments}

The detectors used to search for \bbonu\ are designed, in general, to measure the energy of the radiation emitted by a \bbonu\ source. In a neutrinoless double beta decay, the sum of the kinetic energies of the two released electrons is always the same, and equal to the mass difference between the parent and the daughter nuclei: $\Qbb \equiv M(Z,A)-M(Z+2,A)$. However, due to the finite energy resolution of any detector, \bbonu\ events would be reconstructed within a given energy range centred around \Qbb\ and typically following a gaussian distribution. Other processes occurring in the detector can fall in that region of energies, thus becoming a background and compromising the sensitivity of the experiment  \cite{GomezCadenas:2010gs}.

All double beta decay experiments have to deal with an intrinsic background, the standard two-neutrino double beta decay (\bbtnu), that can only be suppressed by means of good energy resolution. Backgrounds of cosmogenic origin force the underground operation of the detectors. Natural radioactivity emanating from the detector materials and surroundings can easily overwhelm the signal peak, and hence careful selection of radiopure materials is essential. Additional experimental signatures, such as event topological information, that allow the distinction of signal and background, are a bonus to provide a robust result.

Besides energy resolution and control of backgrounds, several other factors such as detection efficiency and scalability to large masses must be taken into consideration in the design of a double beta decay experiment. The simultaneous optimisation of all these parameters is most of the time conflicting, if not impossible, and consequently many different experimental techniques have been proposed. In order to compare them, a figure of merit, the experimental sensitivity to \mbb, is normally used  \cite{GomezCadenas:2010gs}:
\begin{equation}
\mbb \propto \sqrt{1/\varepsilon}\, \left(\frac{b\ \delta E}{M\ t} \right)^{1/4}, \label{eq:sensi}
\end{equation}
where $\varepsilon$ is the signal detection efficiency, $M$ is the \bb\ isotope mass used in the experiment, $t$ is the data-taking time, $\delta E$ is the energy resolution and $b$ is the background rate in the region of interest around \Qbb\ (expressed in counts per kg of \bb\ isotope, year and keV).

 The status of the field has been the subject of several recent reviews  \cite{GomezCadenas:2011it, Cremonesi:2012av, Sarazin:2012ct, Giuliani:2012zu, Zuber:2006hv}. Among the on-going and planned experiments, many different experimental techniques are utilised, each with its pros and cons. The time--honored approach of emphasising energy resolution and detection efficiency is currently spearhead by germanium calorimeters like GERDA  \cite{Cattadori:2012fy}  and {\sc Majorana} \cite{Wilkerson:2012ga}, as well as tellurium bolometers such as CUORE \cite{Gorla:2012gd}. 
 
 A different and powerful approach proposes the use of xenon-based experiments. Two of them, EXO-200 \cite{Auger:2012gs} and KamLAND-Zen \cite{KamLANDZen:2012aa} are already operating, while NEXT \cite{Alvarez:2012haa} is in the initial stages of construction, and foresees to start taking data in 2015.

Other experiments that will operate in the next few years are the SuperNEMO demonstrator \cite{Sarazin:2012ct}, a tracker-calorimeter approach which provides a powerful topological signal (the observation of the two electrons emitted in a \bb\ decay) but is hard to extrapolate to larger masses (the demonstrator itself will have a mass of less than 10 kg of isotope), and SNO+, a large liquid scintillator calorimeter (the same approach as KamLAND-Zen), in which the isotope is dissolved in the scintillator. While neodymium has been the choice of the collaboration so far, in the last months $^{130}$Te has become the default option, due to a higher isotopic abundance and a lower rate of $\beta\beta 2\nu$ \cite{SNOplus}. As a drawback, the lower Q-value increases the potentially dangerous background, such as external gammas, thus limiting the fiducial volume.


\subsection{Xenon experiments} \label{sec:xenon}
Xenon is an almost-optimal element for \bbonu\ searches, featuring many desirable properties, both as a source and as a detector. It has two naturally-occurring isotopes that can decay via the \bb\ process, $^{134}$Xe ($\Qbb=825$~keV) and \XE\ ($\Qbb=2458$~keV). The latter, having a higher $Q$ value, is preferred since the decay rate is proportional to $\Qbb^{5}$ and the radioactive backgrounds are less abundant at higher energies. Moreover, the \bbtnu\ mode of \XE\ is slow ($\sim2.3\times10^{21}$~years \cite{KamLANDZen:2012aa, Albert:2013gpz}) and hence the experimental requirement for good energy resolution to suppress this particular background is less stringent than for other \bb\ sources. The process of isotopic enrichment in the isotope \XE\ is relatively simple and cheap compared to that of other \bb\ isotopes. Xenon has no long-lived radioactive isotopes and is intrinsically clean.

As a detector, xenon is a noble gas, therefore one can build a time projection chamber (TPC) with pure xenon as detection medium. Both a liquid xenon (LXe) TPC and a (high-pressure) gas (HPXe) TPC are suitable technologies, chosen by the EXO-200 and the NEXT experiment respectively. Nevertheless, energy resolution is much better in gas than in liquid, since, in its gaseous phase, xenon is characterized by a small Fano factor, meaning that the fluctuations in the ionization production have a sub-poissonian behaviour.  Being a noble gas, xenon can also be dissolved in liquid scintillator. This is the approach of the KamLAND-Zen experiment.

\subsubsection{KamLAND-Zen}
The KamLAND-Zen experiment is a modification of the well-known KamLAND neutrino detector  \cite{KamLANDZen:2012aa}. A transparent balloon, with a $\sim3$~m diameter, containing 13 tons of liquid scintillator loaded with 320 kg of xenon (enriched to 91\% in \XE) is suspended at the centre of KamLAND. The scintillation light generated by events occurring in the detector is recorded by an array of photomultipliers surrounding it. The position of the event vertex is reconstructed with a spatial resolution of about $15~\mathrm{cm}/\sqrt{E(\mathrm{MeV})}$. The energy resolution is $(6.6\pm0.3)\%/\sqrt{E(\mathrm{MeV})}$, that is, 9.9\% FWHM at the $Q$ value of \XE. The signal detection efficiency is $\sim0.42$ due to the tight fiducial cut introduced to reject backgrounds originating in the balloon. The achieved background rate in the energy window between 2.2~MeV and 3.0~MeV is $10^{-3}$~\ckky.

KamLAND-Zen
has searched for \bbonu\ events with an exposure of 89.5 kg$\cdot$year. They have published a limit on the half-life of \bbonu\ of $T_{1/2}^{0\nu}(\XE) > 1.9 \times 10^{25}$ years \cite{Gando:2012zm}. 

\subsubsection{EXO}

The EXO-200 detector \cite{Auger:2012gs} is a symmetric LXe TPC deploying 110 kg of xenon (enriched to 80.6\% in \XE).

In EXO-200, ionisation charges created in the xenon by charged particles drift under the influence of an electric field towards the two ends of the chamber. There, the charge is collected by a pair of crossed wire planes which measure its amplitude and transverse coordinates. Each end of the chamber includes also an array of avalanche photodiodes (APDs) to detect the 178-nm scintillation light. The sides of the chamber are covered with teflon sheets that act as VUV reflectors, improving the light collection. The simultaneous measurement of both the ionisation charge and scintillation light of the event may in principle allow to reach a detector energy resolution as low as 3.3\% FWHM at the \XE\ $Q$ value, for a sufficiently intense drift electric field \cite{Conti:2003av}. 

The EXO-200 detector achieves currently an energy resolution of 4\% FWHM at \Qbb, and a background rate measured in the \emph{region of interest} (ROI) of $ 1.5 \times 10^{-3}~\ckky$. The experiment has also searched for \bbonu\ events. The total exposure used for the published result is 32.5 kg$\cdot$year. They have published a limit on the half-life of \bbonu\ of $T_{1/2}^{0\nu}(\XE) > 1.6 \times 10^{25}$ years \cite{Auger:2012ar}.

The combination of the KamLAND-Zen and EXO results yields a limit $T_{1/2}^{0\nu}(\XE) > 3.4 \times 10^{25}$ years (120--250 meV, depending on the NME) \cite{Gando:2012zm}, which essentially excludes the long-standing claim of Klapdor-Kleingrothaus and collaborators \cite{Bergstrom:2012nx}.

\subsubsection{NEXT: a preview}

\begin{figure}
\centering
\includegraphics[width=8cm]{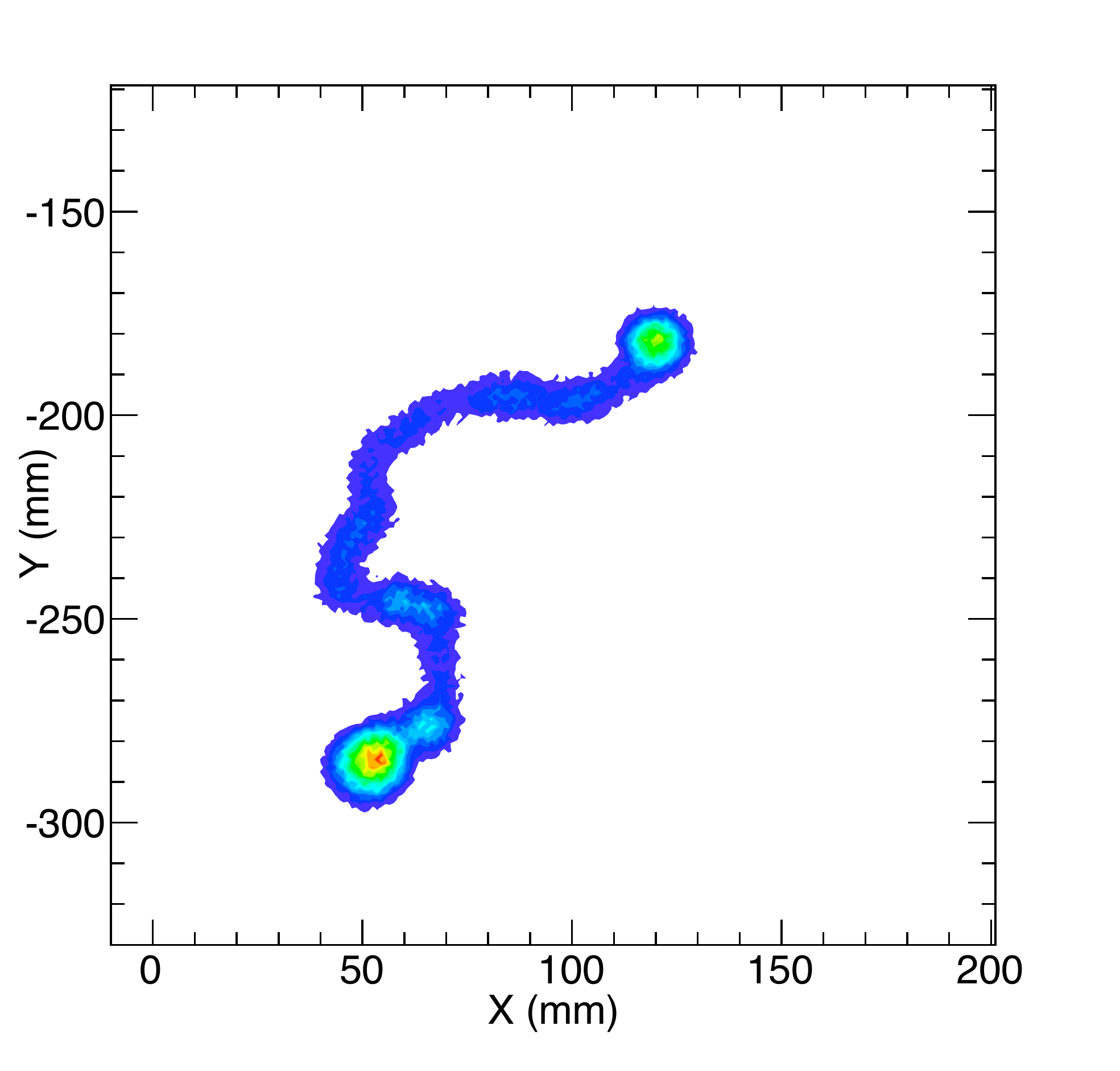}
\caption{Monte-Carlo simulation of a \XE\ \bbonu\ event in xenon gas at 10 bar: the ionization track, about 30 cm long, is tortuous because of multiple scattering, and has larger depositions or \emph{blobs} in both ends.}
\label{fig:track}
\end{figure}

The NEXT experiment will search for \bbonu\ in \XE\ using a high-pressure xenon gas (HPXe) time projection chamber (TPC) containing 100 kilogram of enriched gas, and called NEXT-100. Such a detector 
offers major advantages for the search of neutrinoless double beta decay, namely: 
\begin{itemize}
\item {\bf Excellent energy resolution}, with an intrinsic limit of about 0.3\% FWHM at \Qbb\ and 0.5--0.7\% demonstrated by the NEXT prototypes. For reference, the best energy resolution in the field is achieved by germanium experiments, such as GERDA and {\sc Majorana}, or bolometers such as CUORE, with typical resolutions in the range of 0.2\% FWHM at \Qbb. NEXT-100 targets a resolution which is a factor two worse than these, but a factor 8 (20) better than that of EXO (KamLAND-Zen), the other xenon experiments. 
\item {\bf Tracking capabilities} that provide a powerful topological signature to discriminate between signal (two electron tracks with a common vertex) and background (mostly, single electrons). Neutrinoless double beta decay events leave a distinctive topological signature in gaseous xenon: an ionization track, about 30 cm long at 10 bar, tortuous due to multiple scattering, and with larger energy depositions at both ends (see Figure \ref{fig:track}). The Gotthard experiment \cite{Luscher:1998sd}, consisting in a small xenon gas TPC (5.3 kg enriched to 68\% in \XE) operated at 5 bar, proved the effectiveness of such a signature to discriminate signal from background. The topological signature results in an expected background rate of the order of $5 \times 10^{-4}$ \ckky, improving EXO and KamLAND-Zen by a factor two, and the germanium calorimeters and tellurium bolometers by a factor five to ten. 
\item {\bf A fully active and homogeneous detector}, with no dead regions. Since 3-dimensional reconstruction is possible, events can be located in a fiducial region away from surfaces, where most of the background arises. This is a common feature with the two other xenon experiments. 
\item  {\bf Scalability} of the technique to larger masses, thanks to the fact that: a) xenon is noble gas, suitable for detection and with no intrinsic radioactivity; b) enriched xenon (in Xe-136) can be procured at a moderately limited cost,  for instance a factor 10 cheaper than the baseline $^{76}$Ge choice. This is also a common feature with the other two xenon experiments. 
\end{itemize}

\subsection{Discovery potential of xenon experiments}
\label{sec:disco}

\begin{figure}[!tbh]
\centering
\includegraphics[width=0.9\textwidth]{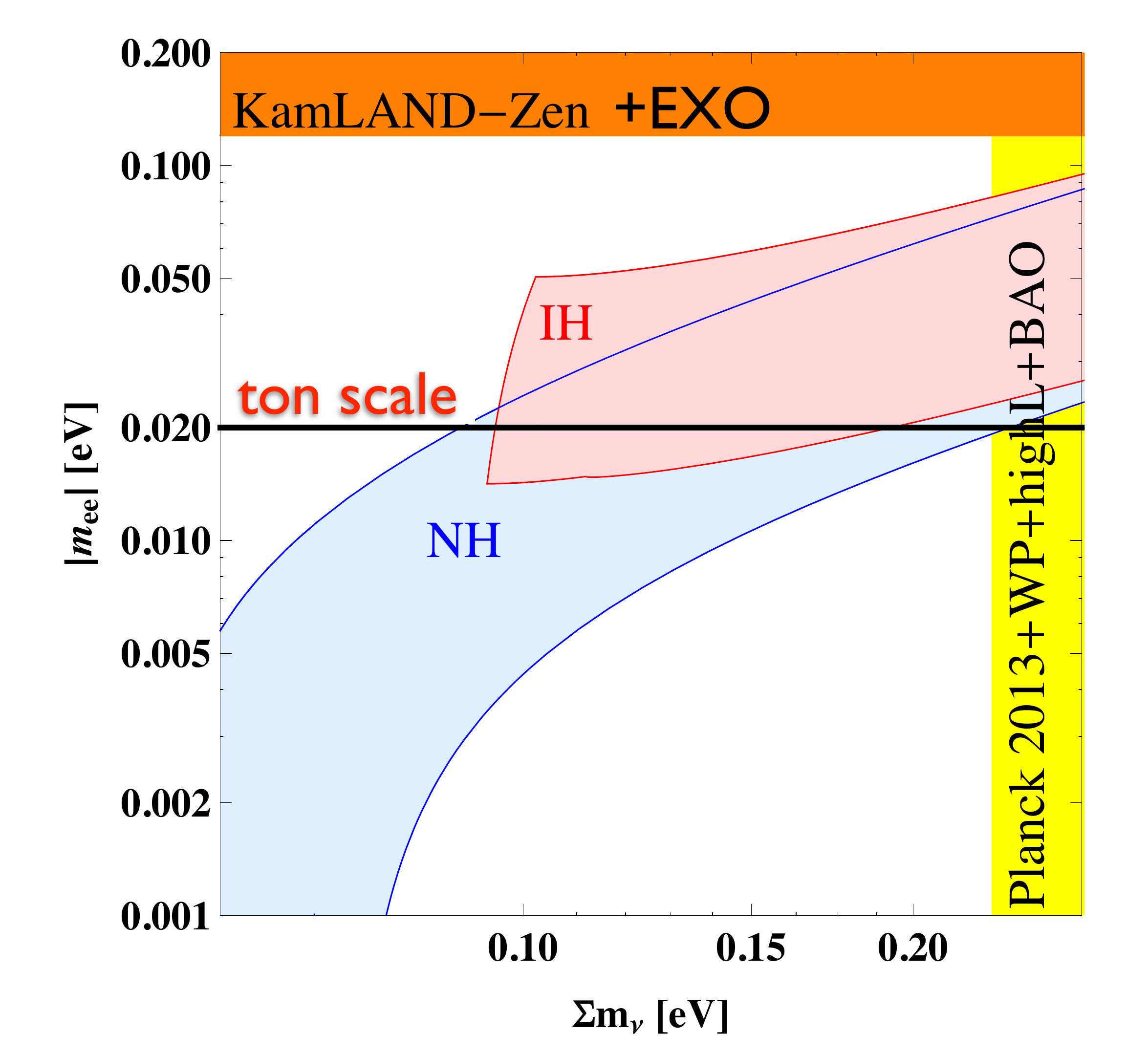}
\caption{The cosmological constraint on the sum of the neutrino mass derived from Planck data, together with the best limits from \bbonu\ experiments (KamLAND-Zen + EXO) and the limit that can be reached by the best experiments in the ton scale, in particular NEXT. Adapted 
from \cite{Haba:2013xwa}.} \label{fig.Planck}
\end{figure}

Recently, an upper limit for the sum of the three light neutrino masses has been reported by Planck measurements of the cosmic microwave background (CMB)\cite{Ade:2013zuv,Haba:2013xwa}:

\begin{equation}
\sum m_\nu= m_1 + m_2 +m_3 < 0.230 {\rm ~eV}~(95\% {\rm CL})
\label{eq:cosmo}
\end{equation}

Figure \ref{fig.Planck} shows the implications of such a measurement, when combined with the current limits from KamLAND-Zen and EXO. As it can be seen, the current sensitivity is not enough to explore significantly the inverse hierarchy, while Planck data exclude most of the so-called degenerate hierarchy. It follows that the next generation of \bbonu\ experiments must aim for extraordinary sensitivities to the effective neutrino mass. In particular, we will show that a sensitivity of 20 meV in \mbb\ is within the reach of a ton-scale HPXe detector. However, with luck, a discovery could be made before, if \mbb\ is near 100 meV. 

 \begin{table}
\centering
\caption{Experimental parameters of the three xenon-based double beta decay experiments: (a) total mass of \XE, $M$; (b) enrichment fraction $f$; (c) signal detection efficiency, $\varepsilon$; (d) energy resolution, $\delta E$, at the $Q$ value of \XE; and (e) background rate, $b$, in the region of interest around \Qbb\ expressed in \ckky\ (shortened as ckky)  \cite{GomezCadenas:2013ue}. } 
\label{tab:ExpParams}
\vspace{0.5cm}
\begin{tabular}{lccccc}
\hline
Experiment & $M$ (kg) & $f$ & $\varepsilon$ & $\delta E$ (\% FWHM) & $b$ ($10^{-3}$~ckky) \\
\hline
EXO-200 		& 110 & 81 & 52 & 3.9  & 1.5 \\
KamLAND-Zen & 330 & 91 & 62 & 9.9 & 1.0  \\
NEXT-100 	& 100 & 91 & 30  & 0.7  & 0.5 \\
\hline
\end{tabular}
\end{table}

\begin{figure}
\centering
\includegraphics[width=0.7\textwidth]{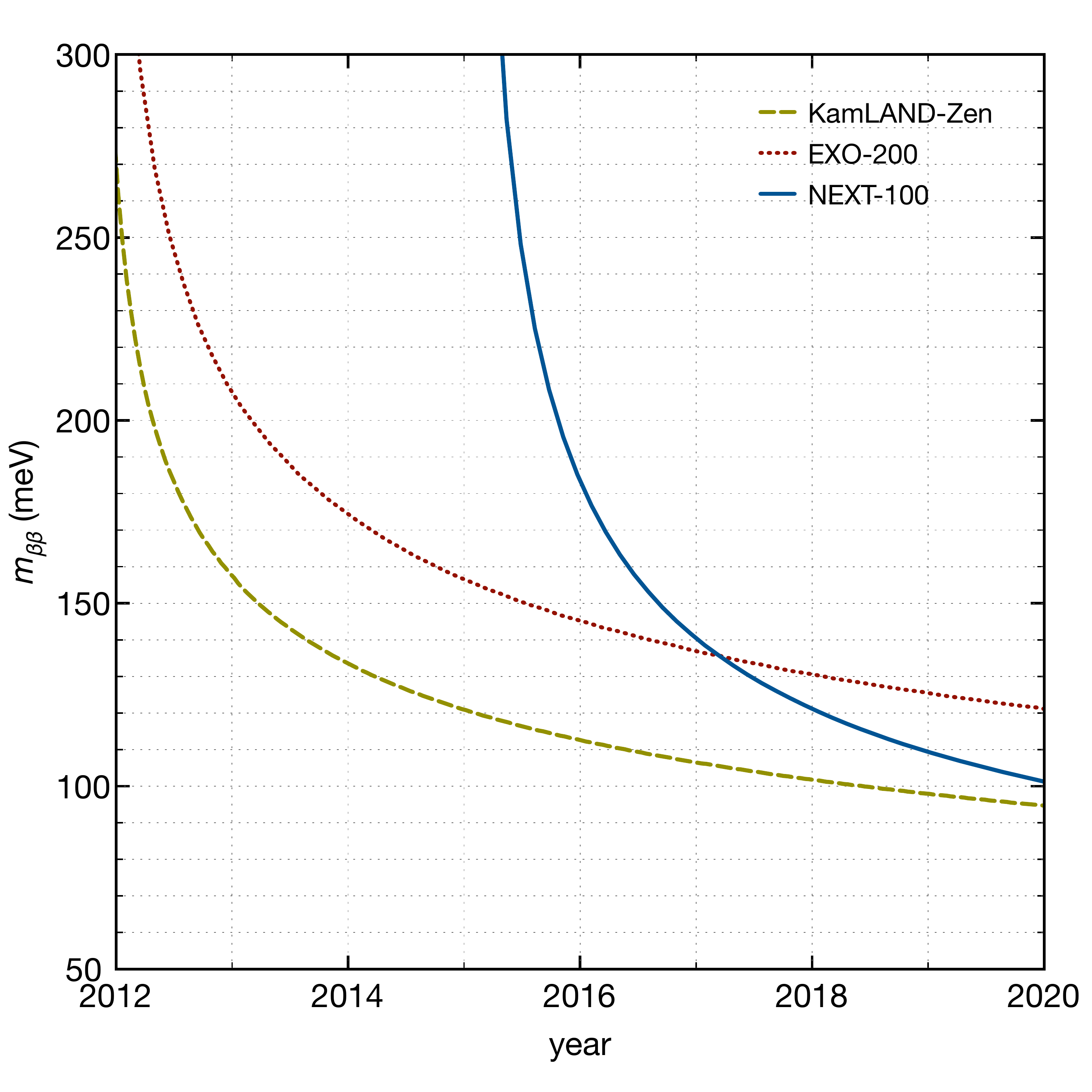}
\caption{Sensitivity of the three xenon experiments as a function of the running time, assuming the parameters described 
in Table \ref{tab:ExpParams}. We consider a run of 8 years for EXO-200 and KamLAND-Zen (2012 to 2020) and a run of 5 years for NEXT (2015 to 2020).} \label{fig.exoNext}
\end{figure}

In order to gain a feeling of the potential of the NEXT technology it is interesting to compare the experimental parameters of the three xenon experiments, which are collected in Table~\ref{tab:ExpParams}. The parameters of EXO-200 and KamLAND-Zen are those published by the collaborations \cite{Auger:2012ar,Gando:2012zm}. The resolution in NEXT corresponds to the most conservative result obtained by the NEXT prototypes  \cite{Alvarez:2012xda}, and the predicted background rate and efficiency comes from the full background model of the collaboration \cite{Alvarez:2012haa, MartinAlbo:2013ve}, assuming a conservative background level for the dominant source of background (the energy--plane PMTs). Notice that the background rate of all the experiments is very good. The HPXe technology offers less efficiency than the other two but a much better resolution. 

Figure \ref{fig.exoNext} shows the expected performance of the three experiments, assuming
the parameters described 
in Table \ref{tab:ExpParams} and the central value of the nuclear matrix elements described in \cite{GomezCadenas:2010gs}.
We consider a run of five years for NEXT (2015 to 2020) and a longer run of eight years for EXO-200 and KamLAND-Zen (2012 to 2020). A total dead-time of 10\% a year for all experiments is assumed.  It follows that all the three experiment will have a chance of making a discovery if \mbb\ is in the range of 100 meV. The fact that the experiments are based in different experimental techniques, with different systematic errors, makes their 
simultaneous running even more attractive. The combination of the three can reach a sensitivity of about 65 meV \cite{GomezCadenas:2013ue}.  Notice that, in spite of its late start, NEXT sensitivity can surpass that of the other xenon experiments.

\subsection{Towards the ton scale}

\begin{table}
\centering
\caption{Expected experimental parameters of the three xenon-based double beta decay technologies in a possible ton-scale experiment: (a) signal detection efficiency, $\varepsilon$; (b) energy resolution, $\delta E$, at the $Q$ value of \XE; and (c) background rate, $b$, in the region of interest around \Qbb\ expressed in \ckky. } \label{tab:FutureParams}
\vspace{0.5cm}
\begin{tabular}{lccc}
\hline
Experiment &  $\varepsilon$ & $\delta E$ (\% FWHM) & $b$ ($10^{-3}$~ckky) \\
\hline
LXe		& 38 & 3.2  & 0.1 \\
XeSci       & 42 & 6.5 & 0.1  \\
HPXe	& 30  & 0.5  & 0.1 \\
\hline
\label{tab:para}
\end{tabular}
\end{table}

\begin{figure}[!htb]
\centering
\includegraphics[width=0.7\textwidth]{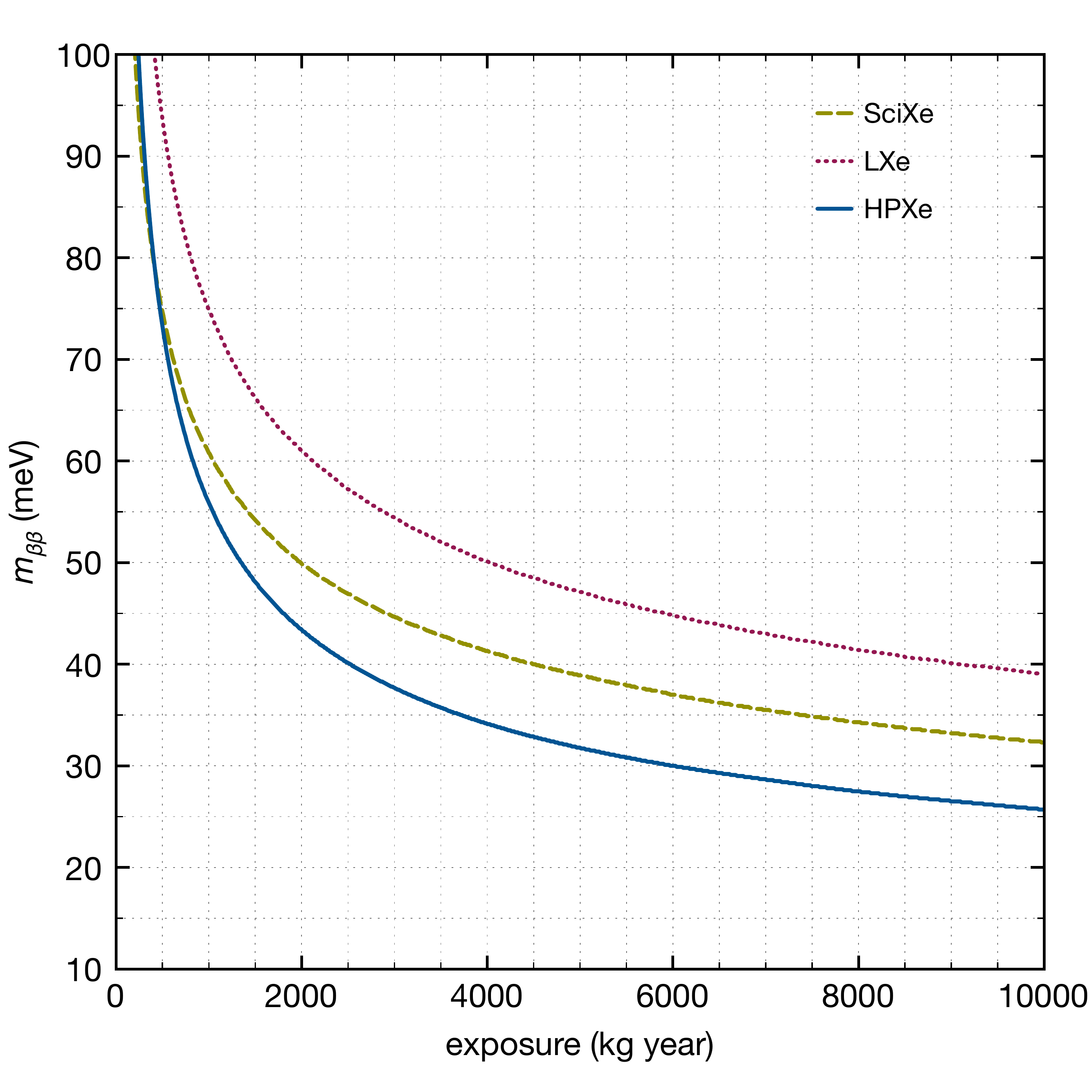}
\caption{Sensitivity of the three technologies experiments as a function of the total exposure, assuming the parameters described 
in Table \ref{tab:FutureParams} \cite{GomezCadenas:2013ue}.} \label{fig.NEXTW}
\end{figure}

To cover the full range allowed by the inverse hierarchy, one needs masses in the range of one ton of isotope. Xenon experiments have the potential to deploy those large masses (for instance, to enrich $^{76}$Ge is 10 times more expensive than $^{136}$Xe). This characteristic, together with the fact that one can build large xenon-based TPCs or calorimeters, make them a preferred choice for the next-to-next generation of experiments.

Table~\ref{tab:FutureParams} summarises a projection \cite{GomezCadenas:2013ue} of the experimental parameters for the three technologies, while Figure \ref{fig.NEXTW} shows the expected performance of xenon experiments assuming
the parameters described 
in Table \ref{tab:FutureParams}, up to a total exposure of 10 ton$\cdot$year. At the maximum exposure, the LXe and XeSci detectors reach a draw at 40 meV, while the HPXe detector reaches 25 meV. 

To summarise, the NEXT experiment has an enormous interest for \bbonu\ searches not only due to its large physics potential --- that is the ability to discover that neutrinos are Majorana particles --- but also as a springboard to the next-to-next generation of very challenging, ton-based experiments.

%% file: src2/NextDetector.tex
\subsection{The SOFT concept}



\begin{figure}
\centering
\includegraphics[width=0.8\textwidth]{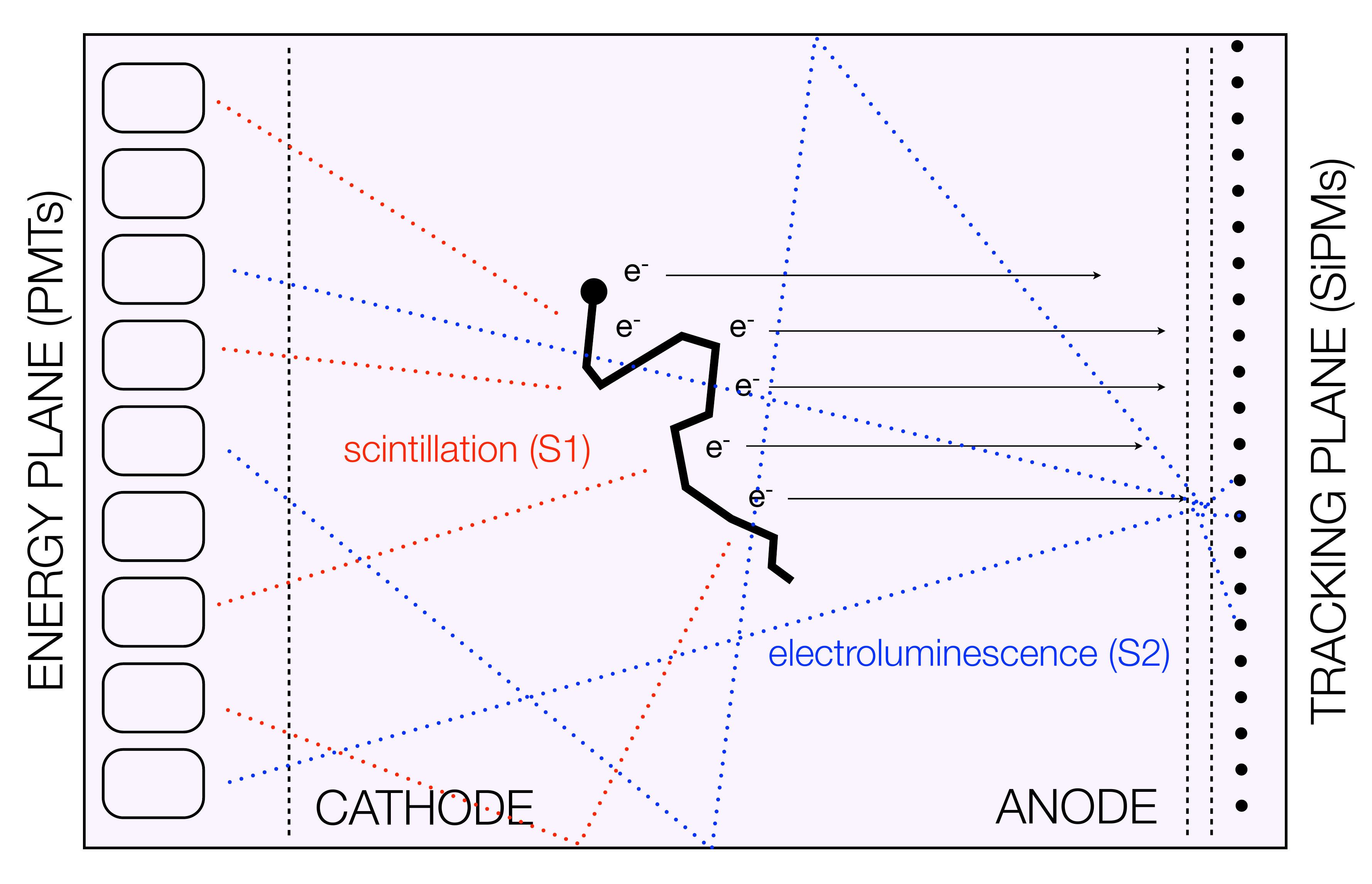}
\caption{The \emph{Separate, Optimized Functions} (SOFT) concept in the NEXT experiment: EL light generated at the anode is recorded in the photosensor plane right behind it and used for tracking; it is also recorded in the photosensor plane behind the transparent cathode and used for a precise energy measurement.} \label{fig:SOFT}
\end{figure}


Xenon, as a detection medium, provides both scintillation and ionization as primary signals. To achieve optimal energy resolution, the ionization signal is amplified in NEXT using electroluminescence (EL). The electroluminescent light provides both a precise energy measurement and tracking. Following ideas introduced in \cite{Nygren:2009zz} and further developed in our CDR \cite{Alvarez:2011my}, the chamber will have separated detection systems for tracking and calorimetry. This is the so-called \emph{SOFT} concept, illustrated in Figure \ref{fig:SOFT}. The detection process is as follows: Particles interacting in the HPXe transfer their energy to the medium through ionization and excitation. The excitation energy is manifested in the prompt emission of VUV ($\sim178$ nm) scintillation light. The ionization tracks (positive ions and free electrons) left behind by the particle are prevented from recombination by an electric field (0.3--0.5 ${\rm kV}/{\rm cm}$). The ionization electrons drift toward the TPC anode, entering a region, defined by two highly-transparent meshes, with an even more intense electric field (3 ${\rm kV}/{\rm cm}/{\rm bar}$). There, further VUV photons are generated isotropically by electroluminescence. Therefore, both scintillation and ionization produce an optical signal, to be detected with a sparse plane of PMTs (the \emph{energy plane}) located behind the cathode. The detection of the primary scintillation light constitutes the start-of-event, whereas the detection of the EL light provides an energy measurement. Electroluminescent light provides tracking as well, since it is detected also a few millimeters away from production at the anode plane, via an array of 1-mm$^{2}$ MPPCs, 1-cm spaced (the \emph{tracking plane}). 

\subsection{The apparatus}\label{sec:apparatus}
\begin{figure}
\centering
\includegraphics[width=0.8\textwidth]{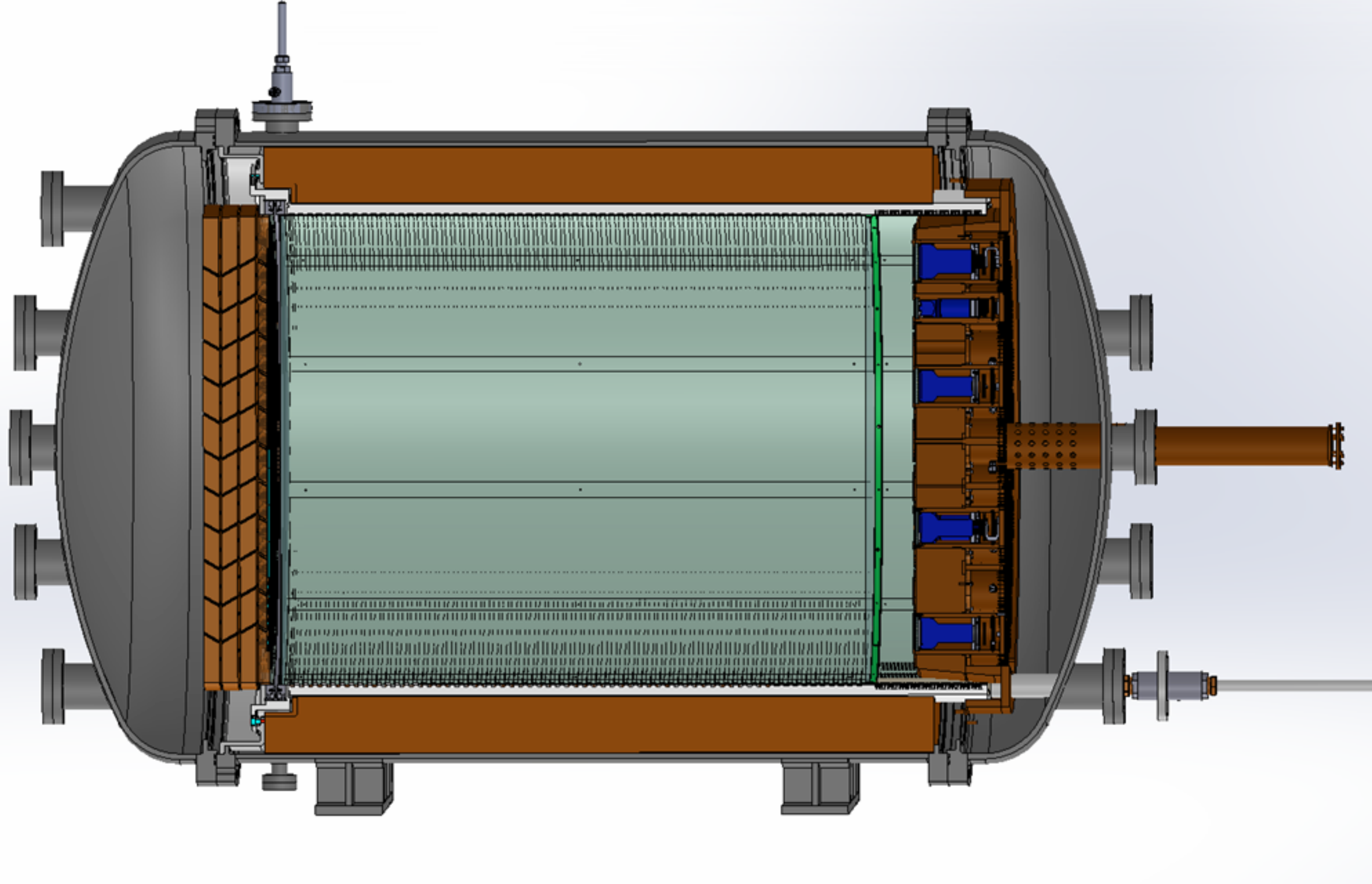}
\caption{A 3D drawing of the NEXT100 detector, showing the pressure vessel (gray), the internal copper shield (brown) and the field cage (green). The PMTs of the energy plane are shown in blue.} \label{fig:NEXT100} 
\end{figure}

Figure \ref{fig:NEXT100} shows a drawing of the NEXT-100 detector, indicating all the major subsystems. These are: 
\begin{itemize}
\item The pressure vessel (PV), built in stainless steel and designed to withstand a pressure of 15 bar, described in Subsection \ref{subsec:pv}. A copper layer on the inside shields the sensitive volume from the radiation originated in the vessel material.   
\item The field cage (FC), electrode grids, HV penetrators and light tube (LT), described in Subsection \ref{subsec:fc}.
\item The energy plane (EP) made of PMTs housed in copper enclosures, described in Subsection \ref{subsec:ep}.
\item The tracking plane (TP) made of MPPCs arranged into dice boards (DB), described in Subsection \ref{subsec:tp}.
\item The gas system, capable of pressurizing, circulating and purifying the gas, described in Subsection \ref{subsec:gs}.
\item The front end electronics, placed outside the chamber, described in Subsection \ref{subsec:elec}.
\item Shielding and other infrastructures, described in Subsection \ref{subsec:shielding}.

\end{itemize}

The NEXT TDR \cite{Alvarez:2012haa} gives the details of the design and components of the detector, which we summarise briefly here.

\subsubsection{The pressure vessel}\label{subsec:pv}
\begin{figure}
\centering
\includegraphics[width=0.8\textwidth]{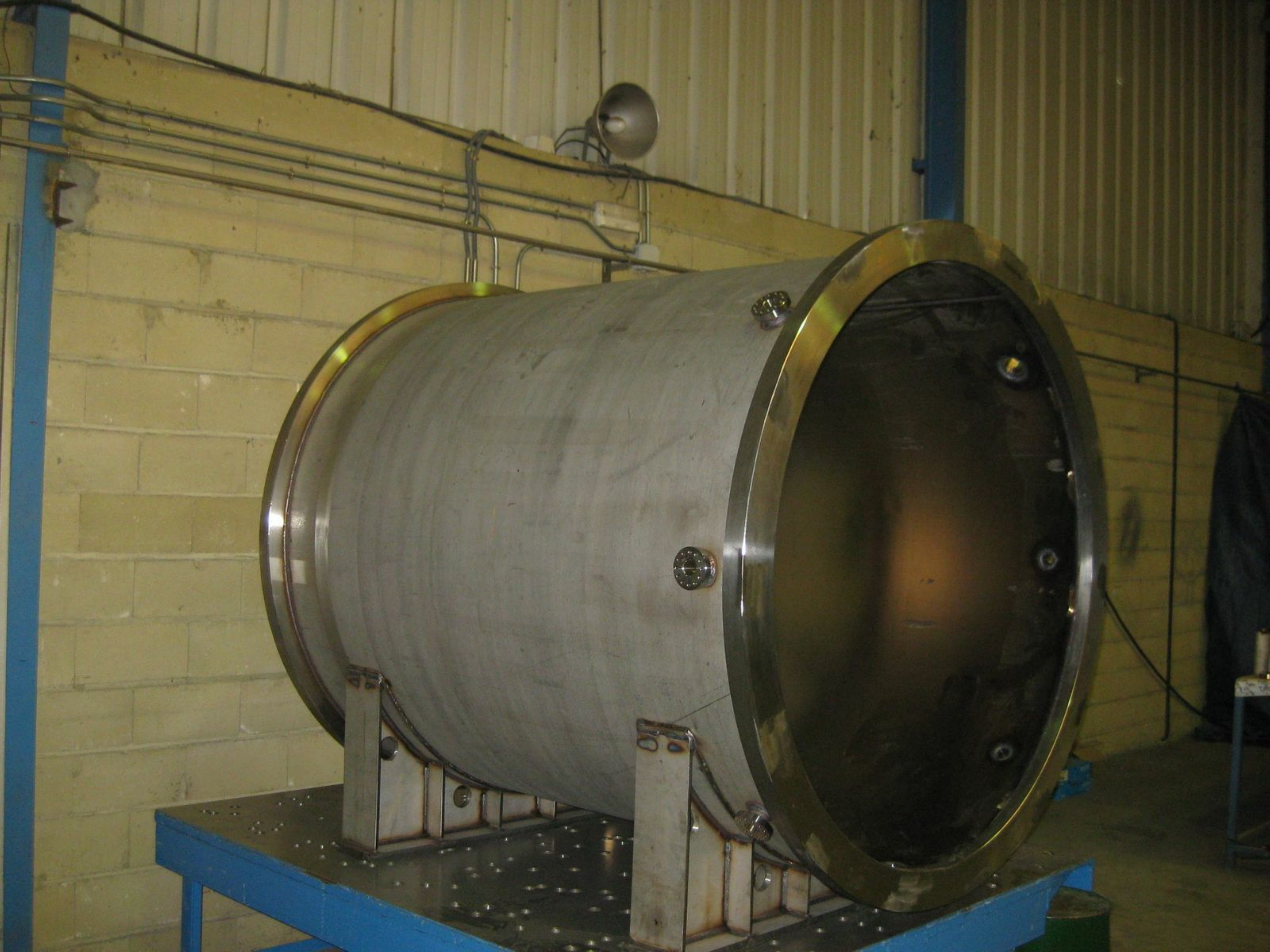}
\includegraphics[width=0.8\textwidth]{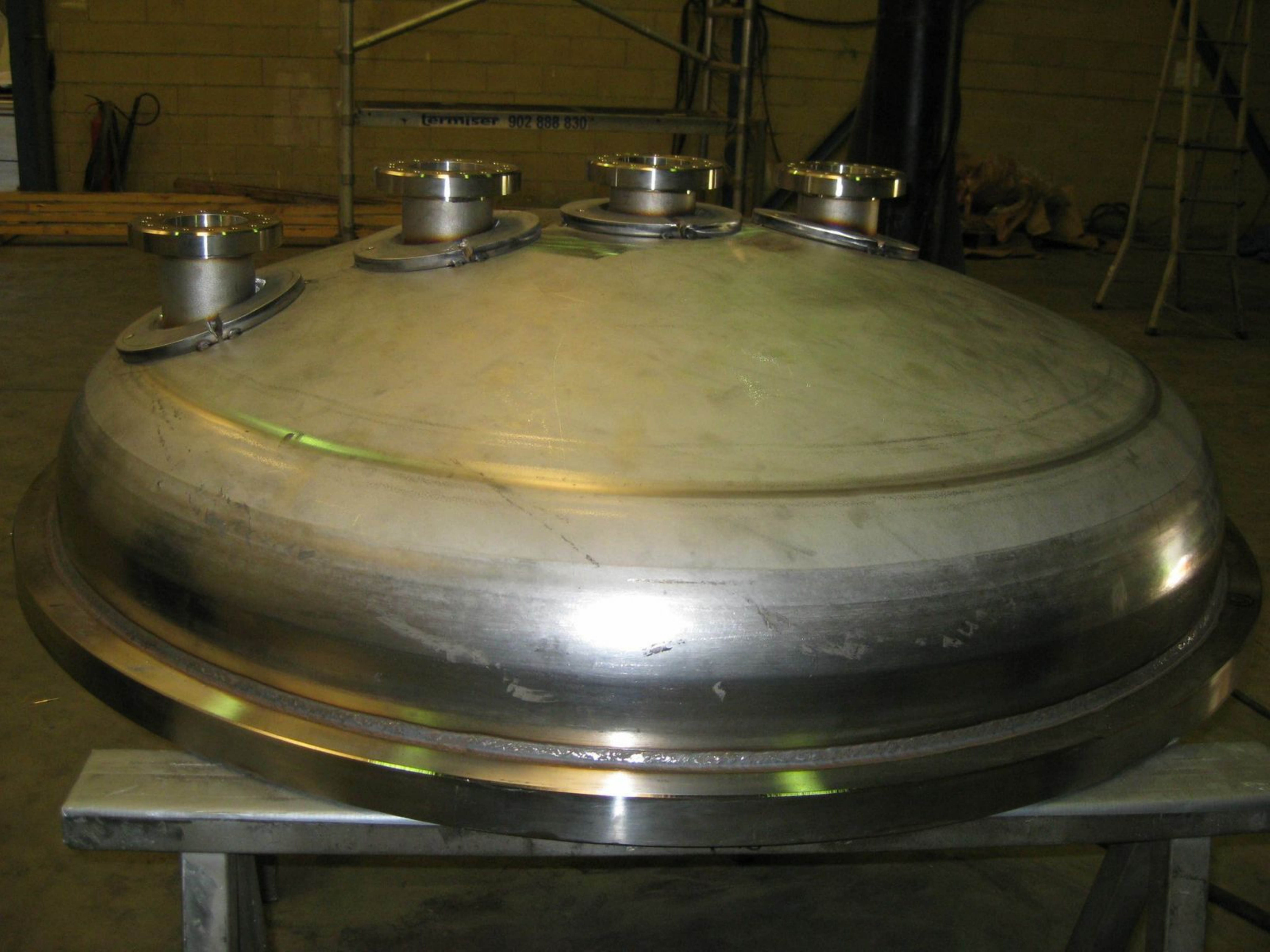}
\caption{NEXT-100 vessel in the finale stages of the production. Main body of the vessel (top), endcap with the port to extract the different signals and circulate the gas (bottom).} \label{fig:PV} 
\end{figure}

The pressure vessel, shown in Figure \ref{fig:PV}, consists of a barrel central section with two identical torispheric heads on each end, their main flanges bolted together, and is made of stainless steel, specifically the low-activity 316Ti alloy. 
In order to shield the activity of the vessel, we introduce an \emph{inner copper shield} (ICS), 12 cm thick and made of radiopure copper. The ICS will attenuate the radiation coming from the external detector (including the PV and the external lead shield) by a factor of 100. 

 The vessel has been built strictly to \emph{ASME Pressure Vessel Design Code, Section VIII} by the Madrid-based company MOVESA. It has been designed almost entirely by the collaboration, under the leadership of the Lawrence Berkeley National Laboratory (LBNL) and the Instituto de F\'isica Corpuscular (IFIC) in Valencia.  
 IFIC is in charge of supervision of fabrication, testing, certification and transport to LSC.

\subsubsection{The field cage}\label{subsec:fc}

\begin{figure}
\centering
\includegraphics[width=0.5\textwidth]{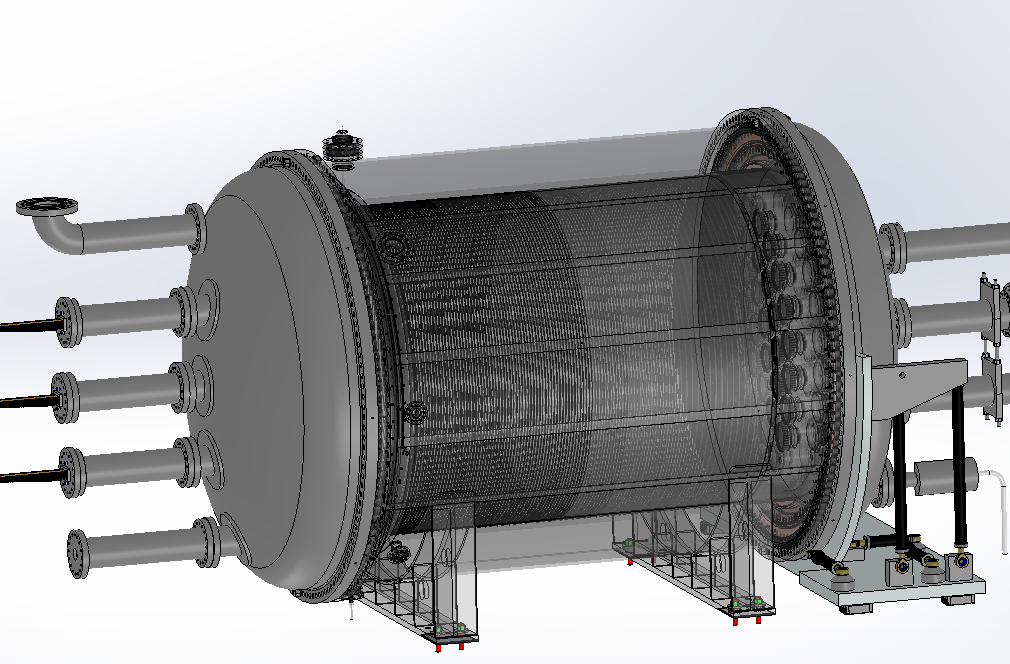}
\includegraphics[width=0.35\textwidth]{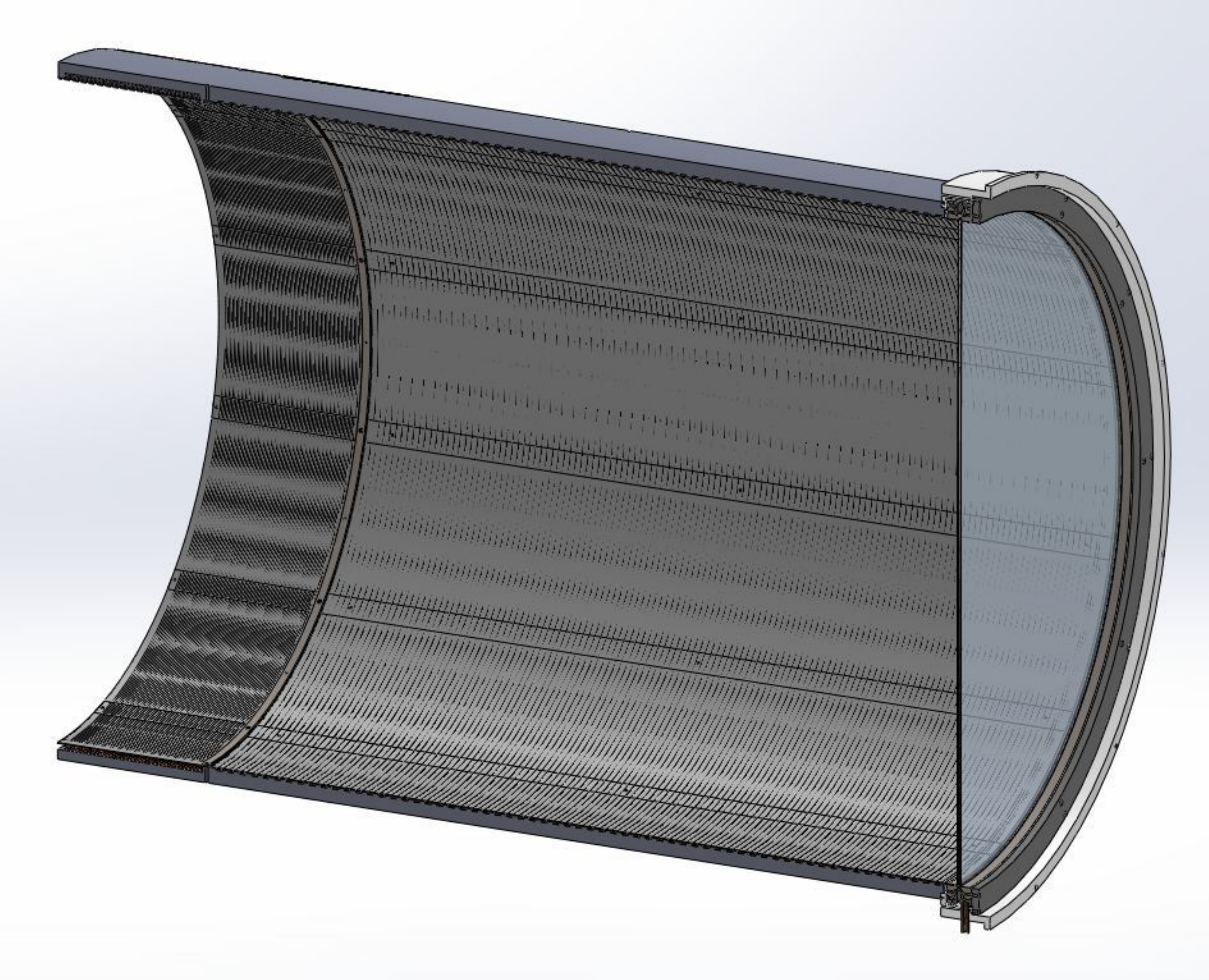}
\caption{A 3D drawing of the detector showing the field cage inside (left). Detailed of the transverse cut of the field cage (right).} \label{fig:FC} 
\end{figure} 

The main body of the field cage (Figure \ref{fig:FC}) will be a high-density polyethylene (HDPE) cylindrical shell, 2.5 cm thick, that will provide electric insulation from the vessel. Three wire meshes --- cathode, gate and anode --- separate the two electric field regions of the detector. The drift region, between cathode and gate, is a cylinder of 107 cm of diameter and 130 cm of length. Copper strips attached to the HDPE and connected with low background resistors grade the high voltage. The EL region, between gate and anode, is 1.0 cm long.

All the components of the field cage have been prototyped with the NEXT-DEMO detector (see Section \ref{sec:DEMO}). 
The NEXT-100 field cage and ancillary systems will 
be built by our USA collaborators. 

\subsubsection{The energy plane}\label{subsec:ep}

\begin{figure}
\centering
\includegraphics[width=0.8\textwidth]{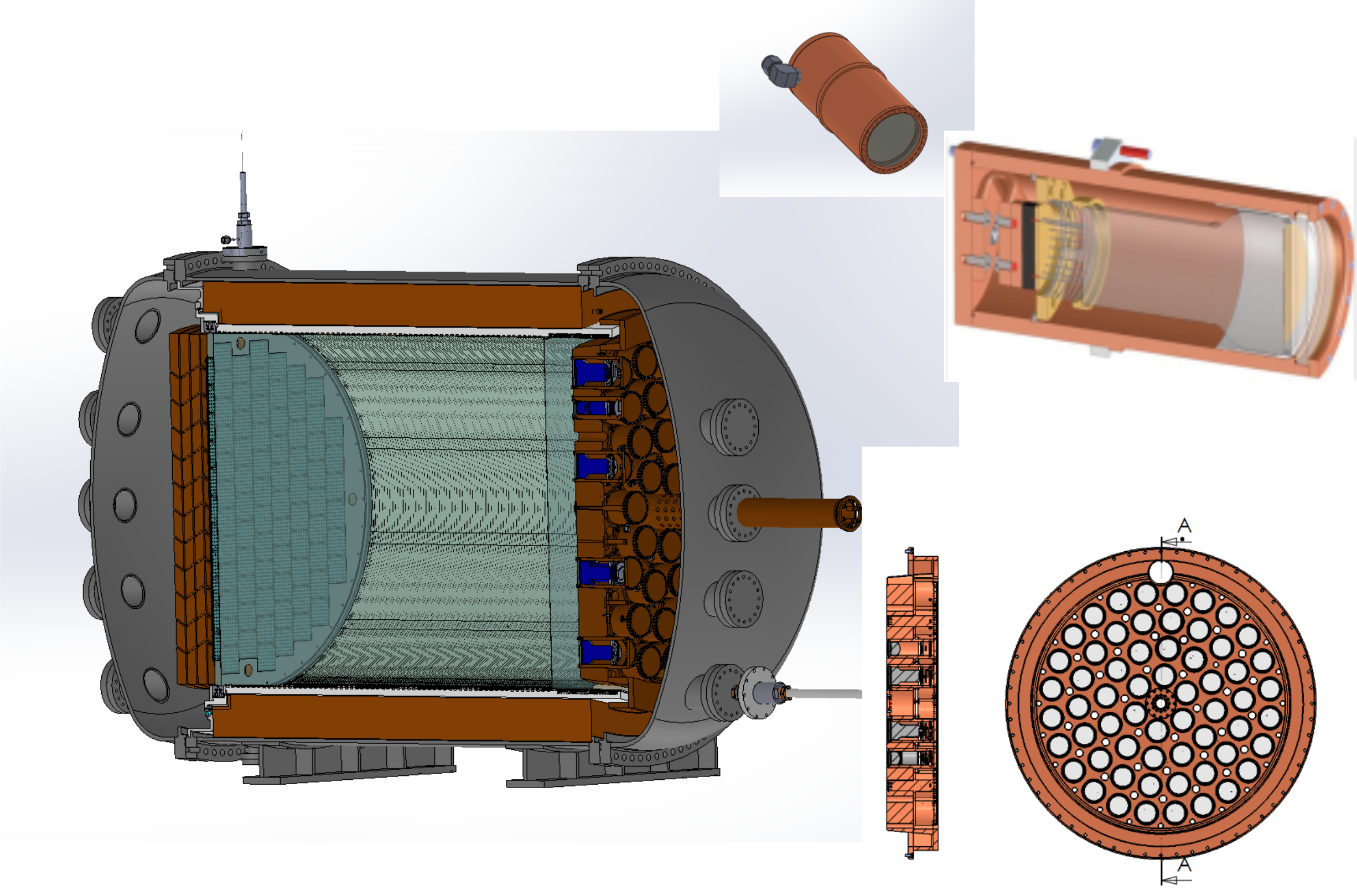}
\caption{A  drawing of the detector showing the energy plane inside the PV.} \label{fig:EP} 
\end{figure} 

The energy measurement in NEXT is provided by the detection of the electroluminescence light by an array of photomultipliers, the \emph{energy plane}, located behind the transparent cathode (Figure \ref{fig:EP}). Those PMTs will also record the scintillation light that indicates the start of the event. 

\begin{figure}
\centering
\includegraphics[height=.4\textwidth]{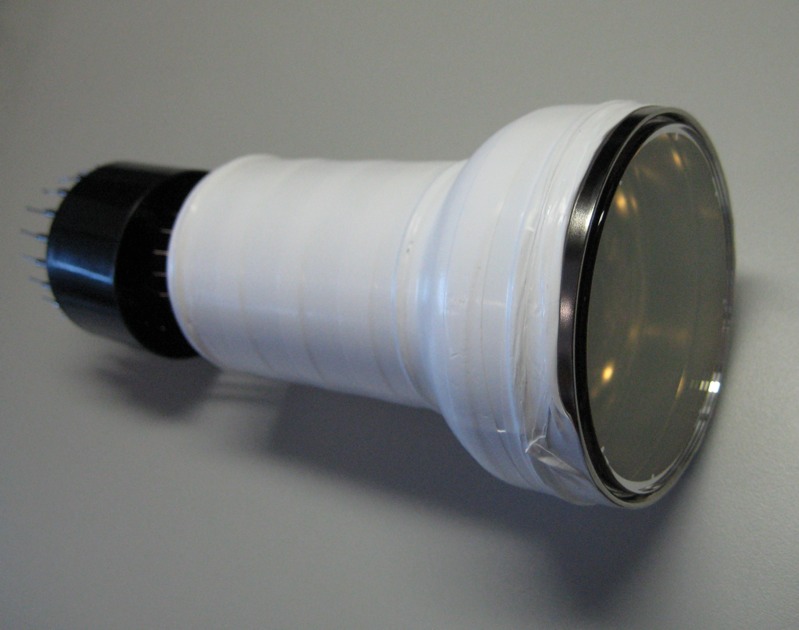}
\caption{The Hamamatsu R11410-10, a 3-inches photomultiplier with high quantum efficiency ($30\%$ in the blue region of the spectrum) at the xenon scintillation wavelengths and low radioactivity.} \label{fig:PMT}
\end{figure}

A total of 60 Hamamatsu R11410-10 photomultipliers (Figure \ref{fig:PMT}) covering 32.5\% of the cathode area constitute the energy plane. This phototube model has been specially developed for radiopure, xenon-based detectors. 
The quantum efficiency of the R11410-10 model is around 35\% in the VUV and 30\% in the blue region of the spectrum, and the dark count rate is 2--3 kHz (0.3 photoelectron threshold) at room temperature \cite{HamamatsuPMTs}.

\begin{figure}
\centering
\includegraphics[width=.8\textwidth]{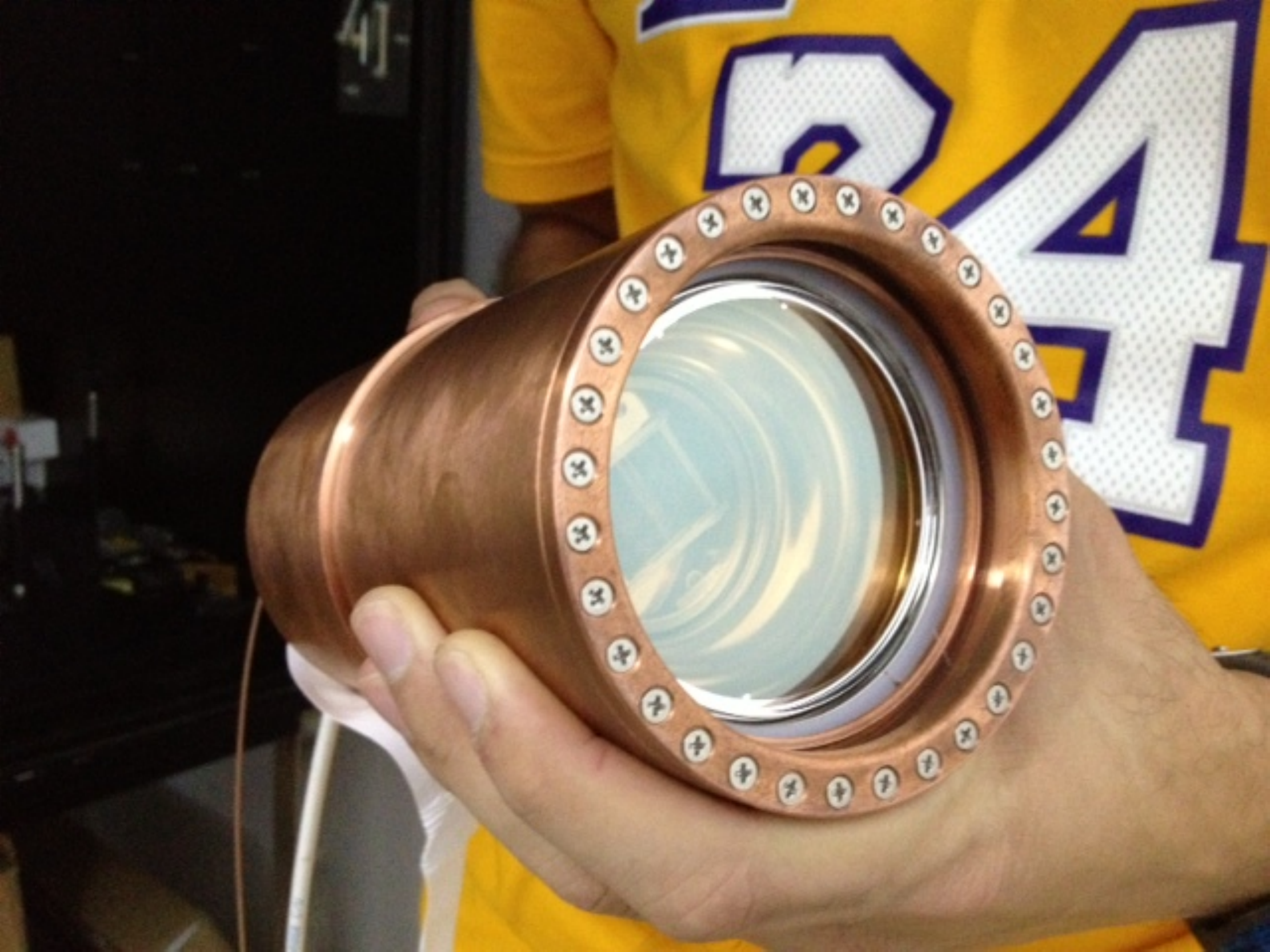}
\caption{The pressure-resistant enclosure, or ``can'' protecting the PMTs inside the PV.} \label{fig:PmtModule}
\end{figure}

Pressure-resistance tests run by the manufacturer showed that the R11410-10 cannot withstand pressures above 6 atmospheres. Therefore, in NEXT-100 they will be sealed into individual pressure resistant, vacuum tight copper enclosures coupled to sapphire windows (see Figure \ref{fig:PmtModule}). The window, 5 mm thick, is secured with a screw-down ring and sealed with an O-ring to the front-end of the enclosure. A similar back-cap of copper seals the back side of the enclosures. The PMT is optically coupled to the window using silicone optical pads of 2--3 mm thickness. A spring on the backside pushes the photomultiplier against the optical pads.

\begin{figure}
\centering
\includegraphics[width=.8\textwidth]{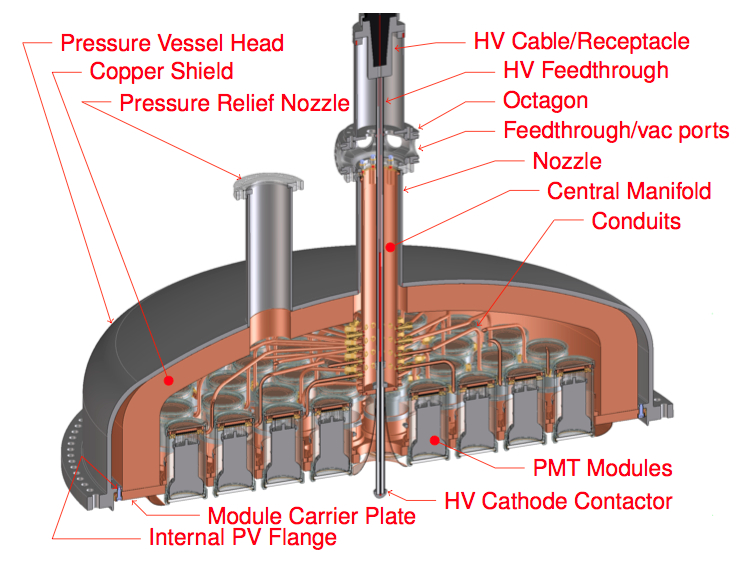}
\caption{The full energy plane of NEXT-100 mounted in the vessel head.} \label{fig:EnergyPlane}
\end{figure}

These PMT modules are all mounted to a common carrier plate that attaches to an internal flange of the pressure vessel head (see Figure \ref{fig:EnergyPlane}). The enclosures are all connected via individual pressure-resistant, vacuum-tight tubing conduits to a central manifold, and maintained at vacuum well below the Paschen minimum, avoiding sparks and glow discharge across PMT pins. The PMT cables route through the conduits and the central manifold to a feedthrough in the pressure vessel nozzle. 

The design of the energy plane has been shared between the IFIC, the UPV (Universidad Polit\'ecnica de Valencia) and the LBNL groups. The PMTs have already been purchased and tested, and are currently being screened for radioactivity at the LSC. Prototype PMT enclosures have been built and a full prototype energy plane including 14 PMTs is under construction and will be tested at LSC in 2013. The full energy plane will be installed in the detector during 2014.

\subsubsection{The tracking plane}\label{subsec:tp}

\begin{figure}
\centering
\includegraphics[width=\textwidth]{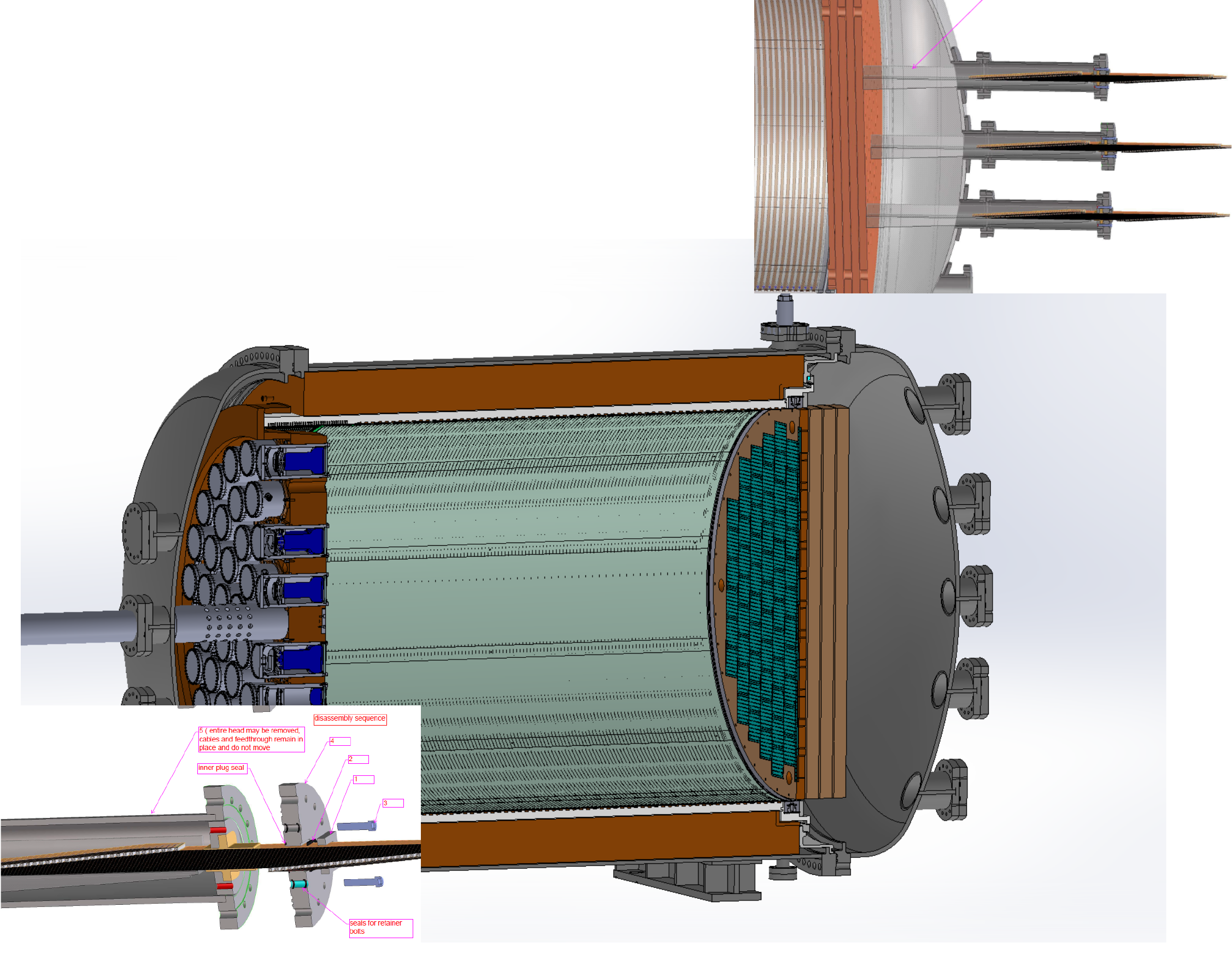}
\caption{A 3D drawing of the detector showing the tracking plane inside, including the feedthroughs to extract the cables carrying the signals from the MPPCs.} \label{fig.ft} 
\end{figure} 

The tracking function in NEXT-100 will be provided by a plane of multi-pixel photon counters (MPPCs) operating as sensor pixels and located behind the transparent EL gap. The chosen MPPC is the S10362-11-050P model by Hamamatsu. This device has an active area of 1 mm$^{2}$, 400 sensitive cells (50 $\mu{\rm m}$ size) and high photon detection efficiency in the blue region (about $\sim50\%$ at 440 nm). MPPCs are very cost-effective and their radioactivity is very low, given its composition (mostly silicon) and very small mass.  

The MPPCs will be mounted in Dice Boards (DB), identical to those prototyped in NEXT-DEMO (see Section \ref{sec:DEMO}). The electronics for the MPPCs will also be an improved version of the electronics for the DEMO detector. 
Also, like in NEXT-DEMO, all the electronics will be outside the chamber. The large number of channels in NEXT-100, on the other hand, requires the design and fabrication of large custom-made feedthroughs (LCFT) to extract the signals, as illustrated in Figure \ref{fig.ft}.

The $\sim$~8\,000 MPPCs needed for the tracking plane have already been purchased. The design of the DBs and the Front End Electronics (FEE) have been made at IFIC and UPV. The DBs have been fully tested in NEXT-DEMO and are ready for production.

A prototype of the LCFT will be tested in 2013. The full tracking plane can be installed in 2014.  

\subsubsection{The gas system} \label{subsec:gs}
The gas system must be capable of pressurizing, circulating, purifying, and depressurizing the NEXT-100 detector with xenon, argon and possibly other gases with negligible loss and without damage to the detector. In particular, the probability of any substantial loss of the very expensive enriched xenon (EXe) must be minimized. A list of requirements, in approximate decreasing order of importance, considered during the design is given below:
\begin{enumerate}
\item Pressurize vessel, from vacuum to 15 bar (absolute).
\item Depressurize vessel to closed reclamation system, 15 bar to 1 bar (absolute), on fault, in 10 seconds maximum.
\item Depressurize vessel to closed reclamation system, 15 bar to 1 bar (absolute), in normal operation, in 1 hour maximum.
\item Relieve pressure (vent to closed reclamation system) for fire or other emergency condition.
\item Allow a maximum leakage of EXe through seals (total combined) of $100\ {\rm g}/{\rm year}$.
\item Allow a maximum loss of EXe to atmosphere of $10\ {\rm g}/{\rm year}$.
\item Accommodate a range of gasses, including Ar and N$_2$\label{gases}. 
\item Circulate all gasses through the detector at a maximum rate of 200 standard liters per minute (slpm) in axial flow pattern.
\item Purify EXe continuously. Purity requirements: $<1$ ppb O$_2$, CO$_2$, N$_2$, CH$_4$.
\end{enumerate}

The most vulnerable component of the gas system is the re-circulation compressor, which must have sufficient redundancy to minimize the probability of failure and leakage. The collaboration has chosen a compressor manufactured by \textsc{sera ComPress GmbH}. This compressor is made with metal-to-metal seals on all the wetted surfaces. The gas is moved through the system by a triple stainless steel diaphragm. Between each of the diaphragms there is a sniffer port to monitor for gas leakages. In the event of a leakage, automatic emergency shutdown can be initiated.

The gas system will be equipped with both room-temperature (SAES MC50) and heated getters (SAES PS4-MT15) that
remove electronegative impurities (O2, H2O, etc.) from the xenon.

An automatic recovery system of the expensive EXe will also be needed to evacuate the chamber in case of an emergency condition. A 30-m$^3$ expansion tank will be placed inside the laboratory to quickly reduce the gas pressure in the system. Additionally, we will implement a similar solution to that proposed by the LUX collaboration, where a chamber permanently cooled by liquid nitrogen will be used. 

The gas system has been designed as a collaboration between IFIC and the University of Zaragoza, taking advantage of the experience gained with our prototypes. The basic gas system needed for the initial operation of the NEXT-100 apparatus has already been purchased and shipped to LSC, but the system must be upgraded during 2014 for the enriched xenon run in 2015. 

\subsubsection{Electronics} \label{subsec:elec}
The NEXT-100 data-acquisition system (DAQ) follows a modular architecture named the Scalable Readout System (SRS), already described in our CDR \cite{Alvarez:2011my}. At the top of the hierarchy, a PC farm running the DAQ software, DATE, receives event data from the DAQ modules via Gigabit Ethernet (GbE) links. The DATE PCs (Local Data Concentrators, LDCs) assemble incoming fragments into sub-events, which are sent to one or more additional PCs (Global Data Concentrators, GDC). The GDCs build complete events and store them to disk for offline analysis.

The DAQ modules used are Front-End Concentrator (FEC) cards, which serve as the generic interface between the DAQ system and application-specific front-end modules. The FEC module can interface different kinds of front-end electronics by using the appropriate plug-in card. The FEC card and the overall SRS concept have been developed within the framework of the CERN RD-51 collaboration. Three different FEC plug-in cards are used in NEXT-100 (energy plane readout digitization, trigger generation, tracking plane readout digitization).

\subsubsection*{Electronics for the energy plane} \label{subsec:fe_energy}
The front-end (FE) electronics for the PMTs in NEXT-100 will be very similar to the system developed for the NEXT-DEMO and NEXT-DBDM prototypes. The first step in the chain is to shape and filter the fast signals produced by the PMTs (less than 5 ns wide) to match the digitizer and eliminate the high frequency noise. An integrator is implemented by simply adding a capacitor and a resistor to the PMT base. The charge integration capacitor shunting the anode stretches the pulse and reduces the primary signal peak voltage accordingly.

Our design uses a single amplification stage based on the fully differential amplifier THS4511, which features low noise (2 ${\rm nV}/\sqrt{\rm Hz}$) and provides enough gain to compensate for the attenuation in the following stage, based on a passive RC filter with a cut frequency of 800 kHz. This filtering produces enough signal stretching to allow acquiring many samples per single photo-electron at 40 MHz.

\subsubsection*{Electronics for the tracking plane} \label{subsec:fe_tracking}
\begin{figure}
\centering
\includegraphics[width=0.5\paperwidth]{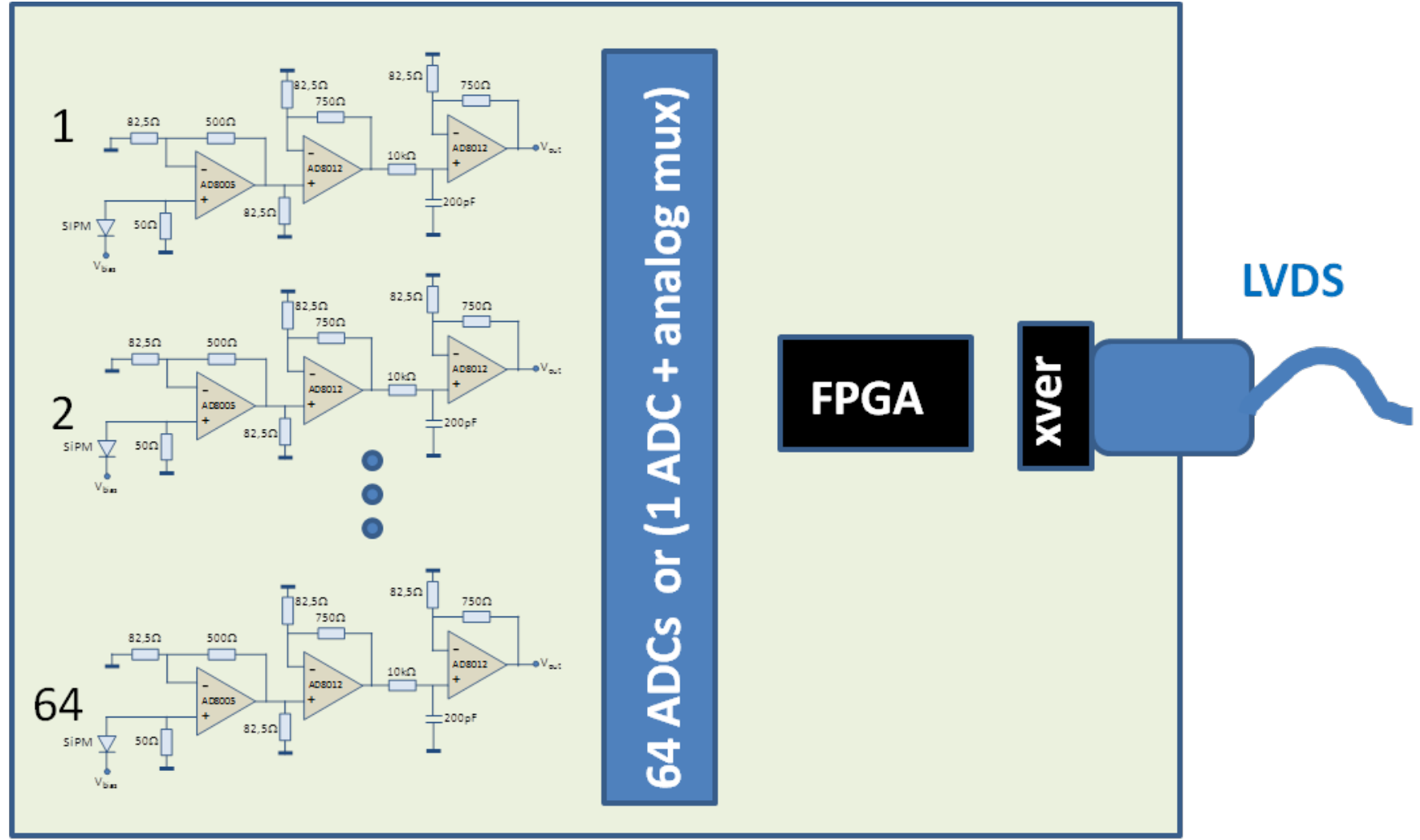}
\caption{\small Functional blocks in the FEB card.}
\label{fig:feb}
\end{figure}

The tracking plane will have $\sim$ 8\,000 channels. On the other hand, the electronics for the MPPCs is simplified given the large gain of these devices. Our design consists of a very simple, 64 channel Front-End Board (FEB, Figure~\ref{fig:feb}). Each FEB takes the input of a single DB (transmitted via low-crosstalk kapton flat cables) and includes the analog stages, ADC converters, voltage regulators and an FPGA that handles, formats, buffers and transmits data to the  DAQ. LVDS clock and trigger inputs are also needed. A total of 110 FEBs are required. The architecture of the  FEB is described in our TDR.

The design of the electronics is a collaboration between UPV and LBNL. It will be an evolution of the electronics currently operational at NEXT-DEMO. The DAQ is responsibility of UPV, and it will also be an improved version of the DEMO DAQ. 

\subsubsection{Shielding and other infrastructures} \label{subsec:shielding}
\begin{figure}
\centering
\includegraphics[width=0.6\textwidth]{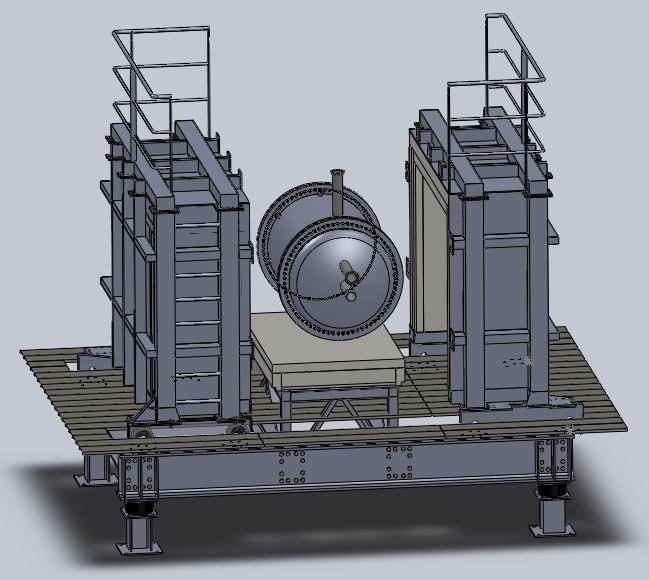}
\caption{Drawing of the NEXT-100 lead castle shield in its open configuration.}\label{fig:Shielding}
\end{figure}

To shield NEXT-100 from the external flux of high-energy gamma rays a relatively simple lead castle, shown in Figure \ref{fig:Shielding}, has been chosen, mostly due to its simplicity and cost-effectiveness. The lead wall has a thickness of 20 cm and is made of layers of staggered lead bricks held with a steel structure. The lead bricks have standard dimensions ($200\times100\times50$ mm$^3$), and, by requirement, an activity in uranium and thorium lower than $0.4\ {\rm mBq}/{\rm kg}$. 

The lead castle is made of two halves mounted on a system of wheels that move on rails with the help of an electric engine. The movable castle has an open and a closed position. The former is used for the installation and service of the pressure vessel; the latter position is used in normal operation. A lock system fixes the castle to the floor in any of the two configurations to avoid accidental displacements. 

The design of the lead castle has been led by the University of Girona, in collaboration with UPV and IFIC. The design is completed and the shield is ready to be built pending the availability of funds.


The construction of the infrastructures needed for the NEXT-100 experiment (working platform, seismic pedestal) is currently underway. They will be fully installed at LSC by the end of 2013.  

\begin{figure}
\centering
\includegraphics[height=7cm]{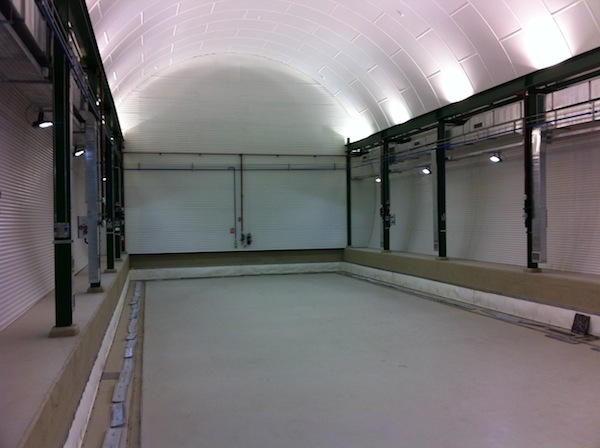}
\caption{View of Hall A of the Laboratorio Subterr\'aneo de Canfranc prior to any equipment installation.} \label{fig:LSC_Hall_A}
\end{figure}

Figure \ref{fig:LSC_Hall_A} shows an image of Hall A, future location of NEXT-100. The pool-like structure is intended to be a catchment reservoir to hold xenon or argon --- a liquid-argon experiment, ArDM, will be neighbouring NEXT-100 in Hall A --- gas in the event of a catastrophic leak. Therefore, for reasons of safety all experiments must preclude any personnel working below the level of the top of the catchment reservoir.

An elevated working platform has already been built. It is designed to stand a uniform load of $1500\ {\rm kg}/{\rm m}^2$ and a concentrated load of $200\ {\rm kg}/{\rm m}^2$. It is anchored to the hall ground and walls. The platform floor tiles are made of galvanized steel and have standard dimension to minimize cost. 

Due to the mild seismic activity of the part of the Pyrenees where LSC is located, a comprehensive seismic study has been conducted as part of the project risk analysis. As a result, an anti-seismic structure that will hold both the pressure vessel and the shielding has been designed. This structure will be anchored directly to the ground and independent of the working platform to allow seismic displacements in the event of an earthquake. 

\begin{figure}
\centering
\includegraphics[height=9cm]{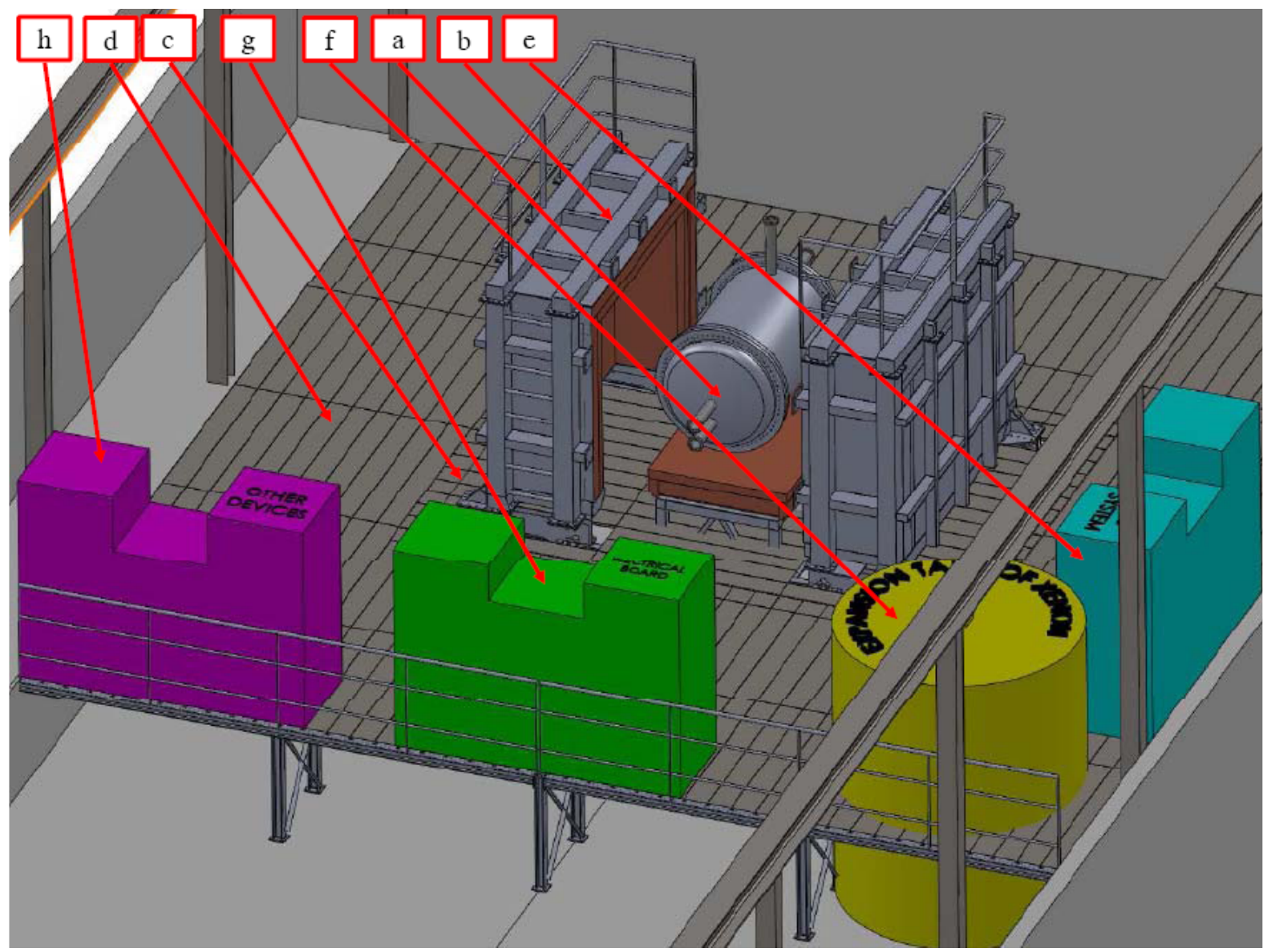}
\includegraphics[height=9cm]{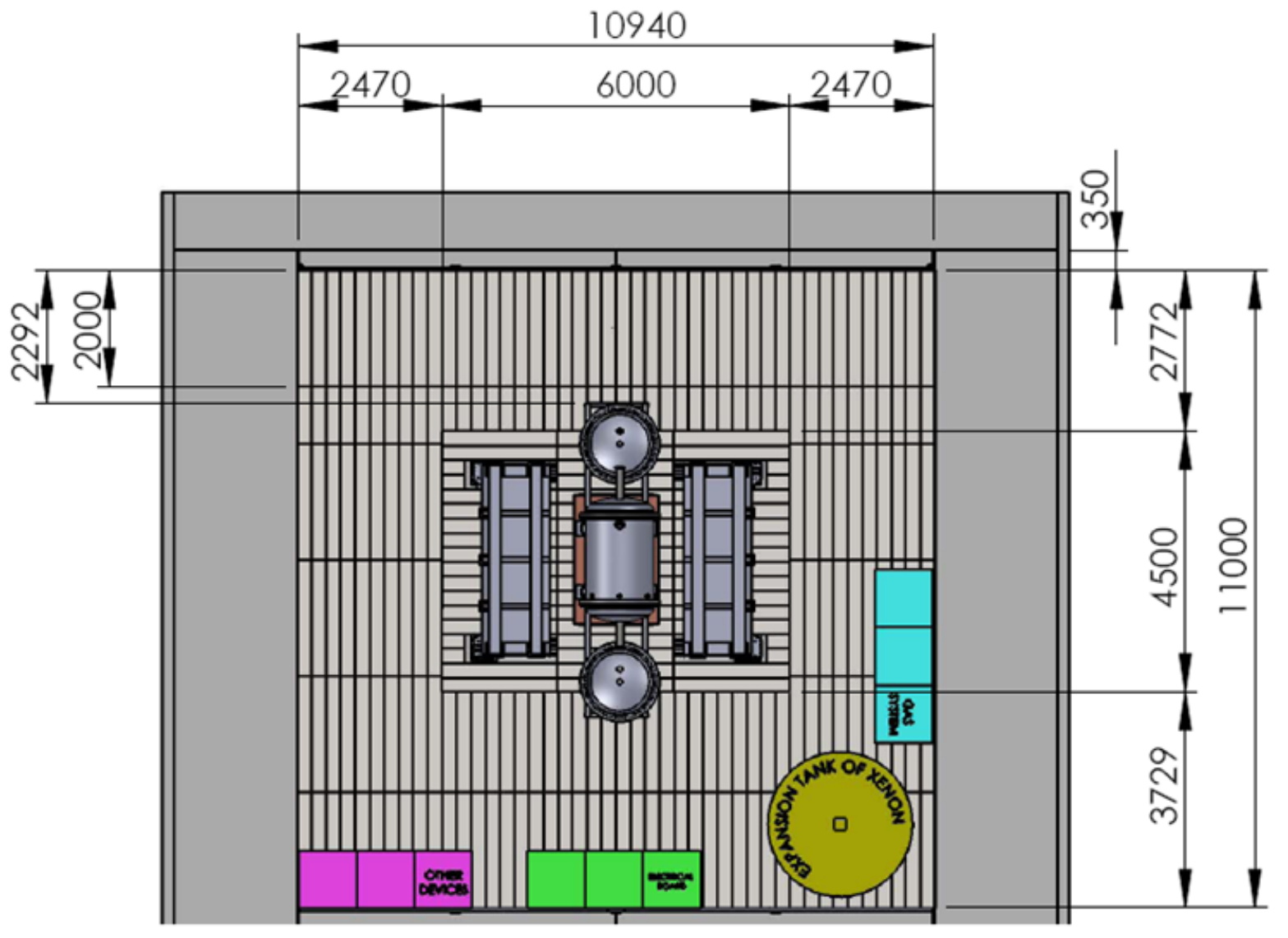}
\caption{Top: Intended location of the components  and subsystems for the operation of NEXT-100 on the working platform: (a) NEXT-100; (b) the lead castle shield in its open configuration; (c) seismic platform; (d) working platform; (e) gas purification system; (f) emergency gas vent tank; (g) data acquisition system; (h) other systems. Bottom: Top view showing the dimensions of the working platform.} \label{fig:Infrastructure}
\end{figure}

Figure \ref{fig:Infrastructure} shows the placement of NEXT-100 and components on the platform as well as the dimensions. 

%% file: src2/NextBKGM.tex

The NEXT background model describes the sources of radioactive contaminants in the detector and their activity. It allows us, via detailed simulation, to predict the background events that can be misidentified as signal.

\subsection{Sources of background}

\subsubsection*{Radioactive contaminants in detector materials}

After the decay of \BI, the daughter isotope, \Po, emits a number of de-excitation gammas with energies above 2.3 MeV. The gamma line at 2447 keV, of intensity 1.57\%, is very close to the $Q$-value of \XE. The gamma lines above \Qbb\ have low intensity and their contribution is negligible. 

The daughter of \TL, \Pb, emits a de-excitation photon of 2614 keV with a 100\% intensity. The Compton edge of this gamma is at 2382 keV, well below \Qbb. However, the scattered gamma can interact and produce other electron tracks close enough to the initial Compton electron so they are reconstructed as a single object falling in the energy region of interest (ROI). Photoelectric electrons are produced above the ROI but can loose energy via bremsstrahlung and populate the window, in case the emitted photons escape out of the detector. Pair-creation events are not able to produce single-track events in the ROI. 

\subsubsection*{Radon}
Radon constitutes a dangerous source of background due to the radioactive isotopes $^{222}$Rn (half-life of 3.8\,d) from the $^{238}$U chain and $^{220}$Rn (half-life of 55\,s) from the $^{232}$Th chain. As a gas, it diffuses into the air and can enter the detector. \BI\ is a decay product of $^{222}$Rn, and \TL\ a decay product of $^{220}$Rn. In both cases, radon undergoes an alpha decay into polonium, producing a positively charged ion which is drifted towards the cathode by the electric field of the TPC.  As a consequence, $^{214}$Bi and $^{208}$Tl contaminations can be assumed to be deposited on the cathode surface. Radon may be eliminated from the TPC gas mixture by recirculation through appropriate filters. There are also ways to suppress radon in the volume defined by the shielding. Radon control is a major task for a \bbonu\ experiment, and will be of uppermost importance for NEXT-100.

\subsubsection*{Cosmic rays and laboratory rock backgrounds}
Cosmic particles can also affect our experiment by producing high energy photons or activating materials. This is the reason why double beta decay experiments are conducted deep underground. At these depths, muons are the only surviving cosmic ray particles, but 
their interactions with the rock produce neutrons and electromagnetic showers. Furthermore, the rock of the laboratory itself is a rather intense source of \TL\ and \BI\ backgrounds as well as neutrons.

The flux of photons emanating from the LSC walls is (see our TDR and references therein):
\begin{itemize}
\item $0.71 \pm 0.12~{\gamma/\mathrm{cm}^2/\mathrm{s}}$~from the  $^{238}$U chain.
\item $0.85 \pm 0.07~{\gamma/\mathrm{cm}^2/\mathrm{s}}$~from the $^{232}$Th chain.
\end{itemize}

These measurements include all the emissions in each chain. The flux corresponding to the \TL\ line at 2614.5 keV and the flux corresponding to the \BI\ line at 1764.5 keV were also measured (from the latter it is possible to deduce the flux corresponding to the 2448 keV line). The results are:
\begin{itemize}
\item $0.13 \pm 0.01~{\gamma/\mathrm{cm}^2/\mathrm{s}}$~from the \TL\ line.
\item $0.006 \pm 0.001~{\gamma/\mathrm{cm}^2/\mathrm{s}}$~from the \BI\ line at 2448 keV. 
\end{itemize}

The above backgrounds are considerably reduced by the shielding. In addition, given the topological capabilities of NEXT, the residual muon and neutron background do not appear to be significant
for our experiment.

\subsection{Radioactive budget of NEXT-100}\label{sec:rabudget}

Information on the radiopurity of the materials expected to be used in
the construction of NEXT-100 has been compiled, performing specific
measurements and also examining data from the literature for materials
not yet screened. A detailed description is presented in \cite{Alvarez:2012as}. A brief summary of the results presented there for the main materials is shown in Table \ref{tab:RA}, where the initials of the subsystems refer to the ones introduced in Subsection \ref{sec:apparatus}. 

\begin{table}
\caption{Activity (in ${\rm mBq}/{\rm kg}$) of the most relevant materials used in NEXT.} \label{tab:RA}
\begin{center}
\begin{tabular}{lllll}
\hline
Material & Subsystem &$^{238}$U & $^{232}$Th & Ref. \\  
\hline
Lead  & Shielding & 0.37 & 0.07  & \cite{Alvarez:2012as}\\

Copper & ICS & $<0.012$ & $<0.004$  & \cite{Alvarez:2012as}\\

Steel (316Ti) & PV  & $<0.57$ & $<0.54$  & \cite{Alvarez:2012as}\\

Polyethylene & FC &  0.23 & $<0.14$ & \cite{Aprile:2011ru} \\

PMT (R11410-MOD per pc) & EP &  $< 2.5$ & $< 2.5$ & \cite{Aprile:2011ru} \\
\hline

\end{tabular}  
\end{center}
\end{table} 

\subsection{Expected background rate}
The only relevant backgrounds for NEXT are the photons emitted by the \TL\ line (2614.5 keV) and the \BI\ line (2448 keV). These sit very near \Qbb\ and the interaction of the photons in the gas can fake the \bbonu\ signal. NEXT-100 has the structure of a Matryoshka (a Russian nesting doll). The flux of gammas emanating from the LSC walls is drastically attenuated by the lead castle, and the residual flux, together with that emitted by the lead castle itself and the materials of the pressure vessel is further attenuated by the inner copper shielding. One then needs to add the contributions of the ``inner elements'' in NEXT: field cage, energy plane, and the elements of the tracking plane not shielded by the ICS.

A detailed Geant4 \cite{Agostinelli2003250} simulation of the NEXT-100 detector was written in order to compute the background rejection factor achievable with the detector. Simulated events, after reconstruction, were accepted as a \bbonu\ candidate if
\begin{enumerate}
\item[(a)] they were reconstructed as a single track confined within the active volume;
\item[(b)] their energy fell in the region of interest, defined as $\pm 0.5$ FWHM around \Qbb; 
\item[(c)] the spatial pattern of energy deposition corresponded to that of a \bbonu\ track (\emph{blobs} in both ends).
\end{enumerate}

The achieved background rejection factor together with the selection efficiency for the signal are shown in Table \ref{tab:RF}. As can be seen, the cuts suppress the radioactive background by more than 7 orders of magnitude. This results in an estimated background rate of about $5\times10^{-4}~\ckky$.

\begin{table}
\caption{Acceptance of the selection cuts for signal and backgrounds.}
\label{tab:RF}
\begin{center}
\begin{tabular}{lccc}
\toprule
 & \multicolumn{3}{c}{Fraction of events} \\
Selection cut & \bbonu\ & \BI\ & \TL\ \\ \midrule 
Confined, single track & 0.48 & $6.0\times10^{-5}$ & $2.4 \times 10^{-3}$ \\
Energy ROI & 0.33 & $2.2\times10^{-6}$ & $1.9 \times 10^{-6}$ \\
Topology \bbonu\ & 0.25 & $1.9\times10^{-7}$ & $1.8 \times 10^{-7}$ \\
\bottomrule
\end{tabular}
\end{center}
\end{table}%

%% file: src2/NextDemo.tex
To prove the innovative concepts behind the NEXT design we have built two EL prototypes:

%
%
\begin{itemize}

\item \emph{NEXT-DEMO}, operating at IFIC. This is a large prototype, which can hold a mass similar to that of the Gotthard experiment. It is conceived to fully test and demonstrate the EL technology. 

\item \emph{NEXT-DBDM}, operating at LBNL. This was our first operative prototype and has demonstrated a superb resolution, which extrapolates to 0.5\% FWHM at \Qbb.

\end{itemize}


The two prototypes are fully operational since 2011 and our initial results and operation experience have recently been published \cite{Alvarez:2012hh, Alvarez:2012xda, Alvarez:2012hu}. 

\subsection{NEXT-DEMO}\label{sec:DEMO}
In this section we describe in more detail the NEXT-DEMO demonstrator and our first results.
The main goal of the prototype was the demonstration of the detector concept to be used in NEXT-100, more specifically: 

\begin{enumerate}
\item To demonstrate good energy resolution (better than 1\% FWHM at \Qbb) in a large system with full spatial corrections.
\item To demonstrate track reconstruction and the performance of MPPCs.
\item  To test long drift lengths and high voltages (up to 50 kV in the cathode and 25 kV in the anode).
\item To understand gas recirculation in a large volume, including operation stability and robustness against leaks.
\item To understand the transmittance of the light tube, with and without wavelength shifter. 
\end{enumerate}

\begin{figure}
\centering
\includegraphics[width=0.7\textwidth]{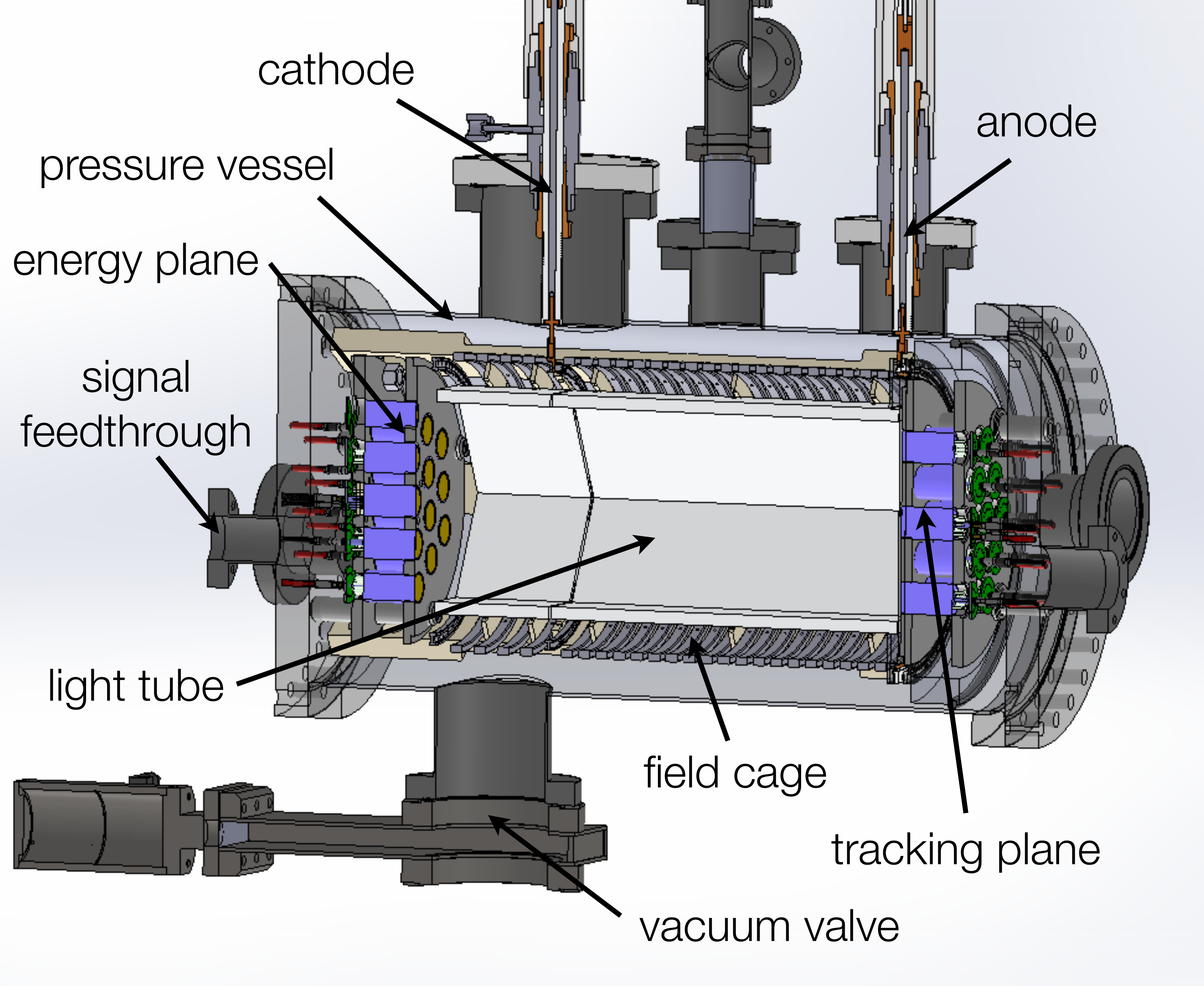}
\caption{Cross-section drawing of the NEXT-DEMO detector with all major parts labelled.} \label{fig:NextDemoXSec}
\end{figure}

\begin{figure}
\centering
\includegraphics[width=0.9\textwidth]{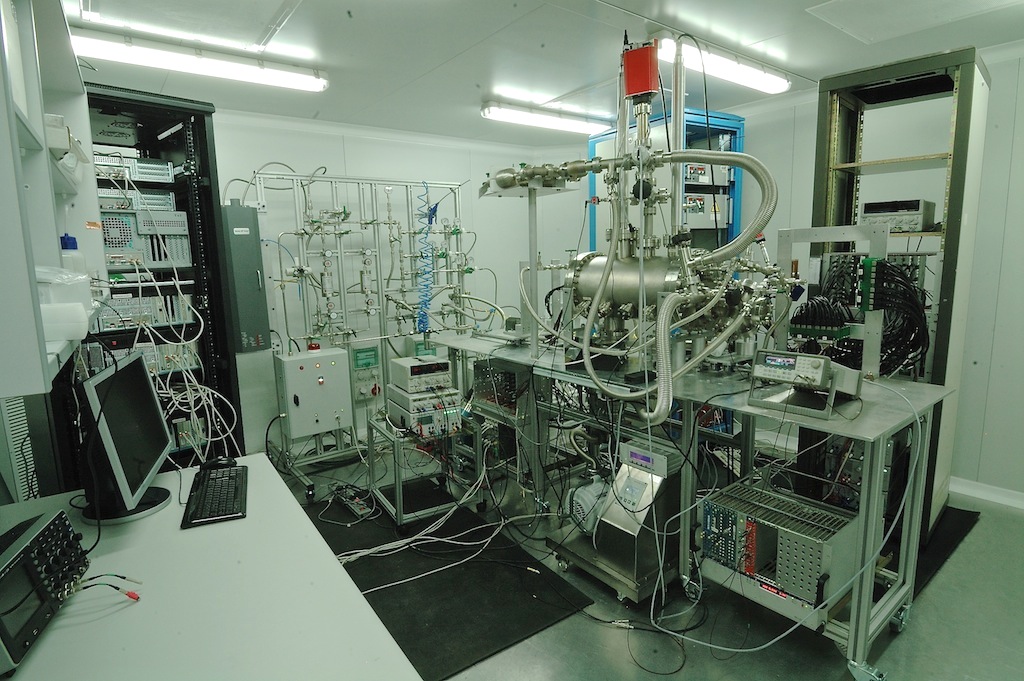}
\caption{The NEXT-DEMO detector and ancillary systems (gas system, front-end electronics and DAQ) in their location in a semi-clean room at IFIC.} 
\label{fig:NextDemoCleanRoom}
\end{figure}

The apparatus, shown in Figure~\ref{fig:NextDemoXSec}, is a high-pressure xenon electroluminescent TPC implementing the SOFT concept. Its active volume is 30 cm long. A tube of hexagonal cross section made of PTFE is inserted into the active volume to improve the light collection. The TPC is housed in a stainless-steel pressure vessel, 60 cm long and with a diameter of 30 cm, that can withstand up to 15 bar. Natural xenon circulates in a closed loop through the vessel and a system of purifying filters. The detector is not radiopure and is not shielded against natural radioactivity. It is installed in a semi-clean room (see Figure~\ref{fig:NextDemoCleanRoom}) at IFIC, in Valencia, Spain.

\begin{figure}
\centering
\includegraphics[width=0.75\textwidth]{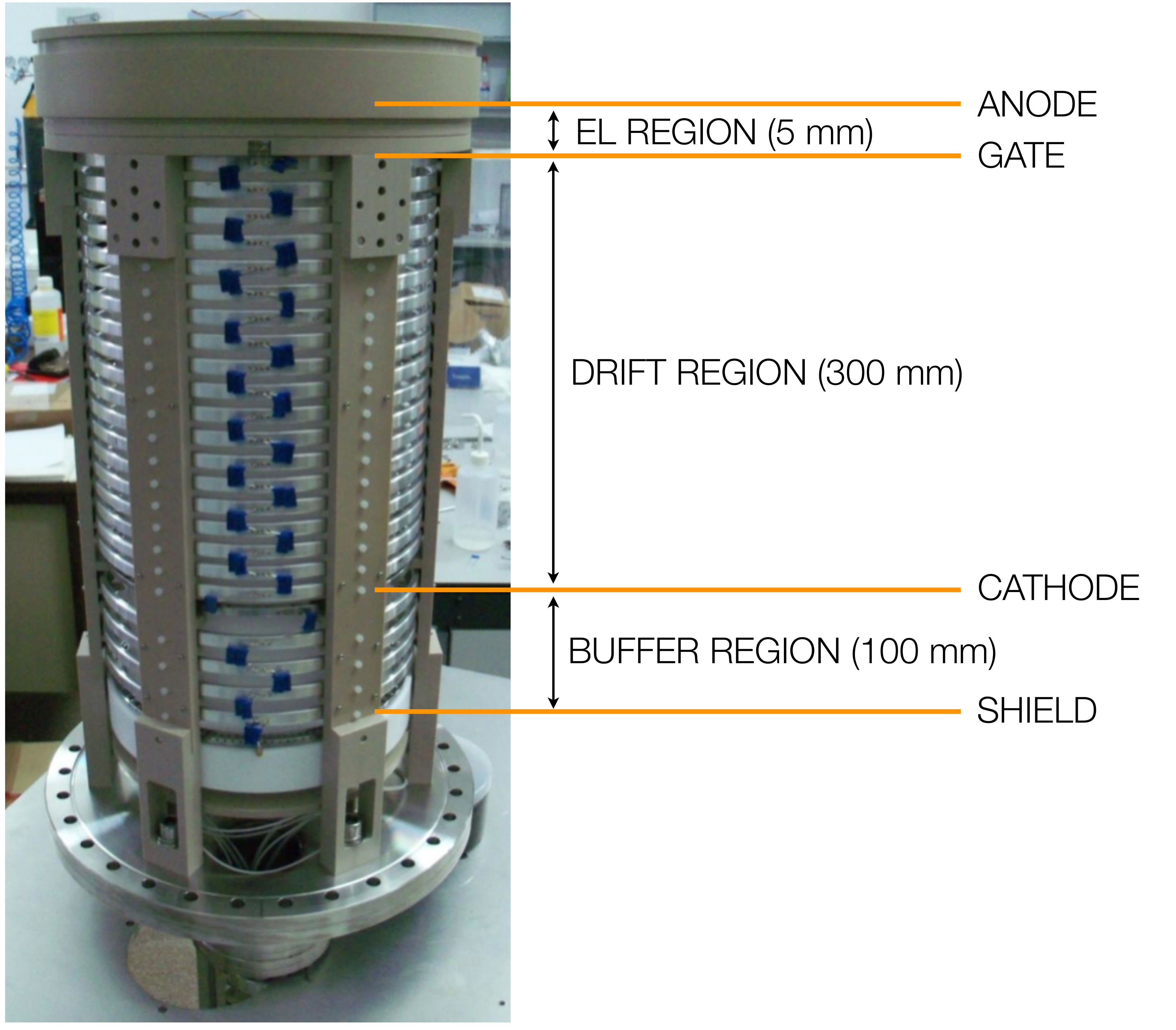}
\caption{External view of the time projection chamber mounted on one end-cap. The approximate positions of the different regions of the TPC are indicated.} \label{fig:TPC}
\end{figure}

The time projection chamber itself is shown in Figure \ref{fig:TPC}.
Three metallic wire grids --- called \emph{cathode}, \emph{gate} and \emph{anode} --- define the two active regions: the 30-cm long \emph{drift region}, between cathode and gate; and the 0.5-cm long \emph{EL region}, between gate and anode. 
The electric field is created by supplying a large negative voltage to the cathode, then degrading it using a series of metallic rings of 30 cm diameter spaced 5 mm and connected via 5~G$\Omega$ resistors.  The gate is at negative voltage so that a moderate electric field --- typically of 2.5 to 3 $\mathrm{kV~cm^{-1}~bar^{-1}}$ --- is created between the gate and the anode, which is at ground.
A \emph{buffer region} of 10 cm between the cathode and the energy plane protects the latter from the high-voltage by degrading it safely to ground potential.

\begin{figure}
\centering
\includegraphics[scale=.7]{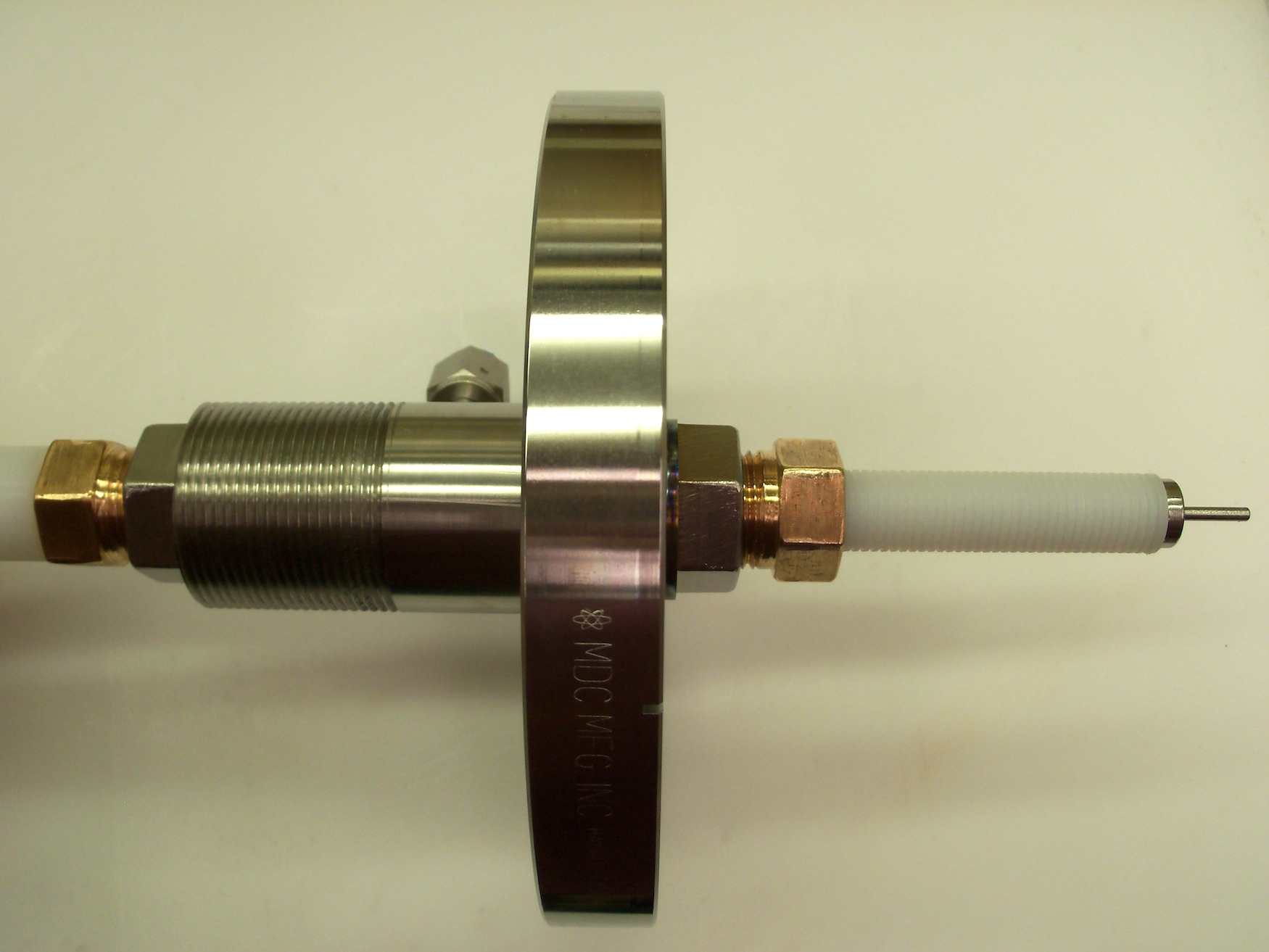} 
\caption{The NEXT-DEMO high-voltage feed-through, designed and built by Texas A\&M.}
\label{fig:HVFT}
\end{figure}

The high voltage is supplied to the cathode and the gate through custom-made high-voltage feed-throughs (HVFT), shown in Figure~\ref{fig:HVFT}, built pressing a stainless-steel rod into a Tefzel (a plastic with high dielectric strength) tube, which is then clamped using plastic ferrules to a CF flange. They have been tested to high vacuum and 100 kV without leaking or sparking. 

\begin{figure}
\centering
\includegraphics[height=5.75cm]{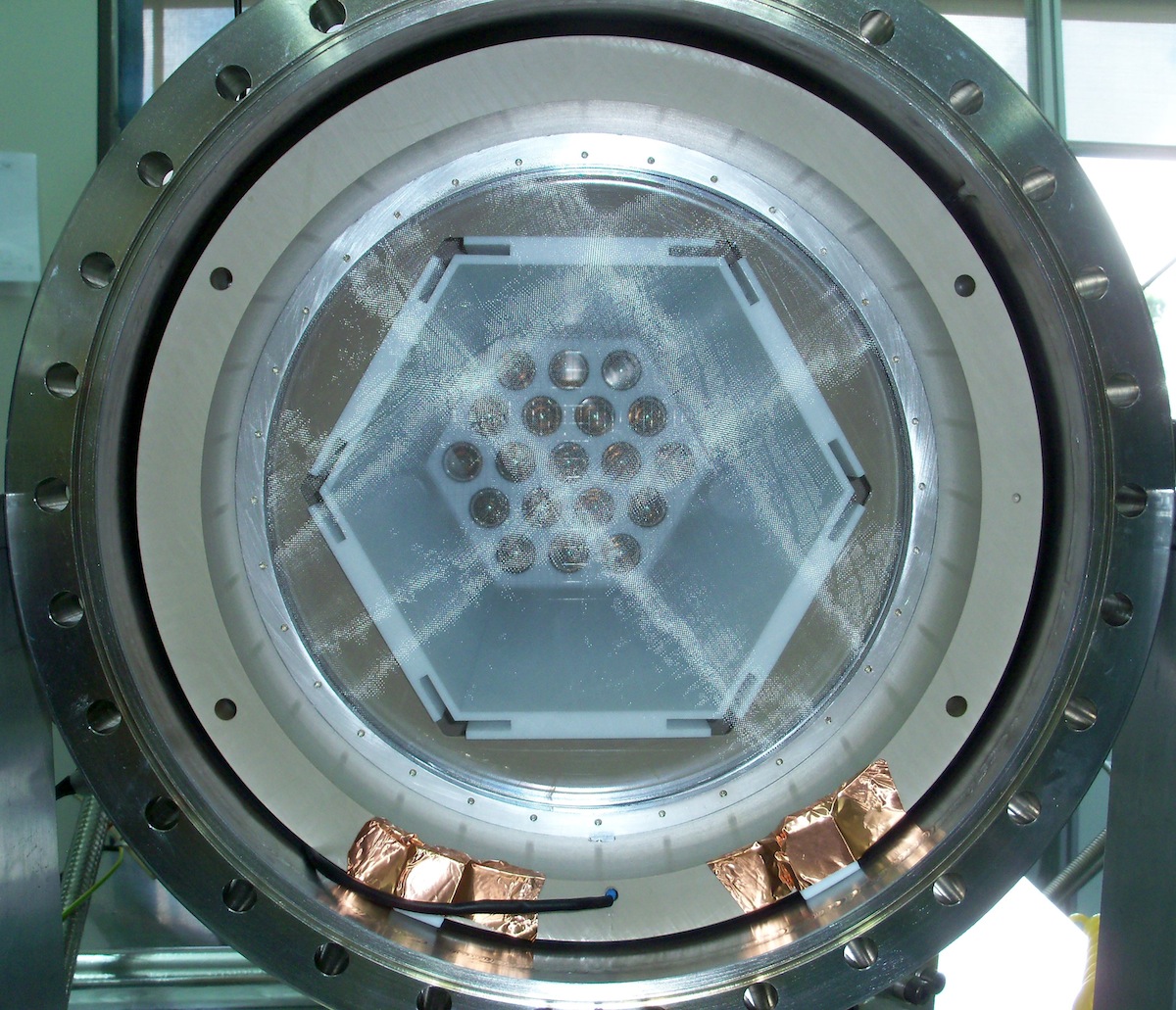}
\includegraphics[height=5.75cm]{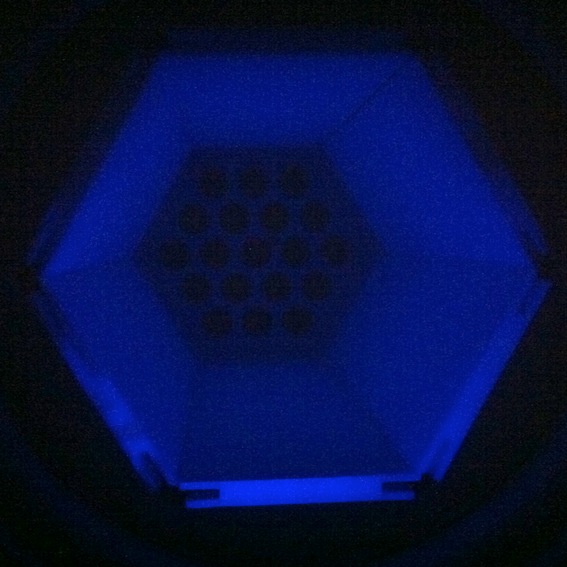}  
\caption{View of the light tube from the position of the tracking plane. Left: The meshes of the EL region can be seen in the foreground, and in the background, at the end of the light tube, the PMTs of the energy plane are visible. Right: The light tube of NEXT-DEMO illuminated with a UV lamp after being coated with TPB.} \label{fig:LightTube}
\end{figure}

A set of six panels made of polytetrafluoroethylene (PTFE)  are mounted inside the electric-field cage forming a \emph{light tube} of hexagonal cross section (see Figure~\ref{fig:LightTube}) with an apothem length of 8 cm. PTFE is known to be an excellent reflector in a wide range of wavelengths \cite{Silva:2009ip}, thus improving the light collection efficiency of the detector. In a second stage, the panels were vacuum-evaporated with TPB --- which shifts the UV light emitted by xenon to blue ($\sim430$~nm) --- in order to study the  improvement in reflectivity and light detection. Figure~\ref{fig:LightTube} (right panel) shows the light tube illuminated with a UV lamp after the coating.

Six bars manufactured from PEEK, a low outgassing plastic, hold the electric-field cage and the energy plane together. The whole structure is attached to one of the end-caps using screws, and introduced inside the vessel with the help of a rail system. All the TPC structures and the HVFT were designed and built by Texas A\&M.

\begin{figure}
\centering
\includegraphics[scale=.7]{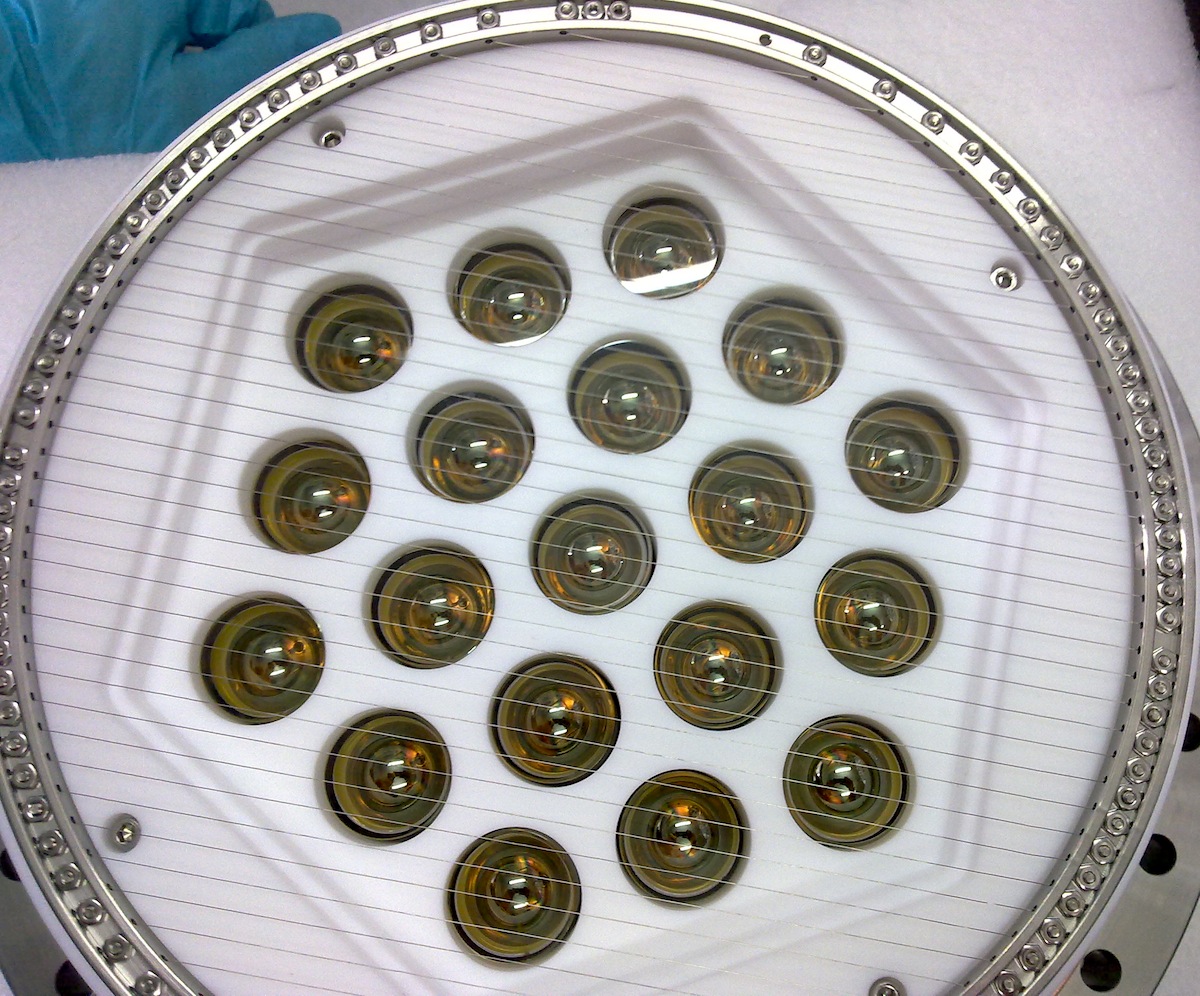}
\caption{The energy plane of NEXT-DEMO, equipped with 19 Hamamatsu R7378A PMTs.} \label{fig:DetPlanes}
\end{figure}

The energy plane (see Figure~\ref{fig:DetPlanes}) is equipped with 19 Hamamatsu R7378A photomultiplier tubes. These are 1-inch, pressure-resistant (up to 20 bar) PMTs with acceptable quantum efficiency ($\sim$~15\%) in the VUV region and higher efficiency at TPB wavelengths ($\sim$~25\%). The resulting photocathode coverage of the energy plane is about 39\%. The PMTs are inserted into a PTFE holder following a hexagonal pattern. A grid, known as \emph{shield} and similar to the cathode but with the wires spaced 0.5~cm apart, is screwed on top of the holder and set to $\sim$ 500 V. As explained above, this protects the PMTs from the high-voltage set in the cathode, and ensures that the electric field in the 10-cm buffer region is below the EL threshold.

\begin{figure}
\centering
\includegraphics[width=0.7\textwidth]{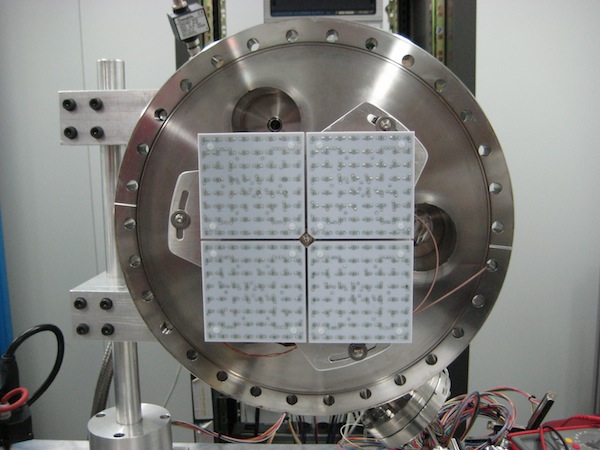}
\caption{Dice Boards installed in NEXT-DEMO, containing 64 (8$\times$8) MPPCs each. There will be about 100 such boards in NEXT-100.} \label{fig.db2}
\end{figure}

The initial operation of NEXT-DEMO implemented a tracking plane made of 19 pressure-resistant photomultipliers, identical to those used in the energy plane but operated at a lower gain. Instrumenting the tracking plane with PMTs  during this period simplified the initial commissioning, debugging and operation of the detector due to the smaller number of readout channels (19 PMTs in contrast to the 256 SiPMs currently operating in the tracking plane) and their intrinsic sensitivity to the UV light emitted by xenon. Since October 2012, NEXT-DEMO has been operating with a full tracking plane made with SiPMs, as shown in Figure \ref{fig.db2}. It consists of four boards, containing 8$\times$8 SiPMs each, 1-cm spaced. Its higher granularity allows for a finer position reconstruction in the plane orthogonal to the drift axis, thus increasing the fiducial volume of the chamber. SiPMs are not sensitive to VUV light, but they are to blue light, therefore they had to be coated with TPB. The isotropical light emission of TPB, together with an improvement of the reflectivity of PTFE for wavelengths in the blue range, produced an overall increase of the collected light.

\begin{figure}
\centering
\includegraphics[width=0.495\textwidth]{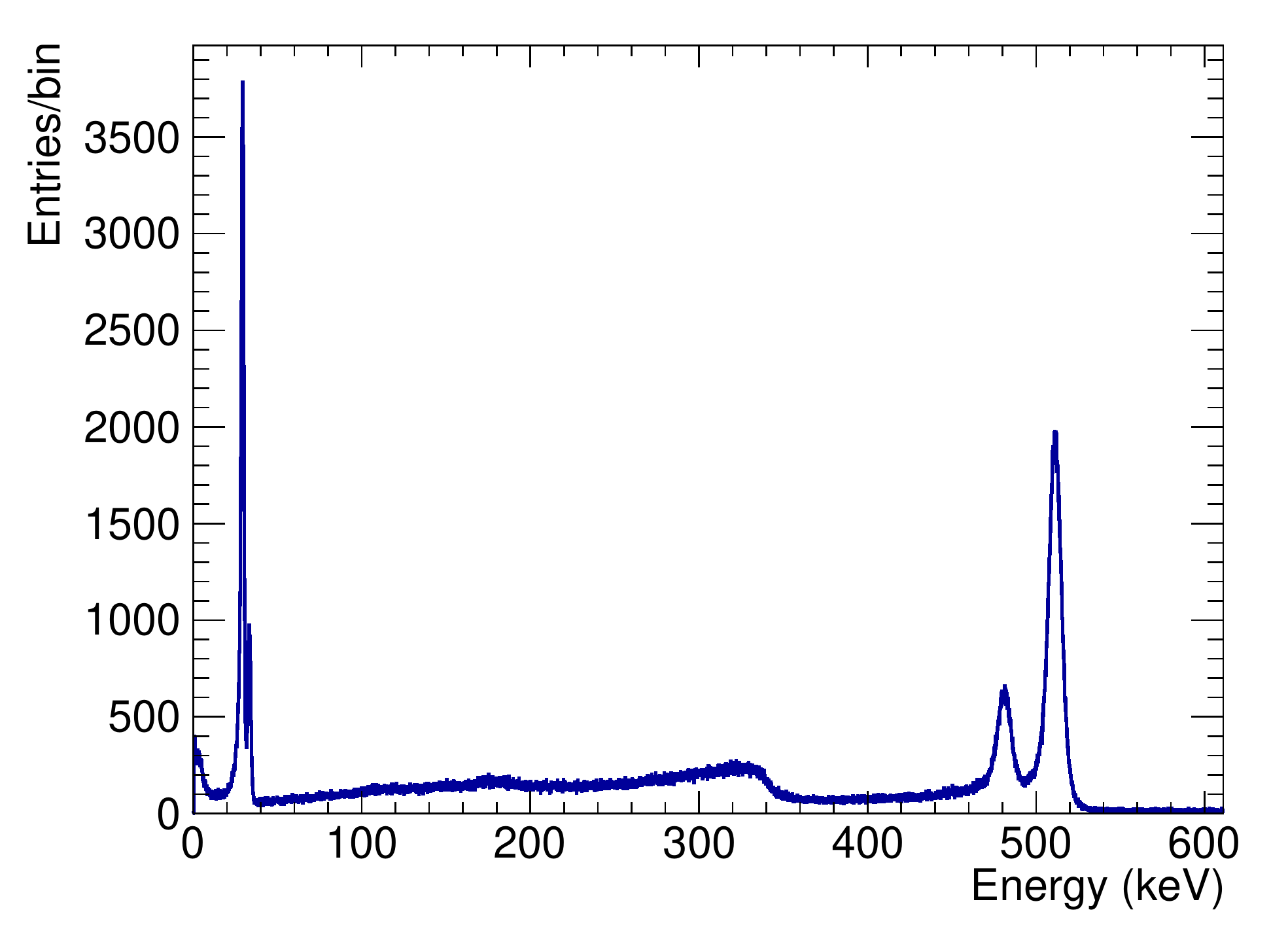}
\includegraphics[width=0.495\textwidth]{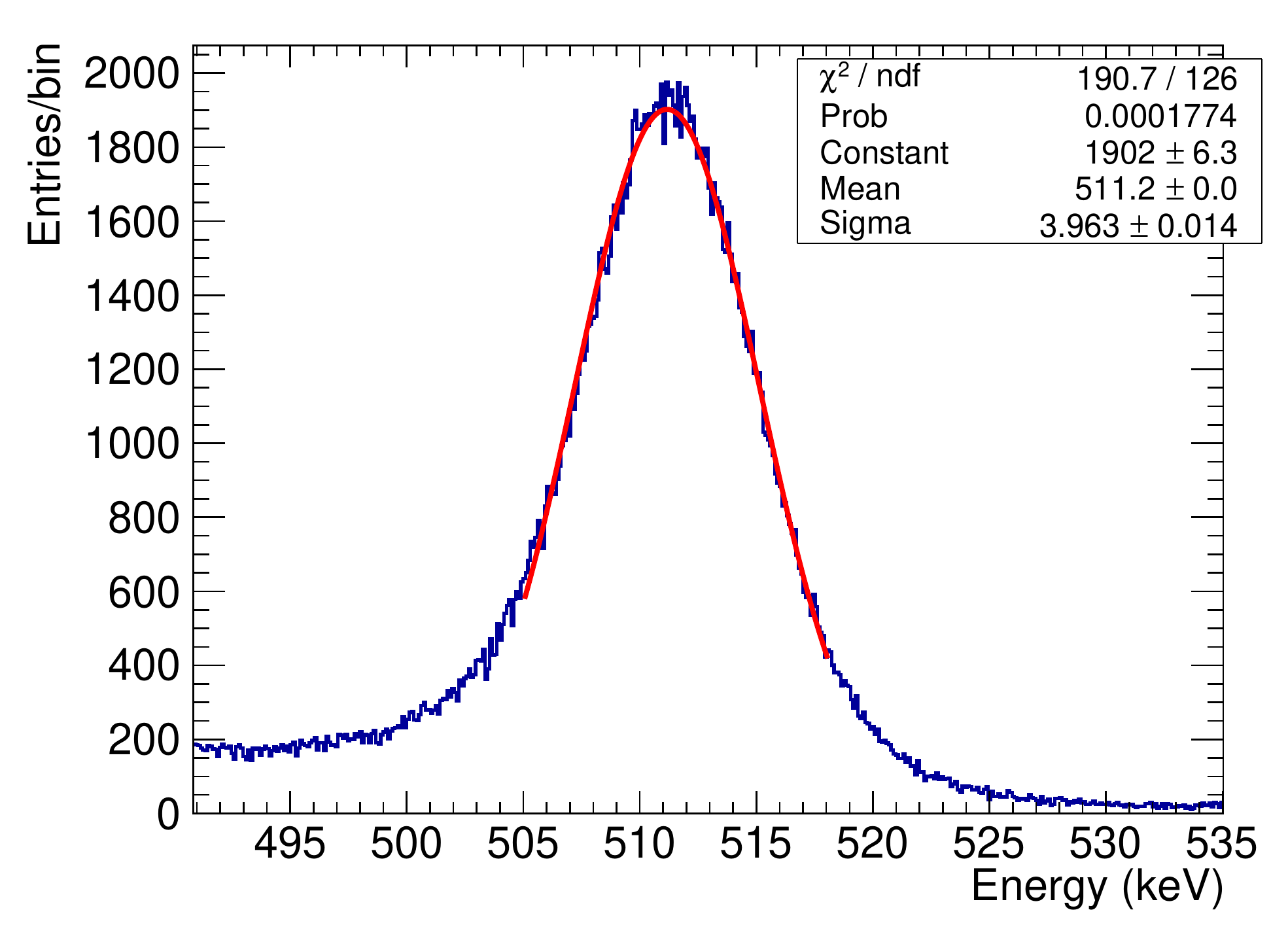}
\caption{Energy spectra for $^{22}$Na gamma-ray events within the fiducial volume of NEXT-DEMO. Left: the whole spectrum. Right: zoom in the photoelectric peak \cite{Alvarez:2013gxa}.} \label{fig:FinalSpectrum}
\end{figure}

Figure~\ref{fig:FinalSpectrum} shows the measured energy spectrum of  511-keV gamma rays from \NA\ in the fiducial volume of NEXT-DEMO. A gaussian fit to the photoelectric peak indicates an energy resolution of 1.82\% FWHM. Extrapolating the result to the $Q$ value of \XE\ (2458~keV) assuming a $E^{-1/2}$ dependence, we obtain a resolution of 0.83\% FWHM, better than the NEXT target resolution of 1\% FWHM at \Qbb. The DEMO apparatus measures electrons in a large fiducial volume, therefore this result can be safely extrapolated to NEXT-100. We believe that an ultimate resolution of 0.5\% FWHM, as found by DBDM (see the next section), can eventually be attained. 

\begin{figure}
  \begin{center}
  \includegraphics[width=0.495\textwidth]{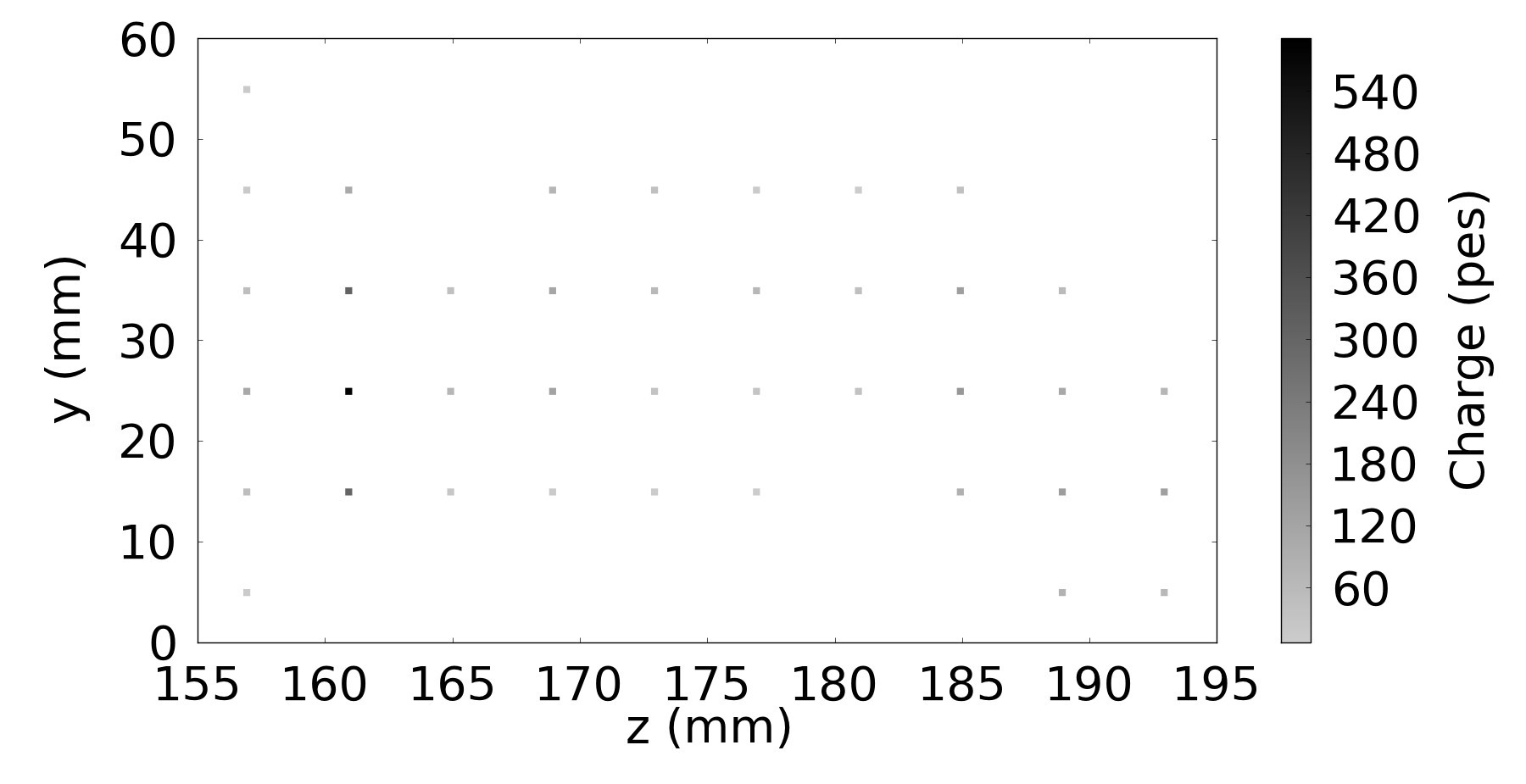}
  \includegraphics[width=0.495\textwidth]{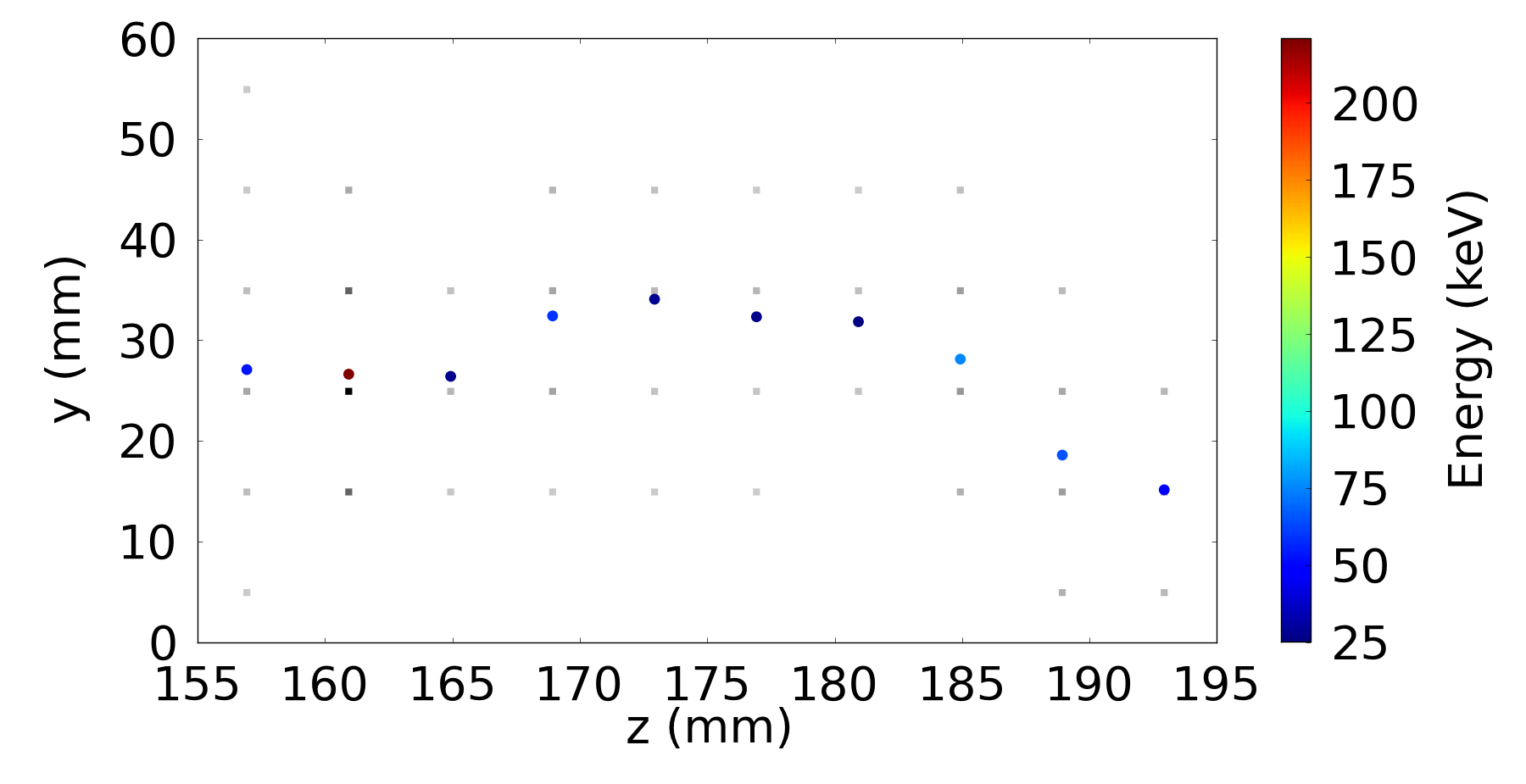}
 \includegraphics[width=0.495\textwidth]{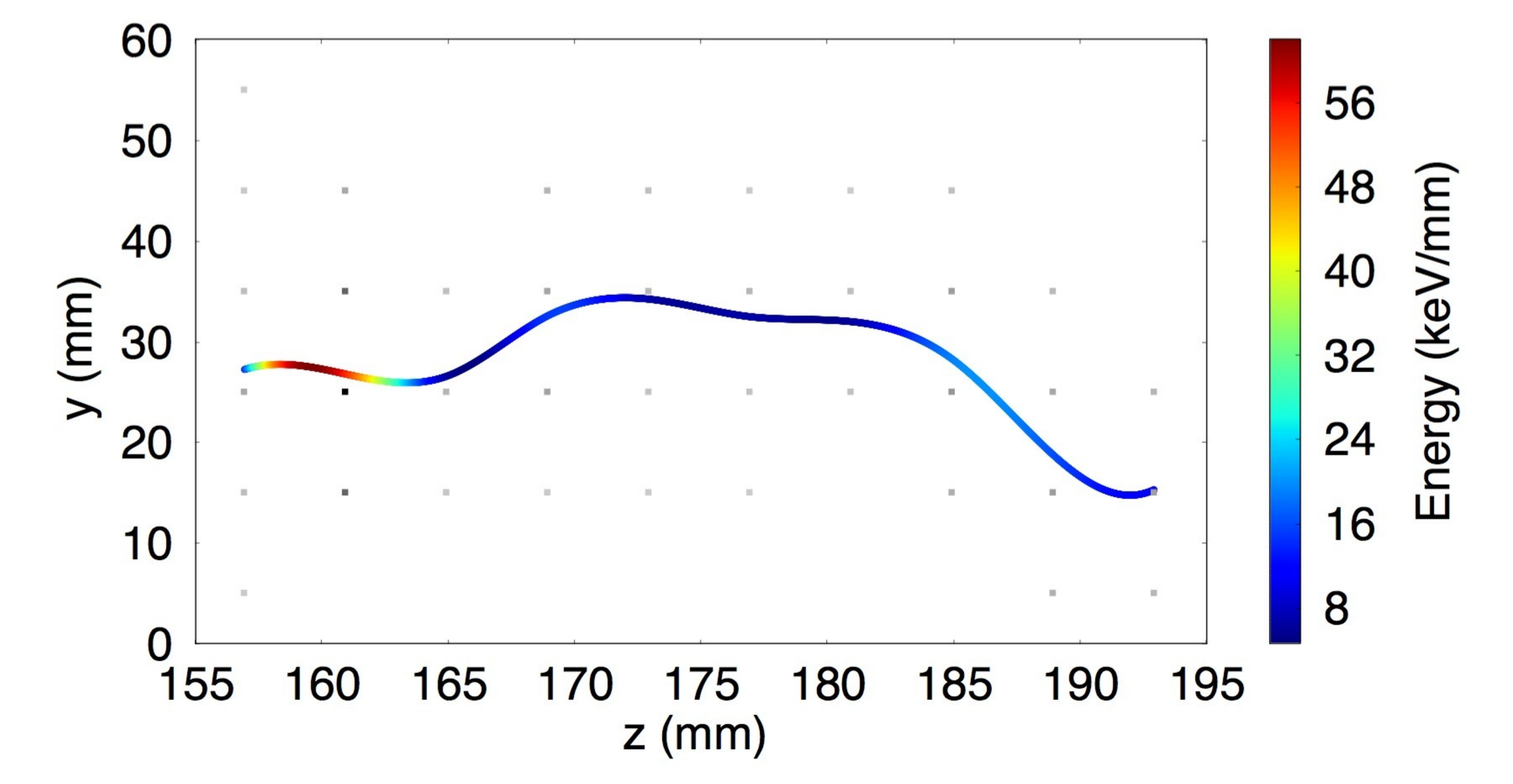}
  \includegraphics[width=0.495\textwidth]{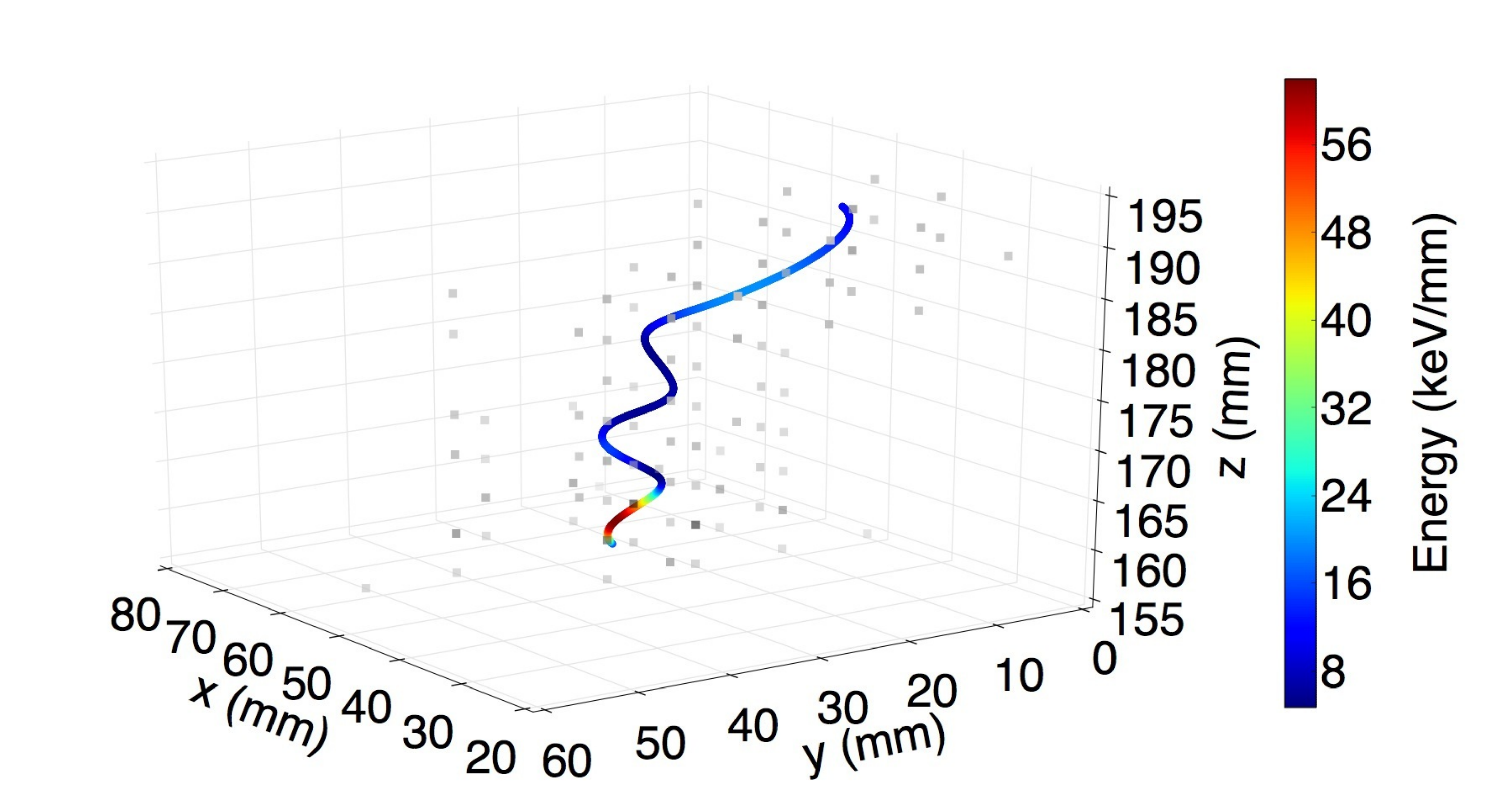}
\end{center}
  \caption{Example of the reconstruction of a  \CS\ track: The charge of the different SiPMs is split into slices of 4~mm width in $z$ (top left). One point is calculated for each slice using the barycentre method and the energy of the points is then associated with the measurement made in the cathode (top right). A cubic spline is used to interconnect the different points. The result is shown in the bottom line: YZ projection (bottom left) and 3D image (bottom right) of the reconstructed track \cite{Alvarez:2013gxa}.}
  \label{fig:Track}
\end{figure}

A first approximation of the event topology reconstruction is performed subdividing the charge in time slices (z-dimension) and reconstructing a single $xy$ point per slice. A further detailed analysis to allow the reconstruction of multiple depositions per slice is being studied. For this analysis, a slice width of 4~$\mu$s is used as it gives enough information in the tracking plane to achieve a reliable \textit{xy} reconstruction and it is also comparable to the time an electron needs to cross the EL gap.
The $xy$ position of a slice is reconstructed using the averaged position of the SiPMs with higher recorded secondary scintillation signal, weighted with their collected integrated charge. The energy associated with this position is recorded in the cathode for the same time interval, so that the $dE/dz$ of the event can be studied. The energy and position information are then used to calculate a cubic spline between the individual points in order to obtain a finer description of the path (see Figure~\ref{fig:Track}).

\begin{figure}
  \begin{center}
    \includegraphics[width=0.495\textwidth]{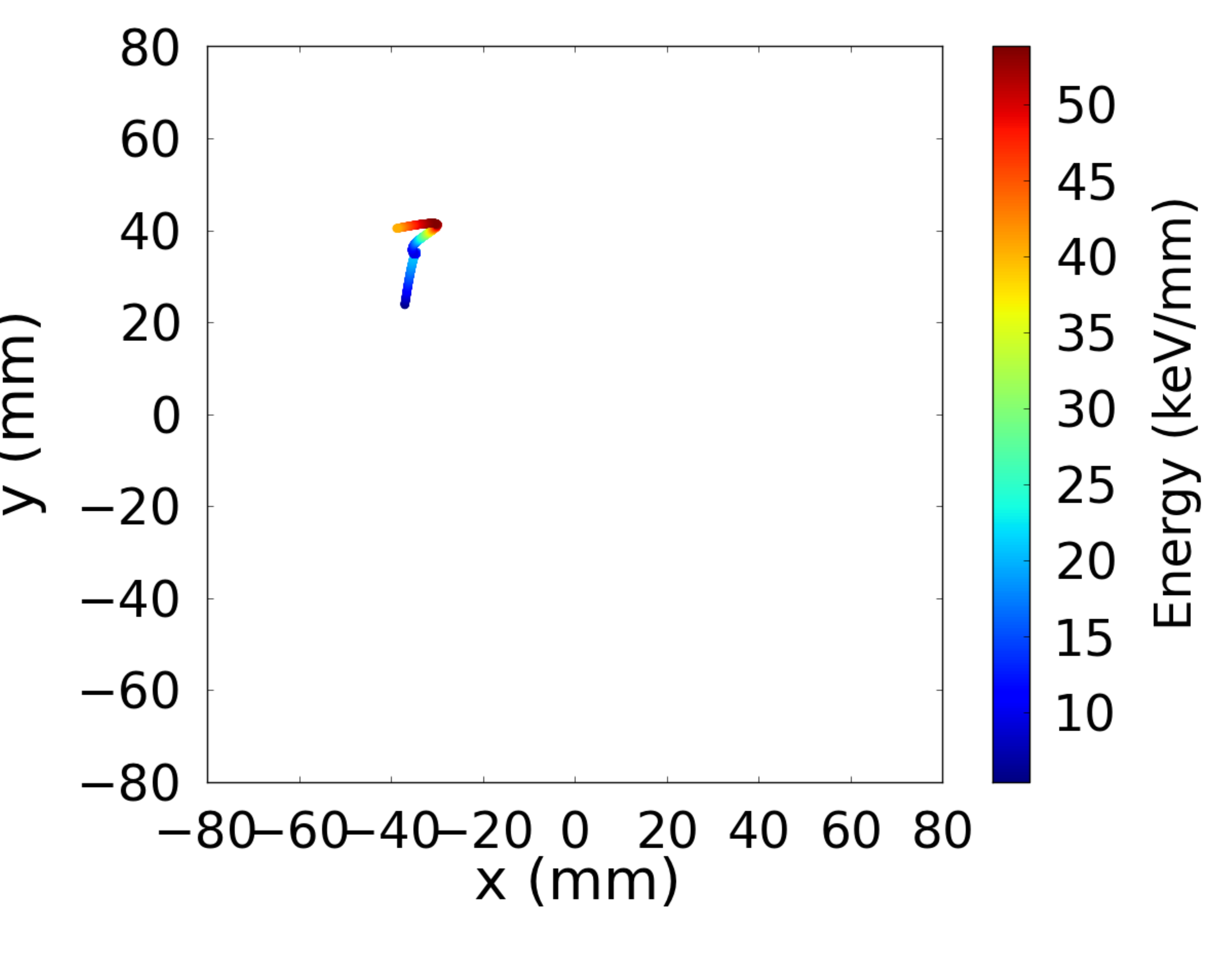}
    \includegraphics[width=0.495\textwidth]{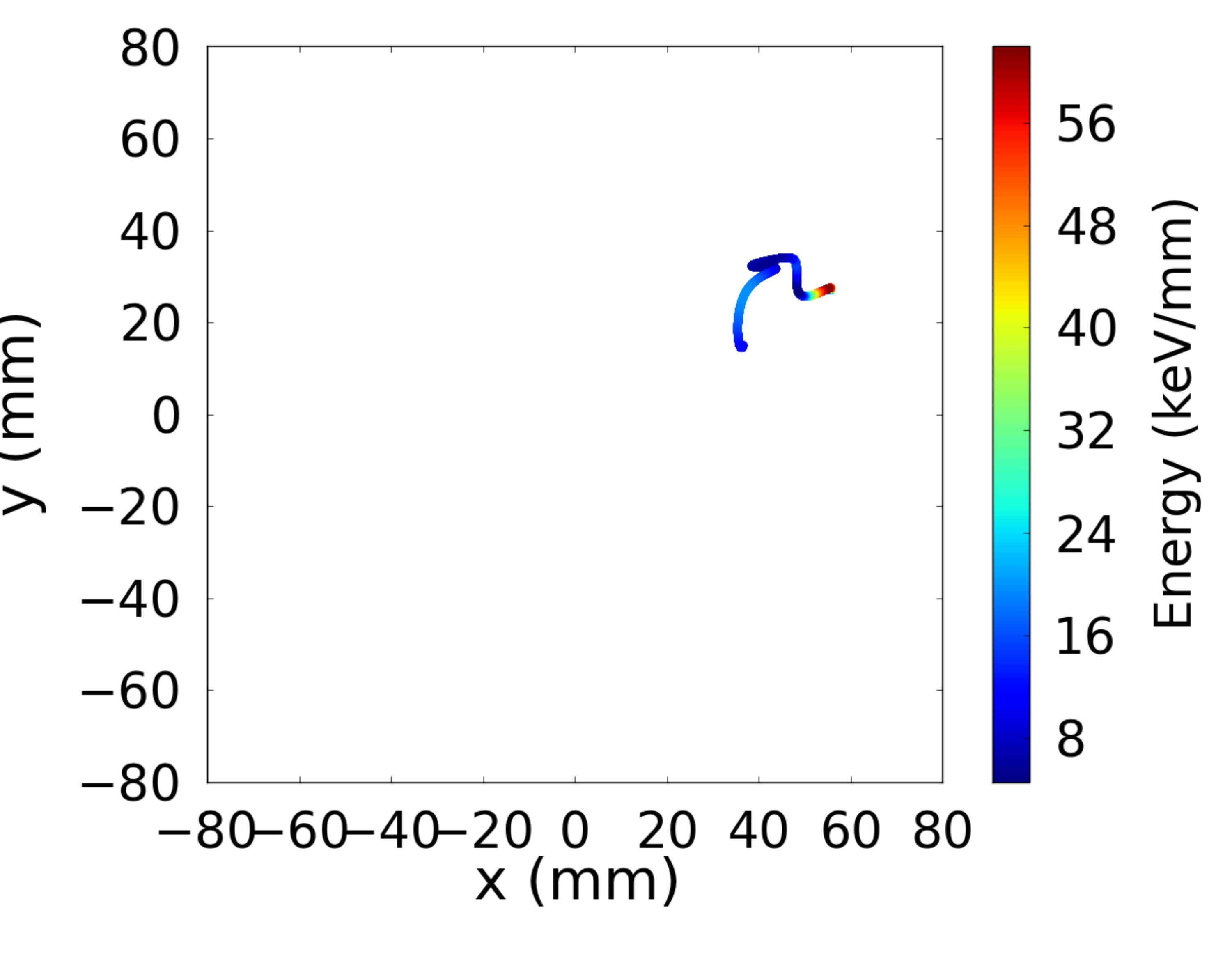}
    \includegraphics[width=0.495\textwidth]{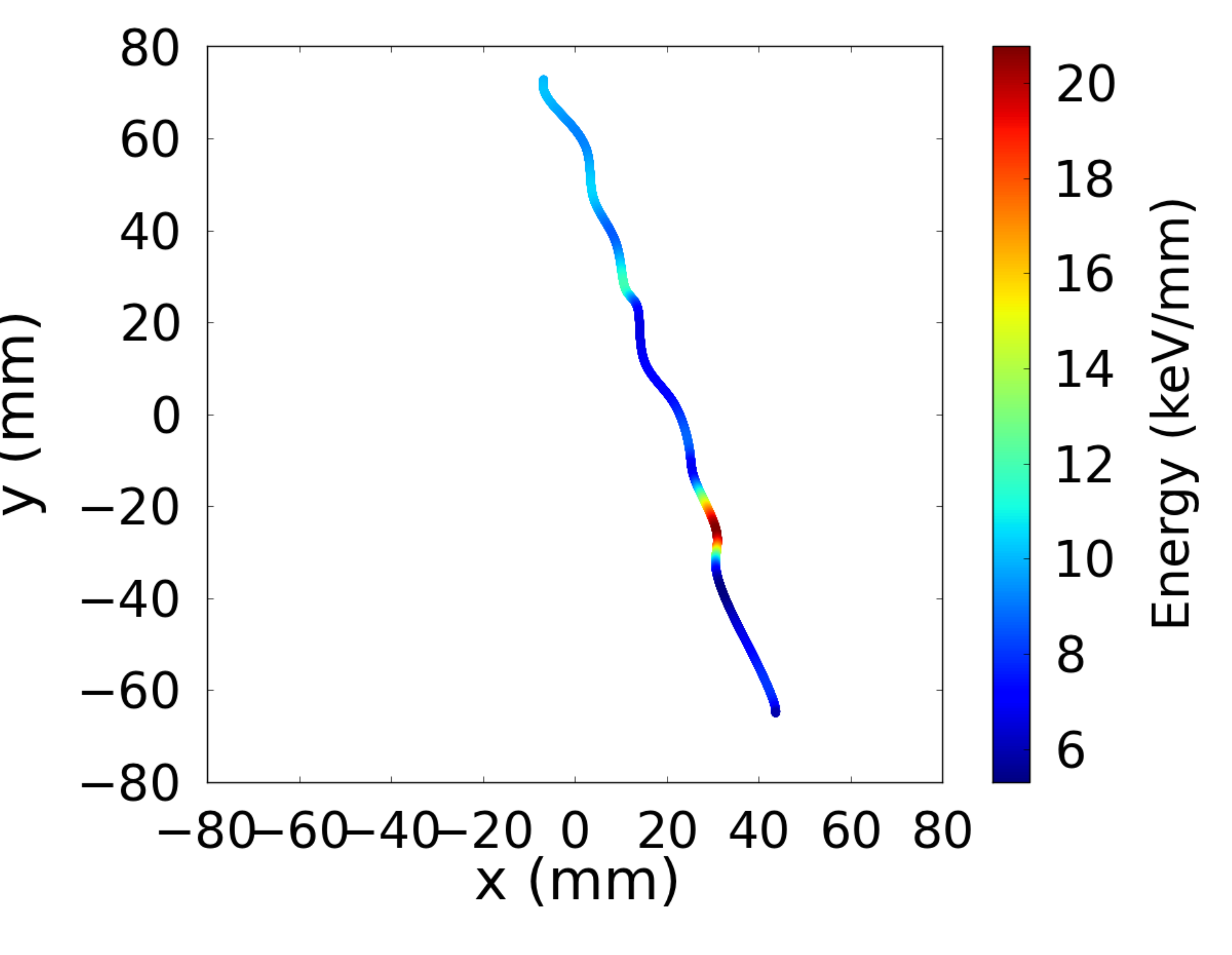}
    \includegraphics[width=0.495\textwidth]{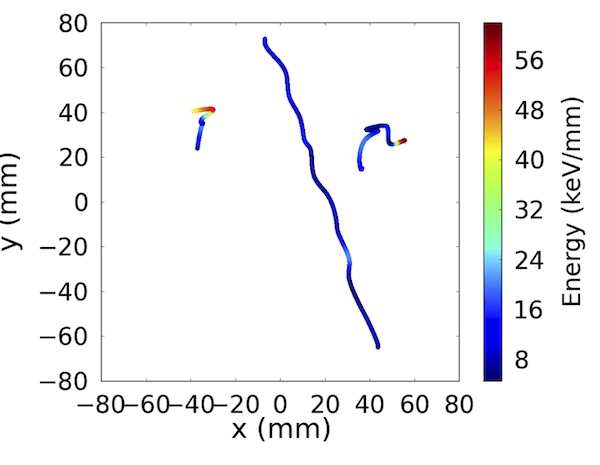}
    \caption{Examples of \NA\ (top left), \CS\ (top right) and muon (bottom left) track $xy$ plane projections. Bottom right: the three events with the same energy scale. The end point of the electron is clearly visible for \NA\ and \CS\ while the energy deposition for the muon is almost constant. Tracks reconstructed from NEXT-DEMO data \cite{Alvarez:2013gxa}.}
    \label{fig:TrackExs}
  \end{center}
\end{figure}

The first reconstructed events (Figure \ref{fig:TrackExs}) demonstrate the topology capabilities of the NEXT technology. The reconstructed electrons show a random walk into the gas with a clearly visible end-point at the end of the track with a higher energy deposition (blob). On the other hand, the reconstruction of a muon track shows a straight line through the detector with a fairly uniform energy deposition.


NEXT-DEMO has been running successfully for two years, proving perfect high voltage operation and a great stability against sparks. The gas system, completed with a hot getter, has demonstrated to be leakproof (less than 0.1$\%$ leakage per day) and has allowed a continuous recirculation and purification of the gas, which resulted in a measured electron lifetime of up to tens of milliseconds.  The light collection efficiency has been thoroughly understood, by studies of both primary and electroluminescent scintillation signals. The TPB coating on the PTFE reflectors in the drift region produced an increase in the EL light collection of a factor of 3 \cite{Alvarez:2012xda}, thus improving light statistics. Data produced with an alpha source have allowed studies of primary scintillation signals along the whole drift length, leading to a better understanding of light reflectance and loss in our detector, through the support of Monte Carlo simulations \cite{Alvarez:2012hu} .

To summarise, the NEXT-DEMO detector is operating continuously at IFIC since 2011. The current configuration, with a SiPM tracking plane, a PMT energy plane and a light tube coated with TPB, demonstrates the design chosen for the NEXT-100 detector, exercises all the technical solutions, and shows excellent energy resolution and electron reconstruction. Further work is currently in progress analysing the many millions of events acquired with the chamber.

%% file: src2/NextDBDM.tex
\subsection{NEXT-DBDM}

\begin{figure*}[!htb]
  \centering
\includegraphics[width=0.7\textwidth]{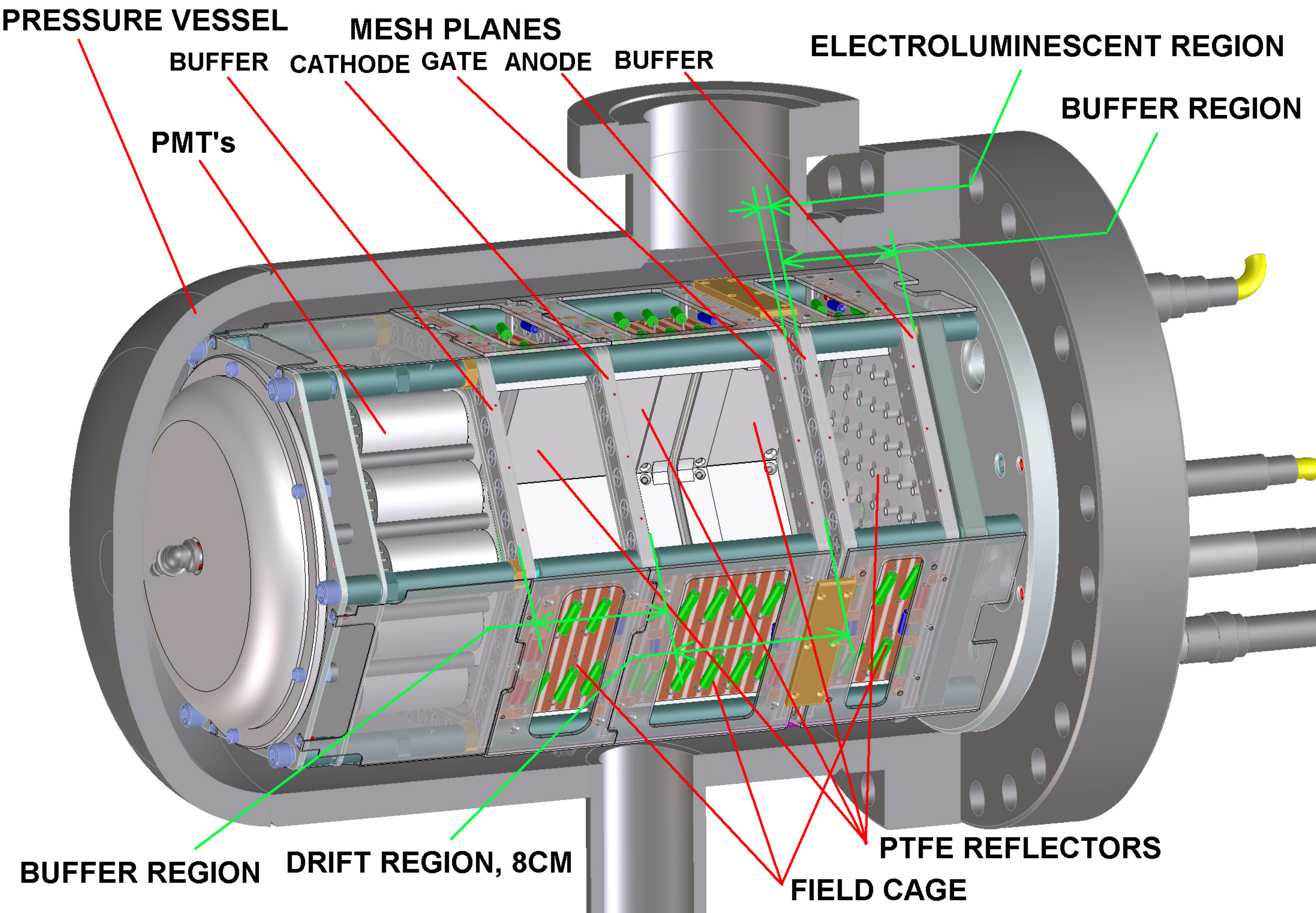}
  \caption{\label{tpc_configuration} The NEXT-DBDM electroluminescent TPC configuration: An array of 19 photomultipliers (PMTs) measures S1 
  primary scintillation light from the 8 cm long drift region and S2 light produced in the 0.5 cm electroluminescence (EL) region. Two 5 cm long buffer regions behind the EL anode mesh and between the PMTs and the cathode mesh grade the high voltages (up to 
  $\pm$17 kV) down to ground potential.}
\end{figure*}

\begin{figure}
\centering
\includegraphics[width=0.7\textwidth]{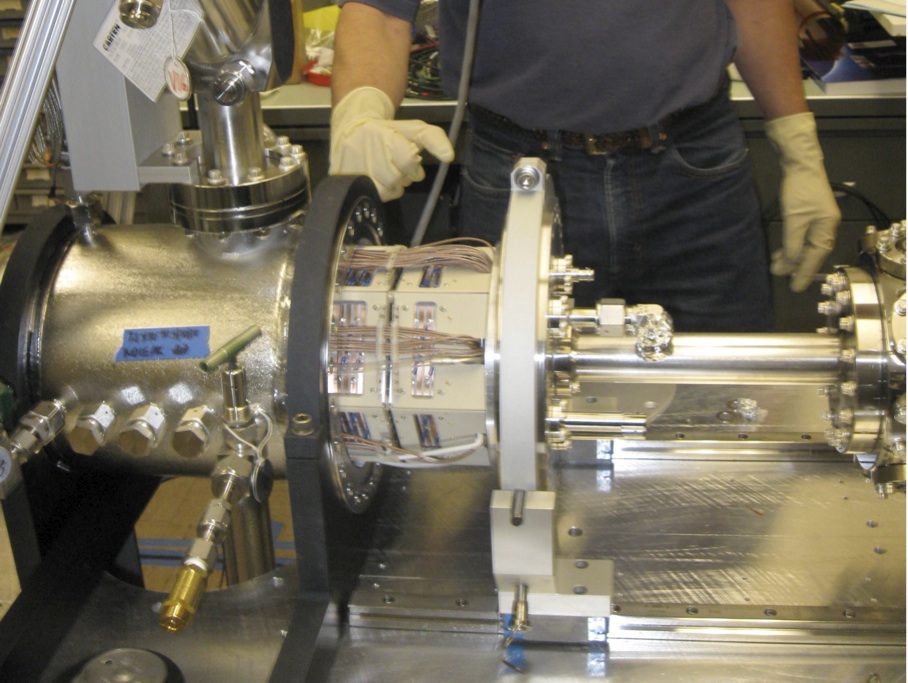}
\caption{The NEXT-DBDM prototype, operating at LBNL. Insertion of the time projection chamber into the stainless-steel pressure vessel.} \label{fig:DBDM}
\end{figure}

The basic building blocks of the NEXT-DBDM xenon electroluminescent TPC are shown in
Figures \ref{tpc_configuration} and \ref{fig:DBDM}: a stainless steel pressure vessel, a gas system 
that recirculates and purifies the xenon at 10-15 atm, stainless steel wire meshes that establish high-voltage equipotential planes in 
the boundaries of the drift and the EL regions, field cages with hexagonal cross sections to establish uniform electric fields 
in those regions,  an hexagonal pattern array of 19 VUV sensitive PMTs inside the pressure vessel and an associated readout 
electronics and data acquisition (DAQ) system.

In the NEXT-DBDM detector the PMT array and the EL region, which are both hexagonal areas with 12.4 cm between opposite 
sides, are 13.5 cm away from each other. Thus, point-like isotropic light produced in the 
EL region illuminates the PMT array with little PMT-to-PMT variation. This geometric configuration also makes the illumination pattern 
and the total light collection only very mildly dependent on the position of the light origin within the EL region. The 
diffuse reflectivity of the TPC walls increases this light collection uniformity further. As a result, the device provides 
good energy measurements with little dependence on the position of the charge depositions. On the other hand, without a light 
sensor array near the EL region precise tracking information is not available. Still, the position reconstruction achievable allows the fiducialization of pulses to select events/pulses within regions of the TPC with uniform light collection efficiencies. 

 The field configuration in the TPC is established by five stainless steel meshes with 88\% open area at a $z$ position of 
0.5 cm (cathode buffer or PMT mesh), 5.5 cm (cathode or drift start mesh), 13.5 cm (field transition or EL-start mesh), 14.0 cm (anode or EL-end mesh) and 19.0 cm (anode buffer or ground mesh) from the PMT windows. Electroluminescence occurs between 13.5 and 14.0 cm. The meshes are supported
and kept tense by stainless steel frames made out of two parts and tensioning screws on the perimeter.   
The TPC side walls, made out of 18 individual rectangular assemblies 7.1 cm wide (and 5 and 8 cm long) connecting adjacent 
meshes (except around the 0.5 cm EL gap), serve the dual purpose of light cage and field cage.
Each side wall assembly is made of a 0.6 cm thick PTFE panel and a ceramic support panel. The PTFE panels are bare on the 
side facing the active volume and have copper stripes  parallel to the mesh planes every 0.6 cm on the other side. The bare PTFE 
serves as reflector for the VUV light. Adjacent copper stripes are linked with 100 M$\Omega$ resistors to grade the potential and 
produce a uniform electric field. The ceramic support panels are connected, mechanically and electrically, to the outer 
perimeter of the mesh support frames and to the first and last copper stripes on their corresponding PTFE panel. High voltage 
connections to establish the TPC fields (HHV) are made directly to the mesh frames.          

\begin{figure}[!htb]
  \centering
  \includegraphics[width=0.7\textwidth]{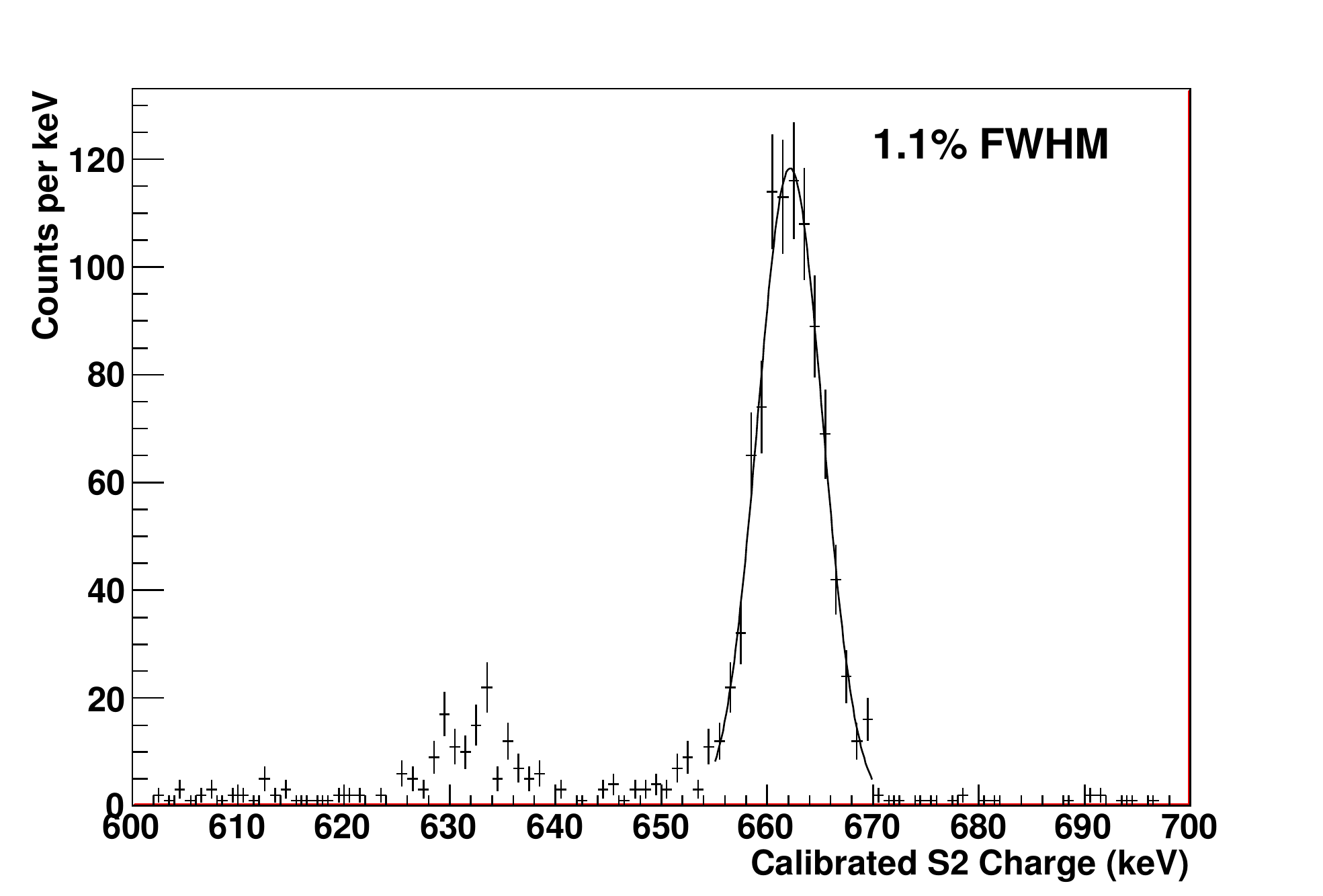}
  \caption{\label{cs137_resolution_10atm} Energy resolution at 10 atm for 662 keV gamma rays \cite{Alvarez:2012hh}.}
\end{figure}

In Figure \ref{cs137_resolution_10atm} the energy spectrum in the 662 keV full energy region obtained at 10 atm is shown. A 1.1\% FWHM energy 
resolution was obtained for events reconstructed in the central 0.6 cm radius region. A small drift-time dependent correction for attachment 
losses with $\tau$ = 13.9 ms was applied. The xenon X-ray escape peak is clearly visible, $\sim$30 keV below the main peak. For the spectrum 
taken at 15 atm a 1\% FWHM resolution was obtained.  This resolution extrapolates to 0.52\% FWHM at 
$Q_{\beta\beta}$=2.458 MeV if the scaling follows a statistical 1/$\sqrt{E}$ dependence and no other systematic effect dominates. 

\begin{figure}[!htb]
  \centering
  \includegraphics[width=0.7\textwidth]{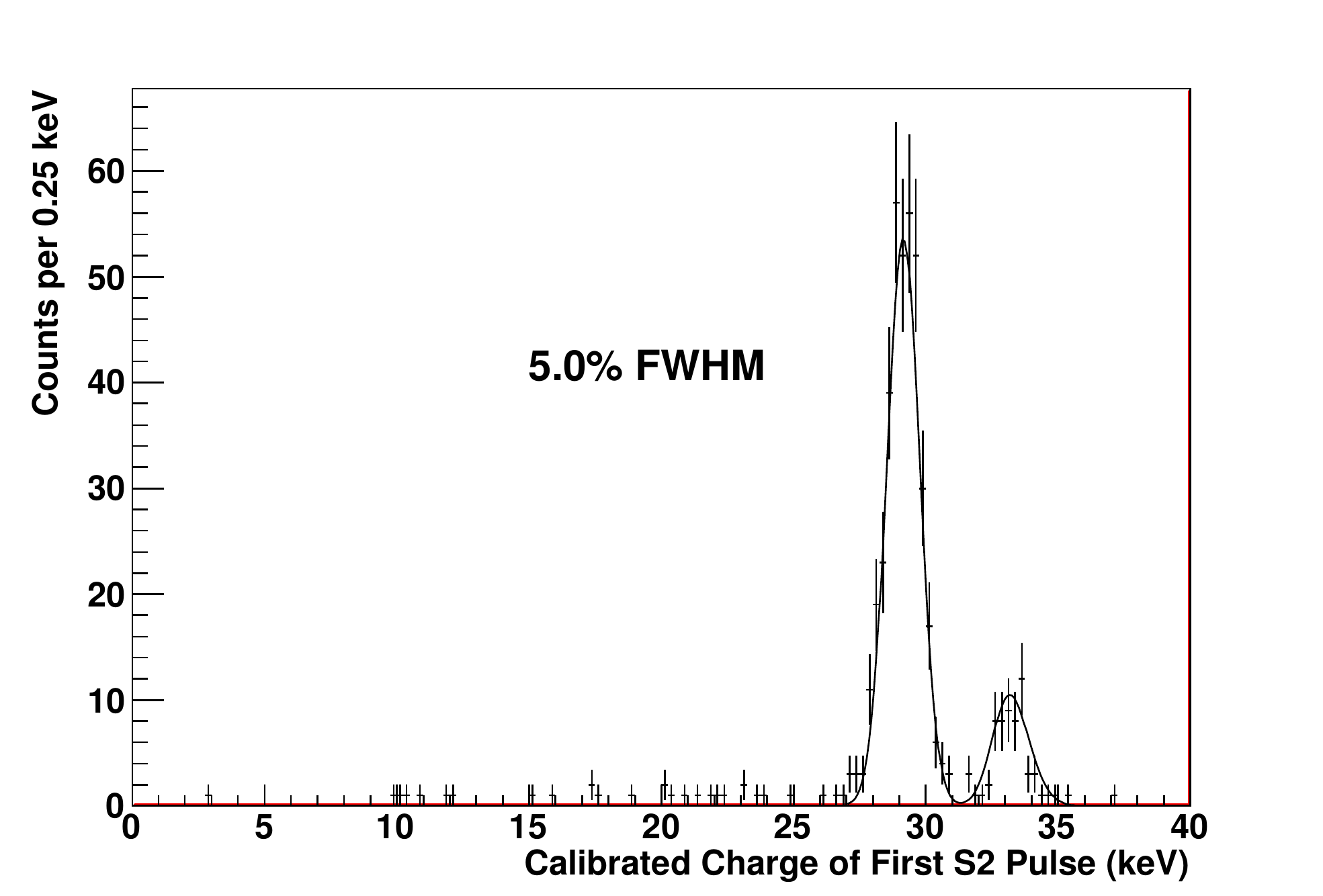}
  \caption{\label{xray_resolution} Energy resolution at 10 atm for 30 keV xenon X-rays \cite{Alvarez:2012hh}.
  }
\end{figure}

In order to study the EL TPC energy resolution at lower energies, full energy 662 keV events that had a well separated X-ray pulse reconstructed in the central 1.5 cm radius region were used.  
Figure \ref{xray_resolution} shows the energy spectrum obtained at 10 atm with a 5\% FWHM resolution. 

\begin{figure*}[!htb]
  \centering
  \includegraphics[width=0.7\textwidth]{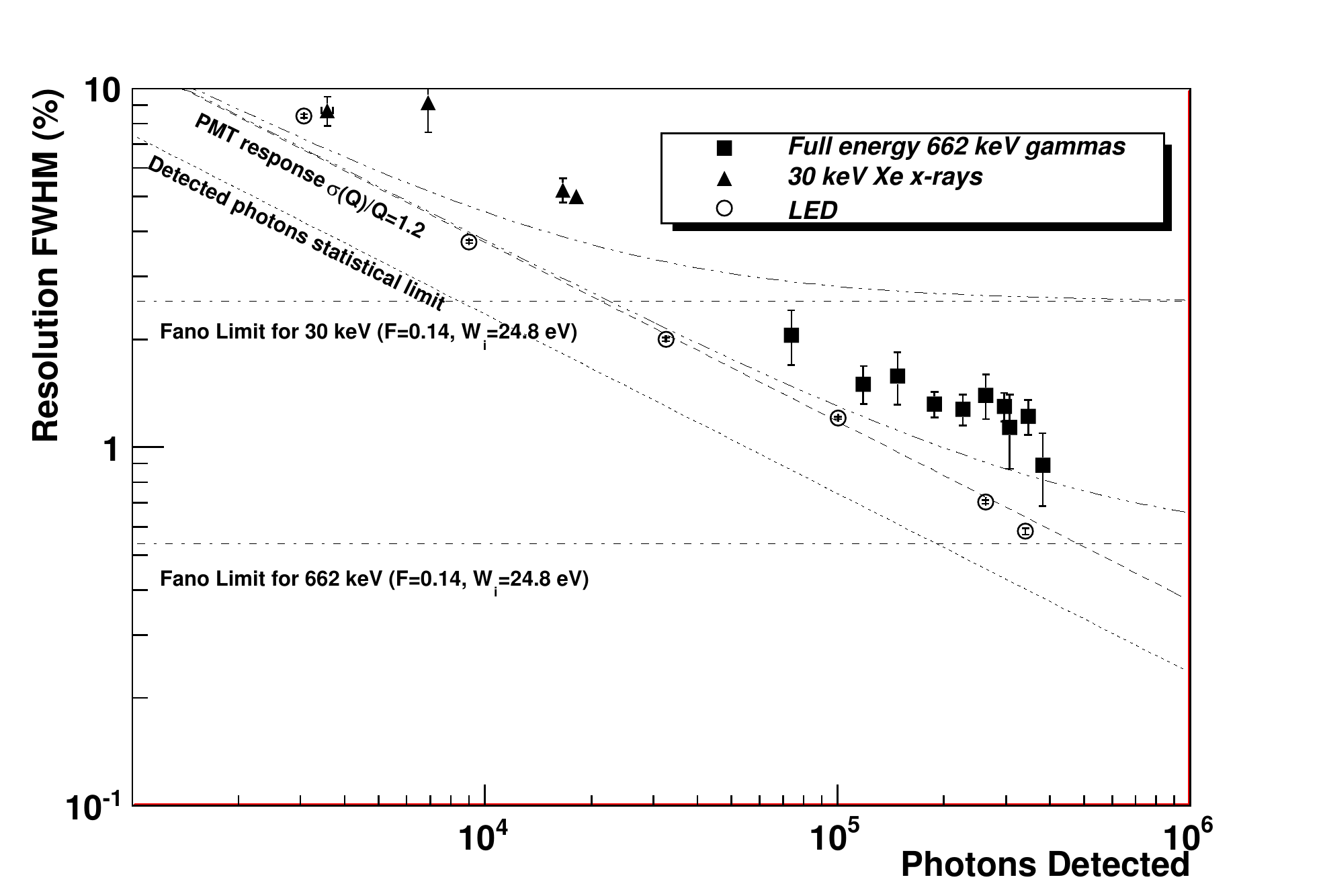}
  \caption{\label{resolution_summary} Energy resolution in the high-pressure xenon NEXT-DBDM electroluminescent TPC: Data points show the 
  measured energy resolution for 662 keV gammas (squares), $\sim$30 keV xenon X-rays (triangles) and LED light pulses (circles) as a function of the 
  number of photons detected. The expected resolution including the intrinsic Fano factor, the statistical fluctuations in the number of detected 
  photons and the PMT charge measurement variance is shown for  X-rays (dot dot dashed) and for 662 keV gammas (dot dot dot dashed). Resolutions 
  for the 662 keV peak were obtained from 15.1 atm data runs while X-ray resolutions we obtained from 10.1 atm runs \cite{Alvarez:2012hh}. }
\end{figure*}

Figure \ref{resolution_summary} summarizes our measurements and understanding of the EL TPC energy resolution. The lower diagonal line represents 
the Poisson statistical limit from the measurement of a small fraction of the photons produced by the EL gain while the upper diagonal line 
includes the degradation (mostly from PMT afterpulsing) due to PMT response. The circle data points show the energy resolutions obtained for 
dedicated LED runs with varying light intensities per LED pulse. The LED points follow the expected resolution over the two decades range studied. 
The two horizontal lines represent the xenon gas nominal intrinsic resolution for 30 and 662 keV, respectively, and the two curved lines are the 
expected EL TPC resolutions with contributions from the intrinsic limit and the photons' measurement. Our 662 keV data (squares) and xenon X-ray 
data (triangles) taken with various EL gains follow the expected functional form of the resolution but are 20-30\% larger possibly due to the $x-y$ 
response non-uniformity. Detailed track imaging from a dense photosensor array near the EL region, such as the one recently commissioned for 
the NEXT-DBDM prototype, will enable the application of $x-y$ position corrections to further improve the energy measurement.    
 
